\documentclass[reprint,
amsmath,amssymb,aps,
prb,
]{revtex4-2}

\usepackage{graphicx}% Include figure files
\usepackage{dcolumn}% Align table columns on decimal point
\usepackage{bm}% bold math
\usepackage{mathtools}
\usepackage{amsmath}
\usepackage{amssymb}
\usepackage{dsfont}
\usepackage{fancyhdr}
\usepackage{svg}

\usepackage[dvipsnames]{xcolor}
\usepackage[hidelinks]{hyperref}
\hypersetup{colorlinks=true, linkcolor=RoyalBlue, urlcolor=RoyalBlue, citecolor=RoyalBlue}

\usepackage[most]{tcolorbox}
\tcbuselibrary{skins,breakable}
\newtcolorbox{mybox}[2][]{breakable,sharp corners, skin=enhancedmiddle jigsaw,parbox=false,
boxrule=0mm,leftrule=1mm,boxsep=0mm,arc=0mm,outer arc=0mm,attach title to upper,
after title={.\ }, coltitle=black,colback=gray!10,colframe=black, title={#2},
fonttitle=\bfseries,#1}
\newcommand{\U}{\mathrm{U}}
\newcommand{\A}{\mathcal{A}}

\newcommand{\F}{\mathcal{F}}

\renewcommand{\L}{\mathcal{L}}

\renewcommand{\O}{\mathcal{O}}

\newcommand{\T}{\mathcal{T}}

\renewcommand{\vec}[1]{\boldsymbol{\mathbf{#1}}}
\newcommand{\angled}[1]{\left\langle #1 \right\rangle}
\newcommand{\inv}{^{-1}}
\usepackage[shortlabels]{enumitem}
\newcommand{\eps}{\varepsilon}
\renewcommand{\phi}{\varphi}

\newcommand{\1}{\mathbb{1}}
\renewcommand{\d}{\mathrm{d}}

\newcommand{\kb}{k_\mathrm{B}}

\usepackage{import}

\begin{document}

\preprint{APS/123-QED}

\title{Shot noise in strongly correlated double quantum spin Hall edges}% Force line breaks with \\

\author{Andreas Tsantilas}
\email{atsant@stanford.edu}

\author{Trithep Devakul}
% \email{tdevakul@stanford.edu }

\author{Julian May-Mann}%
\email{maymann@stanford.edu}

\affiliation{%
Department of Physics, Stanford University, Stanford, California 94305, USA
}%

\begin{abstract}
We consider the effects of interactions on the edges of ``double" quantum spin Hall insulators (DQSHIs), motivated by recent experiments on moir\'e twisted metal dichalcogenides. Without interactions, a DQSHI can be understood as two copies of a conventional quantum spin Hall insulator. If interactions are present and $s^z$-spin is conserved, we show that there are two possible phases for the DQSHI edge. First is a weakly correlated edge which has two pairs of helical modes and is adiabatically equivalent to two conventional quantum spin Hall edges. Second is a strongly correlated edge with only one pair of helical modes. The strongly correlated edge also has a gap to single electrons, but is gapless to pairs of electrons. In a quantum point contact geometry, this single-electron gap leads to a Fano factor of $2$ in shot noise measurements, compared to a Fano factor of $1$ for a weakly correlated edge.

\end{abstract}

\maketitle

%\tableofcontents

\section{\label{intro}Introduction.}

Quantum spin Hall insulators (QSHIs)~\cite{konig2008quantum, sinova2004universal, kane2005quantum, bernevig2006quantum, bernevig2006quantum2,konig2007quantum} are prototypical examples of symmetry-protected topological phases in two dimensions. The edges of QSHIs characteristically host one-dimensional helical modes~\cite{wu2006helical, giamarchi2003quantum, haldane1981luttinger}, which are guaranteed to be gapless as long as time-reversal symmetry ($\mathsf{T}$-symmetry) is present and/or the $s^z$-component of spin is conserved. 
If only $\mathsf{T}$-symmetry is present, QSHIs have a $\mathbb{Z}_2$ topological index~\cite{kane2005z,kitaev2009periodic,chiu2016classification}, such that two copies of a QSHI form a trivial insulator. %This can be directly understood from the fact that two pairs of helical modes can be gapped out without breaking $\mathsf{T}$-symmetry. 
However, QSHIs that conserve the $s^z$-component of spin instead have a $\mathbb{Z}$ topological index~\cite{maciejko2011quantum}, which is simply the $s^z$-weighted Chern number. In this case, two copies of a QSHI form a topologically non-trivial double quantum spin Hall insulator (DQSHI) \cite{PhysRevB.74.085308, PhysRevB.80.125327, PhysRevB.110.195142}.

Recently, there has been experimental evidence of DQSHIs in twisted homobilayers of the transition metal dichalcogenides (TMDs) WSe$_2$~\cite{kang2024observationdoublequantumspin,kang2024double} and MoTe$_2$~\cite{ kang2024evidence,xu2024interplay,kang2025time}. The TMD-based DQSHIs can be understood as band insulators, where one fills the two lowest bands of the $K$ valley (which both have spin-up and Chern number $C = +1$) and the two lowest bands of the $K'$ valley (which both have spin-down and $C = -1$). Since electronic interactions are not necessary for forming the bulk state, it is tempting to conclude that interactions have no effect at all. If this were true, TMD-based DQSHI edges would always have two pairs of helical edge modes, one per QSHI copy. However, twisted TMDs are strongly correlated systems, and gapless topological edge modes can be heavily modified by interactions  even when the bulk is stable~\cite{schulz1996phases, wu2003competing, tanaka2009two, liu2011charge,sela2011majorana, schmidt2012inelastic, keselman2015gapless, oreg2014fractional,werman2015exciton, santos2015interaction, santos2016phase, maymann2024theoryhalfintegerfractionalquantum, PhysRevB.95.205120, yurkevich2021superconducting, 9n85-r2xw}. One therefore needs to carefully consider the effects of interactions in order to fully characterize the edges of TMD-based DQSHIs.

\begin{figure}[t]
    \centering
    \def\svgwidth{\columnwidth}  % Scale the figure to text width
    \includegraphics[width=\columnwidth]{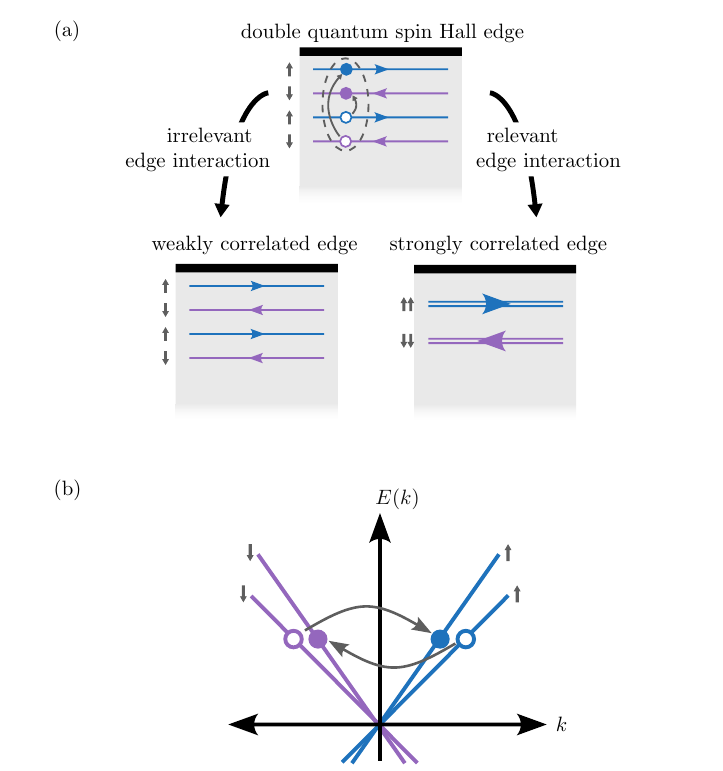}
    \caption{(a) Effect of the interactions on the DQSHI edge. Without interactions, there are two pairs of helical edge modes (blue arrows indicate the direction of propagation, and gray arrows indicate spin of the mode). If the edge interactions (depicted in the dashed circle) are irrelevant, the number of helical modes is unchanged. If the interactions are relevant, the edge enters a strongly correlated phase with only a single pair of helical modes that carry twice the charge and twice the spin of the original helical modes. (b) Edge interaction depicted in energy-momentum space.}
    \label{fig:1}
\end{figure}

%With interactions, QSHI edges can be understood as helical Luttinger liquid . 

In this work, we show that DQSHI edges have two symmetry-preserving phases if we include interactions: a weakly correlated phase and a strongly correlated phase. In the weakly correlated phase, the edge has two pairs of helical modes, and is adiabatically connected to the non-interacting limit. The strongly correlated phase, however, only has a single pair of helical modes. This edge also has a gap to adding or removing single electrons, but is gapless to adding or removing \emph{pairs} of electrons. The phase transition between the weakly and strongly correlated phases occurs when a certain ``double spin-flip" interaction term becomes relevant, as shown schematically in Fig.~\ref{fig:1}.

Given that the DQSHI edge can realize two distinct phases, the next immediate question to ask is how one can distinguish them in an experimental setting. This is \textit{not} possible via simple electrical transport, as the strongly and weakly correlated edges have the same quantized edge conductance. It is, however, possible to distinguish between the two phases using shot noise measurements~\cite{landauer1993solid, PhysRevLett.123.137701, PhysRevB.94.045425, niyazov2024shot, niyazov2024shotnoiseaharonovbohminterferometers}. 
In a quantum point contact (QPC) geometry, the Fano factor of the current noise is $\mathcal{F} = 1$ for the weakly correlated edge (same as for a conventional QSHI or integer quantum Hall state~\cite{fendley1995exact, ponNag1999,trauzettel2004appearance, dolcini2005transport}). However, for the strongly correlated edge, we instead find a doubled Fano factor of $\mathcal{F} = 2$. %This holds for both the weak backscattering and strong backscattering regimes of the QPC. 
This doubled Fano factor cannot occur in non-interacting systems of electrons and is a clear indicator of strong edge correlations. A similar strongly correlated edge and doubled Fano-factor was also predicted to occur in bilayer graphene in a strong magnetic field \cite{bi2017bilayer,zhang2017fingerprints}. We do not require an external magnetic field, since TMDs have strong Ising spin-orbit coupling \cite{wang2021ising, PhysRevX.4.011034, xiao2012coupled}.

%Hence, due to the strong Ising spin-orbit coupling of TMDs, we do not require an external magnetic field.

\begin{figure}[t]
    \centering
    \includegraphics[width=\columnwidth]{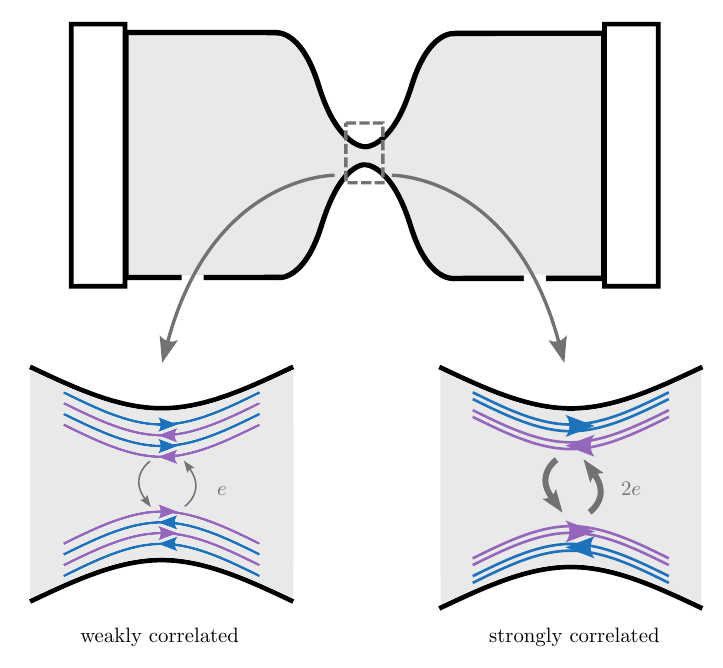}
    \caption{Depiction of the quantum point contact geometry. In the weakly correlated edge, the lowest possible charge that may tunnel across the point contact is $e$, via any of the terms in Eq. \ref{eq:electronTun}. However, for the strongly correlated edge the minimal charge that can tunnel across the point contact is $2e$.}
    \label{fig:2}
\end{figure}

Although we formally establish these results using the out-of-equilibrium Schwinger-Keldysh formalism, the doubled Fano factor of the strongly correlated edge can be informally understood from two simple observations. First, due to the single-electron gap of the strongly correlated edge, charge can only tunnel across the QPC in multiples of $2e$, as shown in Fig.~\ref{fig:2}. Second, the Fano factor is equal to the smallest amount of charge (in units of $e$) that can tunnel across the QPC~\cite{kane1994nonequilibrium,kane1997shot, sandler1999noise, kane2003shot, bid2009shot, hashisaka2015shot, manna2023experimentallymotivatedorderlength, manna2024diagnosticsanomalousconductanceplateaus, PhysRevLett.132.136502, 10.1063/10.0034344}.
%In comparison, individual electrons can be added to the edge of the weakly-coupled edge, leading to the Fano factor of $1$, same as a single QSHI edge or a Luttinger liquid with an impurity.
While the main focus of our work is on the double quantum spin Hall insulators realized in twisted TMDs, it is straightforward to generalize our arguments to other $N$-fold QSHIs, such as the triple ($N = 3$) QSHIs, which has been observed in twisted MoTe$_2$~\cite{kang2024evidence,kang2025time}.

This paper is organized as follows. In Section~\ref{sec:DQSHedge}, we introduce and bosonize the non-interacting DQSHI edge. In Section~\ref{sec:SCedge} we consider the effects of interactions on the DQSHI edge and show that, if certain interaction terms are relevant, the edge enters a strongly correlated phase with a single-electron gap. 
In Section~\ref{sec:shotnoise}, we calculate the shot noise for the two possible DQSHI edges. We conclude our work in Section~\ref{sec:discussion}, and discuss relevance to twisted TMD systems.

\section{Bosonization of the double quantum spin Hall edge.}\label{sec:DQSHedge}

Without additional interactions, the DQSHI edge consists of two pairs of helical edge modes. We define the annihilation operators for the edge mode electrons as $\psi_{\text{1R$\uparrow$}}$, $\psi_{\text{1L$\downarrow$}}$, $\psi_{\text{2R$\uparrow$}}$, and $\psi_{\text{2L$\downarrow$}}$.  Here, without loss of generality, we have chosen the two right-moving modes to have spin-up, and the two left-moving modes to have spin-down. 

To analyze the effects of interactions on the edge modes, we follow the usual prescription of bosonizing the fermionic edge modes \cite{fradkin2013field, gogolin2004bosonization},
\begin{equation}
    \psi_{a\chi}(\bm{x}) \propto e^{i \phi_{a\chi}(\bm{x}) + i k_{a\chi} x}
    \label{bosonizedef}
\end{equation}
where $\bm{x} = (t,x)$ is the space-time coordinate, $a = 1,2$ labels the two pairs of helical modes, and $\chi = \text{R$\uparrow$, L$\downarrow$}= +1,-1$ refers to the chirality/spin of the mode, and $k_{a\chi}$ is the Fermi-wavevector of the mode.

The bosonized edge Lagrangian can be written as 
\begin{equation}
    \L = -\frac{1}{4\pi} \partial_x \vec{\phi}^\intercal \vec{K} \partial_t \vec{\phi} + \frac{1}{4\pi} \partial_x \vec{\phi}^\intercal \vec{V} \partial_x \vec{\phi},
    \label{initLagrangian}
\end{equation}
where $\vec{\phi}$ is
\begin{equation}
    \vec{\phi} \equiv \begin{pmatrix}
        \phi_{\text{1R$\uparrow$}}\\
        \phi_{\text{1L$\downarrow$}}\\
        \phi_{\text{2R$\uparrow$}}\\
        \phi_{\text{2L$\downarrow$}}
    \end{pmatrix},
\end{equation}
and
\begin{equation}
    \vec{K} \equiv \begin{pmatrix}
        1 & & & \\
        & -1 & & \\
        & & 1 & \\
        & & & -1
    \end{pmatrix}.
\end{equation}
The matrix $\vec{V}$ is the velocity matrix. For a non-interacting system, $\vec{V}$ is a diagonal matrix with positive entries on the diagonal. As we shall show in the next section, $\vec{V}$ generically has both diagonal and off-diagonal elements in interacting systems. Here, and throughout, we set $e = \hbar = 1$, unless needed for clarity. 

The DQSHI is guaranteed to have a gapless edge as long as charge and the $s^z$-component of spin are conserved. In twisted TMDs, first-principles calculations show that $s^z$ is indeed a good quantum number,  since the spin-orbit coupling is dominantly Ising-like \cite{wang2021ising, PhysRevX.4.011034, xiao2012coupled, devakul2021magic}. First-principles calculations also indicate that $s^z$ is a good quantum number even when the out of plane mirror symmetry is broken by an external displacement field\cite{zhang2025twist}. The conservation of charge and the $s^z$-component of spin defines two $\U(1)$-symmetries, which we will refer to as $\U(1)_\text{c}$-charge conservation symmetry and $s^z$-spin conservation symmetry, respectively \cite{chubukov2025quantum}. The charge-conservation symmetry acts on the fermion operators as
\begin{equation}
    \psi_{a\chi} \to e^{i\theta_c}\psi_{a\chi},
\end{equation}
whereas spin-conservation symmetry acts axially on the different fermion species,
\begin{equation}
    \psi_{a\chi} \to e^{i\chi\theta_s}\psi_{a\chi}.
\end{equation}
Using the correspondence in equation (\ref{bosonizedef}), one finds that the symmetry operators act on the bosonic modes as
\begin{equation}
   \begin{pmatrix}
        \phi_{\text{1R$\uparrow$}}\\
        \phi_{\text{1L$\downarrow$}}\\
        \phi_{\text{2R$\uparrow$}}\\
        \phi_{\text{2L$\downarrow$}}
    \end{pmatrix} \xrightarrow{\U(1)_c} \begin{pmatrix}
        \phi_{\text{1R$\uparrow$}} + \theta_c\\
        \phi_{\text{1L$\downarrow$}}+ \theta_c\\
        \phi_{\text{2R$\uparrow$}}+ \theta_c\\
        \phi_{\text{2L$\downarrow$}}+ \theta_c
    \end{pmatrix},
\label{eq:U1Trans}\end{equation}
for some phase $\theta_c$. Spin conservation symmetry acts as 
\begin{equation}
    \begin{pmatrix}
        \phi_{\text{1R$\uparrow$}}\\
        \phi_{\text{1L$\downarrow$}}\\
        \phi_{\text{2R$\uparrow$}}\\
        \phi_{\text{2L$\downarrow$}}
    \end{pmatrix} \xrightarrow{s^z} \begin{pmatrix}
        \phi_{\text{1R$\uparrow$}} + \theta_s\\
        \phi_{\text{1L$\downarrow$}} - \theta_s\\
        \phi_{\text{2R$\uparrow$}}+ \theta_s\\
        \phi_{\text{2L$\downarrow$}}-\theta_s
    \end{pmatrix}
\label{eq:szTrans}\end{equation}
where $\theta_s$ is a spin-dependent phase.  

In addition to the $\U(1)_c$ and $s^z$ symmetries, we will also consider time-reversal symmetry, $\mathsf{T}$. This acts on the fermionic operators as 
\begin{equation}
    \psi_{a\chi} \to \chi \psi_{a\bar{\chi}}
\end{equation}
and sends $i$ to $-i$, where $\bar{\chi}$ stands for the opposite chirality/spin label. 
Hence, the bosons transform as 
\begin{equation}
    \begin{pmatrix}
        \phi_{\text{1R$\uparrow$}}\\
        \phi_{\text{1L$\downarrow$}}\\
        \phi_{\text{2R$\uparrow$}}\\
        \phi_{\text{2L$\downarrow$}}
    \end{pmatrix} \xrightarrow{\mathsf{T}} \begin{pmatrix}
        -\phi_{\text{1L$\downarrow$}  }\phantom{+ \pi \ \, }\\
        -\phi_{\text{1R$\uparrow$}} + \pi\\
        -\phi_{\text{2L$\downarrow$} }\phantom{+ \pi \ \, }\\
        -\phi_{\text{2R$\uparrow$}}+ \pi
    \end{pmatrix}.
\end{equation}
It should be noted that while $\mathsf{T}$ is present in TMD DQSHIs, it is not a protecting symmetry, and the edge will be gapless even if $\mathsf{T}$ is broken, provided that the $\U(1)_c$- and $s^z$-conservations are preserved.

\section{\label{sec:SCedge}The strongly correlated edge.}
We now consider the effects of interactions on the DQSHI edge. In general, there are two types of interactions to consider. First, there are the density-density interactions of the form
\begin{equation}
\psi^\dagger_{a\chi}(\bm{x})\psi_{a\chi}(\bm{x})\psi^\dagger_{a'\chi'}(\bm{x})\psi_{a'\chi'}(\bm{x}) \propto \partial_x \phi_{a\chi} \partial_x \phi_{a'\chi'}. 
\end{equation}
where we have suppressed the dependence of the bosons on the space-time coordinate $\bm{x}$.
Since these density-density terms are quadratic in the bosonic fields, we can absorb them into the redefinition of $\vec{V}$. Here we will consider a velocity matrix of the form
\begin{equation}
\vec{V} \rightarrow \begin{pmatrix}
        v & -v' & u & -u'\\
        -v' & v & -u' & u\\
        u & -u' & v & -v'\\
        -u' & u & -v' & v
    \end{pmatrix}, \label{eq:VmatrixDef}\end{equation}
where $v$ is the renormalized velocity of the edge modes, $v'$, $u$, $u'$ correspond to various density interactions. This is not the most general velocity matrix, as we have explicitly omitted certain symmetry allowed density-density interactions. However, this form of $\vec{V}$ will greatly simplify our forthcoming analysis. We provide an analysis of a general symmetric $\vec{V}$ in Appendix~\ref{seclowenergyasymp}. 

In addition to the density-density interactions, there is also a non-trivial ``double spin-flip" interaction
\begin{equation}\begin{split}
    \O_{\text{dsf}}(\bm{x}) &= \psi^\dagger_{1\text{R}\uparrow} (\bm{x})\psi_{2\text{L}\downarrow}(\bm{x}) \psi_{1\text{L}\downarrow}^\dagger(\bm{x}) \psi_{2\text{R}\uparrow}(\bm{x}) + \text{H.c.}\\
    &\propto \cos(\phi_{\text{1R$\uparrow$}} + \phi_{\text{1L$\downarrow$}} - \phi_{\text{2R$\uparrow$}} - \phi_{\text{2L$\downarrow$}}).
\end{split}\label{eq:DSF_fermions}\end{equation}
 %Note that this term does not carry any momentum, as $k_{a\text{R}\uparrow} = -k_{a\text{L}\downarrow}$ due to $\mathsf{T}$-symmetry. 
As we shall show, if this interaction is relevant, the DQSHI edge will enter a non-trivial strongly correlated phase. If the interaction is irrelevant then the DQSHI edge will remain adiabatically connected to the fully non-interacting limit. 

To analyze Eq.~\ref{eq:DSF_fermions} we will change to the following basis
\begin{equation}
    \begin{split}
    \phi_+ & \equiv \frac{1}{2}\left[(\phi_{\text{1R$\uparrow$}} + \phi_{\text{2R$\uparrow$}}) - (\phi_{\text{1L$\downarrow$}} + \phi_{\text{2L$\downarrow$}})\right],\\
        \vartheta_+ & \equiv \frac{1}{2}\left[(\phi_{\text{1R$\uparrow$}} + \phi_{\text{2R$\uparrow$}}) + (\phi_{\text{1L$\downarrow$}} + \phi_{\text{2L$\downarrow$}})\right],\\
        \phi_- & \equiv \frac{1}{2}\left[(\phi_{\text{1R$\uparrow$}} - \phi_{\text{2R$\uparrow$}}) - (\phi_{\text{1L$\downarrow$}} - \phi_{\text{2L$\downarrow$}})\right],\\
        \vartheta_- & \equiv \frac{1}{2}\left[(\phi_{\text{1R$\uparrow$}} - \phi_{\text{2R$\uparrow$}}) + (\phi_{\text{1L$\downarrow$}} - \phi_{\text{2L$\downarrow$}})\right],
    \end{split}
    \label{newbasis}
\end{equation}
where $\phi_+$, and $\vartheta_+$ are the ``channel-symmetric" bosons and $\phi_-$, and $\vartheta_-$ are the ``channel-antisymmetric" bosons \cite{PhysRevB.110.155117}. 

In the new basis, the double spin-flip interaction takes on the simple form 
\begin{equation}
    \O_{\text{dsf}} \propto \cos(2\vartheta_-),
\end{equation}
and the DQSHI edge Lagrangian is described by two fully decoupled Luttinger liquids,
\begin{equation}
    \begin{split}
        \L =& \frac{v_+}{4\pi}\left[g_+ (\partial_x \vartheta_+)^2 + \frac{1}{g_+} (\partial_x \phi_+)^2\right]\\
        &+ \frac{v_-}{4\pi}\left[g_- (\partial_x \vartheta_-)^2 + \frac{1}{g_-} (\partial_x \phi_-)^2\right]\\
        &+ 2\alpha\cos(2\vartheta_-).
    \end{split}
    \label{fulltheory}
\end{equation}
where we have included the double slip-flip interaction, with coupling constant $\alpha$. The velocities $v_\pm$ and Luttinger parameters $g_{\pm}$ are defined as
\begin{equation}
\begin{split}
        v_\pm &= \sqrt{[(v - v') \pm (u - u') ][(v + v') \pm (u + u') ]}\\
    g_\pm &= \sqrt{\frac{(v - v') \pm (u - u') }{(v + v') \pm (u + u') }}.
    \end{split}
\end{equation}
When $v' = u = u' = 0$, the $g_{\pm} = 1$, which is the Luttinger parameter of non-interacting electrons. 

Clearly, the channel-symmetric $\phi_+$ and $\theta_+$ bosons are free and gapless in Eq.~\ref{fulltheory}. The fate of the channel-antisymmetric $\phi_-$ and $\vartheta_-$ bosons depends on if the cosine term, $\cos(2\vartheta_-)$, is a relevant perturbation or not. %If it is irrelevant, then the cosine term flows to weak coupling, and the channel asymmetric bosons are free. In this regime, there will be two pairs of gapless bosons, as in the fully decoupled case. On the other hand, if the cosine term is relevant, then $\phi_-$ will be pinned to its classical expectation value, and the channel asymmetric sector will be gapped out. The resulting strongly correlated edge only contains the channel symmetric bosons, $\phi_+$ and $\theta_+$. 

\subsection{Renormalization Group Analysis.}
We now perform a renormalization group (RG) analysis of the DQSHI edge. At weak coupling, the beta function for $\alpha$ is 
\begin{equation}
    \begin{split}
    \beta(\alpha) &= (2 - \Delta_\alpha)\alpha,
\end{split}
\end{equation}
where $\Delta_\alpha = 2g_-$ is the scaling dimension of the cosine term. When $g_- < 1$, the cosine term is irrelevant and flows to weak coupling. The low energy Lagrangian is then simply Eq.~\ref{fulltheory} with $\alpha = 0$,
\begin{equation}
    \begin{split}
        \L =& \frac{v_+}{4\pi}\left[g_+ (\partial_x \theta_+)^2 + \frac{1}{g_+} (\partial_x \phi_+)^2\right]\\
        &+ \frac{v_-}{4\pi}\left[g_- (\partial_x \vartheta_-)^2 + \frac{1}{g_-} (\partial_x \phi_-)^2\right].
    \end{split}
    \label{decoupledtheory}
\end{equation}
The values of $g_{-}$ and $v_{-}$ may also be renormalized, but this effect is not important for our present analysis. The values of $g_{+}$ and $v_+$ do not renormalize, as the channel-symmetric sector is a free bosonic theory, which is an RG fixed point. We will refer to this edge as the ``weakly correlated edge."

For $g_- > 1$, the cosine term is relevant, and the cosine term will flow to strong coupling. In this case, $\vartheta_{-}$ will become pinned to the bottom of the cosine potential, and the channel-antisymmetric sector will be gapped out. The low energy Lagrangian is simply
\begin{equation}
    \begin{split}
        \L_{\text{eff}} &= \frac{v_+}{4\pi}\left[g_+ (\partial_x \vartheta_+)^2 + \frac{1}{g_+} (\partial_x \phi_+)^2\right],
    \end{split}
    \label{interactingL}
\end{equation}
which consists of a single pair of bosons. We will refer to this as the ``strongly correlated edge". Again, the values of $g_+$ and $v_+$ do not renormalize. It is worth explicitly noting that the $\phi_{-}$ and $\vartheta_{-}$ bosons, do not carry $\U(1)_c$-charge or $s^z$-spin (see Eqs.~\ref{eq:U1Trans},\ref{eq:szTrans} and~\ref{newbasis}), and so we have not broken any symmetries by gapping them out.

In principle, there is also a second double spin-flip term that one can consider, viz.
\begin{equation}\begin{split}
    \O'_{\text{dsf}}(\bm{x}) &= \psi^\dagger_{1\text{R}\uparrow}(\bm{x}) \psi_{1\text{L}\downarrow}(\bm{x}) \psi_{2\text{L}\downarrow}^\dagger(\bm{x}) \psi_{2\text{R}\uparrow}(\bm{x}) + \text{H.c.}\\
    &\propto \cos(\phi_{\text{1R$\uparrow$}} - \phi_{\text{1L$\downarrow$}} - \phi_{\text{2R$\uparrow$}} + \phi_{\text{2L$\downarrow$}}\\  & \phantom{\propto===}+ 2k_{1 \text{R}\uparrow}x- 2k_{2 \text{R}\uparrow}x).
\end{split}\label{eq:DSF_fermions2}\end{equation}
In general, $k_{1 \text{R}\uparrow}\neq k_{2\text{R}\uparrow}$ without fine-tuning. Because of this, the argument of the cosine term oscillates in space, and the interaction is RG-irrelevant. However, if $k_{1 \text{R}\uparrow} =  k_{2\text{R}\uparrow}$, $\O'_{\text{dsf}}$ can be relevant, depending on the Luttinger parameter. Since the terms in Eq.~\ref{eq:DSF_fermions2} and Eq.~\ref{eq:DSF_fermions} do not commute with each other, both terms cannot reach strong coupling simultaneously. Rather, there are two distinct strong coupling fixed points. The first fixed point is the one discussed above, where Eq.~\ref{eq:DSF_fermions} flows to strong coupling, and Eq.~\ref{eq:DSF_fermions2} flows to weak coupling. In the second fixed point, Eq.~\ref{eq:DSF_fermions2} flows to strong coupling, and Eq.~\ref{eq:DSF_fermions} flows to weak coupling. However, these two fixed points are equivalent under exchanging $\psi_{1\text{L}\downarrow} \leftrightarrow \psi_{2\text{L}\downarrow}$. Since one can arbitrarily assign labels to the two channels in the fine-tuned scenario, we will not need to consider Eq.~\ref{eq:DSF_fermions2} in detail.

\subsection{Single-electron gap.}
An important feature of the strongly correlated edge is that, despite being gapless, it is gapped to single electron operators. By contrast, the weakly correlated edge is gapless to single electrons, as in the decoupled limit. 

To show that there is a gap to single electrons, let us first consider the $\psi_{1\text{R}\uparrow}$ electron. The analysis of the other electrons follows from the same logic. In the channel-symmetric/antisymmetric basis, the electron operator is, ignoring momentum dependence for simplicity,
\begin{equation}
    \psi_{1\text{R}\uparrow} \propto e^{i (\phi_+ + \vartheta_+ + \phi_{-} + \vartheta_-)}.
\end{equation}
The correlation function for this electron is then
\begin{equation}\begin{split}
    &\langle \psi^\dagger_{1\text{R}\uparrow}(\bm{x})\psi_{1\text{R}\uparrow}(\bm{x}') \rangle  \\ &\phantom{==}\propto \langle e^{-i \phi_+(\bm{x})}e^{i \phi_+(\bm{x}')} \rangle  \langle e^{-i \vartheta_+(\bm{x})}e^{i \vartheta_+(\bm{x}')} \rangle  \\ &\phantom{===\propto}\times \langle e^{-i \phi_-(\bm{x})}e^{i \phi_-(\bm{x}')} \rangle\langle e^{-i \vartheta_-(\bm{x})}e^{i \vartheta_-(\bm{x}')} \rangle.
\end{split}\end{equation}
For the weakly correlated edge, all bosons are free, and so $\psi^\dagger_{1\text{R}\uparrow}$ has power-law correlations \cite{fradkin2013field}, indicating that it is gapless. However, for the strongly correlated edge, $\vartheta_-$ is pinned to its minimum, and $\phi_-$ has short-ranged correlations. Therefore, $\psi^\dagger_{1\text{R}\uparrow}$ has short-ranged electronic correlations in the strongly correlated edge. By the same reasoning, the three other electron operators from Eq.~\ref{bosonizedef} also have short-ranged correlations. Individual electrons are therefore gapped in the strongly correlated regime. 

While individual electrons are gapped, the following electron bilinear operators are gapless,
\begin{equation}\begin{split}
    &\psi^\dagger_{1\text{R}\uparrow} \psi^\dagger_{2\text{R}\uparrow} \propto e^{i [\phi_+ + \vartheta_+]},\\
    &\psi^\dagger_{1\text{L}\downarrow} \psi^\dagger_{2\text{L}\downarrow} \propto e^{-i [\phi_+ - \vartheta_+]}
\label{eq:FermBL1}\end{split}\end{equation}
\begin{equation}\begin{split}
    &\psi^\dagger_{1\text{R}\uparrow} \psi^\dagger_{1\text{L}\downarrow} \sim \psi^\dagger_{2\text{R}\uparrow} \psi^\dagger_{2\text{L}\downarrow}  \propto e^{i \vartheta_+},
\label{eq:FermBL2}\end{split}\end{equation}
\begin{equation}\begin{split}
    &\psi^\dagger_{1\text{R}\uparrow} \psi_{2\text{L}\downarrow} \sim \psi^\dagger_{2\text{R}\uparrow}\psi_{1\text{L}\downarrow}  \propto e^{i \phi_+ }
\label{eq:FermBL3}\end{split}\end{equation}
where we have used that $\vartheta_-$ acquires a classical expectation value in the strongly correlated phase ($\vartheta_- = \langle \vartheta_- \rangle $).
By inspection, the operators in Eq.~\ref{eq:FermBL1} are bosons with charge $2e$ and spin $\pm 1$, the operator in Eq.~\ref{eq:FermBL2} is a boson with charge $2e$ and spin $0$, and the operator in Eq.~\ref{eq:FermBL3} is a boson with charge $0$ and spin $1$. Since the operators are all independent of $\phi_-$, they are all gapless with power law correlations. 

Furthermore, all other local operators in the strongly correlated edge can be written as combinations of the bosons in Eqs.~\ref{eq:FermBL1}-\ref{eq:FermBL3}. This can be directly argued from the fact that any gapless degree of freedom must be independent of $\phi_-$, and that any local operator (i.e., one made out of a finite number of electron operators) must be invariant under the transformation $\phi_{a\chi}\rightarrow \phi_{a\chi} + 2\pi$, for all $a$ and $\chi$ individually. 

\section{Distinguishing the strongly and weakly correlated edges via shot noise.}\label{sec:shotnoise}
We now ask how one can differentiate between the two possible edge phases of a TMD DQSHI. Using the Kubo formula, one finds that both the weakly and strongly correlated edges have edge conductance $2e^2/h$, (assuming the leads are Fermi liquid--like)~\cite{PhysRevB.52.R5539}. However, as we shall show in this section, the edges can be distinguished by the current shot noise in a quantum point contact (QPC) geometry. In the QPC geometry, the edge states are brought together at a constriction which separates the source from the drain, allowing for tunneling between the top and bottom edges (see Fig.~\ref{fig:2}). The strength of tunneling can be tuned by via a gate voltage~\cite{PhysRevLett.102.076602, PhysRevB.79.235321}. Here, we will assume the QPC is in the weak backscattering limit \cite{kane1994nonequilibrium, strom2009tunneling}. The analysis in the weak tunneling limit is nearly identical, as the bulk of the DQSHI is not topologically ordered. Specifically, the Fano factor is the same in the weak backscattering and weak tunneling limits.

In the weak backscattering limit, we will focus on the backscattering current $I_{\text{b}}$, which can be informally expressed as the difference between the direct current in the QPC geometry, $I$, and the current one would have in a two-terminal geometry, without a QPC,
\begin{equation}
    \angled{I_{\text{b}}} = \angled{I} - \sigma_0 V
\label{eq:backscatterCurDef}\end{equation}
where $\sigma_0 = g_\infty2 e^2/h$ is the universal conductance of a DQSHI edge, and $V$ is the applied voltage. The quantity $g_\infty$ is the Luttinger parameter at the leads. For Fermi liquid--like leads, where the quasi-particles behave like free electrons, we have $g_\infty = 1$. 

Consider a system at temperature $T$ which is driven with a voltage $V$ at frequency $\omega$. In the $T,V,\omega\rightarrow 0$ limit, the fluctuations of the measured current, $\delta I^2$,  become proportional to the average backscattering current $\langle I_{\text{b}}\rangle $ 
\cite{ponNag1999,PhysRevB.71.165309},
\begin{equation}\begin{split}
    &\delta I^2 \overset{\omega \rightarrow 0}{=} \mathcal{F}2e \langle I_{\text{b}}\rangle,
\end{split}
\end{equation}
where $\mathcal{F}$ is the Fano factor. For topological systems dominated by edge transport, $\mathcal{F}$ is expected to be equal to the smallest charge (in units of $e$) that can tunnel across the QPC at low voltages. Since the strongly correlated edge is gapped to all individual single electrons, charge must necessarily be backscattered in units of $2e$. For the weakly correlated edge, there are gapless electrons and charges can be backscattered in units of $e$. Based on this, we expect that $\mathcal{F} = 2$ for the strongly correlated edge, and $\mathcal{F}=1$ for the weakly correlated edge. In the following, we will explicitly show that this is indeed the case, and that shot noise is a robust, experimentally-accessible signature that can distinguish between the two possible DQSHI edges.

\subsection{\label{foldedpicture}Folded picture of the Luttinger liquid.}
Since the top edge of the DQSHI is brought into proximity with the bottom edge at a QPC, we need to introduce 4 additional electron operators for the bottom edge: $\psi_{\text{1L$\uparrow$}}$, $\psi_{\text{1R$\downarrow$}}$, $\psi_{\text{2L$\uparrow$}}$, and $\psi_{\text{2R$\downarrow$}}$. Using all 8 electrons, we can map the four pairs of helical Luttinger liquids of the two edges onto two spinful Luttinger liquid~\cite{PhysRevLett.102.076602, PhysRevB.79.235321}. This is the ``folded'' picture of the helical Luttinger liquids.

We again bosonize the electron operators as
\begin{equation}
    \psi_{a\eta\sigma} \sim e^{i\phi_{a\eta\sigma} + i k_{a\eta\sigma} x}.
\end{equation}
where $a = 1,2$, $\eta = \text{R},\text{L}$, and $\sigma = \uparrow, \downarrow$, and $k_{a\eta\sigma}$ is the Fermi-wavevector of the relevant mode.
Here, we will use the following basis for the 8 bosonic fields,
\begin{equation}
    \begin{split}
        \phi_{\text{c}} &\equiv \frac{1}{2}\left[(\phi_{\text{R}\uparrow+} - \phi_{\text{L}\uparrow+}) +  (\phi_{\text{R}\downarrow +} - \phi_{\text{L}\downarrow +})\right]\\
        \vartheta_{\text{c}} &\equiv \frac{1}{2}\left[(\phi_{\text{R}\uparrow+} + \phi_{\text{L}\uparrow+}) +  (\phi_{\text{R}\downarrow +} + \phi_{\text{L}\downarrow +})\right]\\
        \phi_{\text{s}} &\equiv \frac{1}{2}\left[(\phi_{\text{R}\uparrow+} - \phi_{\text{L}\uparrow+}) -  (\phi_{\text{R}\downarrow +} - \phi_{\text{L}\downarrow +})\right]\\
        \vartheta_{\text{s}} &\equiv \frac{1}{2}\left[(\phi_{\text{R}\uparrow+} + \phi_{\text{L}\uparrow+}) -  (\phi_{\text{R}\downarrow +} + \phi_{\text{L}\downarrow +})\right]\\
        \phi_{\text{n}\pm} &\equiv \frac{1}{2}\left[(\phi_{\text{R}\uparrow-} - \phi_{\text{L}\uparrow-}) \pm (\phi_{\text{R}\downarrow -} - \phi_{\text{L}\downarrow -})\right]\\
        \vartheta_{\text{n}\pm} &\equiv \frac{1}{2}\left[(\phi_{\text{R}\uparrow-} + \phi_{\text{L}\uparrow-}) \pm  (\phi_{\text{R}\downarrow -} + \phi_{\text{L}\downarrow -})\right]\\
    \end{split}
    \label{defchargespin}
\end{equation}
where $\phi_{\eta\sigma \pm} \equiv \frac{1}{2}(\phi_{1 \eta\sigma} \pm \phi_{2 \eta\sigma})$. The $\phi_{\text{c}}$ and $\vartheta_{\text{c}}$ bosons are dual to each other, and carry charge. The $\phi_{\text{s}}$, and $\vartheta_{\text{s}}$ bosons carry spin. The $\phi_{\text{n}\pm}$ and $\vartheta_{\text{n}\pm}$ bosons are neutral.

In terms of the new bosonic fields, the Lagrangian density for the weakly correlated edge is:
\begin{equation}
    \L_0 = \L_{0,\text{cs}} + \L_{0,\text{n}}
\end{equation}
where 

\begin{equation}
    \begin{split}
        \L_{0,\text{cs}} =& \frac{v_{\text{c}}}{4\pi} \left[g_{\text{c}}(\partial_x \vartheta_{\text{c}})^2 + \frac{1}{g_{\text{c}}}(\partial_x \phi_{\text{c}})^2\right]\\
        +& \frac{v_{\text{s}}}{4\pi} \left[ g_{\text{s}}(\partial_x \vartheta_{\text{s}})^2 + \frac{1}{g_{\text{s}}}(\partial_x \phi_{\text{s}})^2\right]\\
        \L_{0,\text{n}} =& \frac{v_{\text{n}+}}{4\pi} \left[g_{\text{n}+}(\partial_x \vartheta_{\text{n}+})^2 + \frac{1}{g_{\text{n}+}}(\partial_x \phi_{\text{n}+})^2\right]\\
        +& \frac{v_{\text{n}-}}{4\pi} \left[g_{\text{n}-}(\partial_x \vartheta_{\text{n}-})^2 + \frac{1}{g_{\text{n}-}}(\partial_x \phi_{\text{n}-})^2\right]\\
    \end{split}
\end{equation}
The Luttinger parameters for the new bosonic fields are 
\begin{equation}\begin{split}
    g_{\text{c}} = g_+ = 1/g_{\text{s}},\\
    g_{\text{n}+} = g_- = 1/g_{\text{n}-}.\\
\end{split}\end{equation}
The velocity of the modes are $v_{\text{c}} =v_{\text{s}} = v_{+}$, and $v_{\text{n}\pm} = v_{-}$. Written in terms of the $\phi$ fields,
\begin{equation}
    \L_{0,\text{cs}} = \phi_{\text{c}} \hat{D}_{\text{c}} \phi_{\text{c}} + \phi_{\text{s}} \hat{D}_{\text{s}} \phi_{\text{s}} 
    \label{Lfreenonint}
\end{equation}
where we have defined the kinetic differential operator 
\begin{equation}
    \hat{D}_{\gamma} \equiv -\frac{ 1}{4\pi g_\gamma v_\gamma}\partial_t^2 + \partial_x \left(\frac{v_\gamma}{4\pi g_{\gamma}}\right)\partial_x
    \label{kinop}
\end{equation}
with $\gamma = \text{c, s}$.  The Lagrangian for the neutral sector, $ \L_{0,\text{n}}$, is defined analogously.
%note to self: double check

The neutral sector is fully gapped for the strongly correlated edge, and so the low energy Lagrangian for the strongly correlated edge is simply $\L_{0,\text{cs}}$.

\subsection{\label{secbackscatteringaction}The backscattering action.}
To model a QPC, we only consider scattering between the top and bottom edges at a single point. Away from this point, the edges are taken to be fully decoupled. As we shall now discuss, we only need to consider processes where one or two electrons are scattered across the QPC. Let us consider $I_{\text{b},m}$, the contribution to the average backscattered current from a process where $m$-particles are backscattered. We will consider the low voltage limit, where the applied voltage $V$ 
is small compared to the characteristic energy scale of the wire, $v_{\text{F}}/L,$
which is the ratio of the plasmon frequency over the length of the wire. 
In this limit, $I_{\text{b},m}$ scales as $V^{m^2(g_\infty + 1/g_\infty)- 1}$, where $g_\infty$ is the value of the Luttinger parameter near the leads~\cite{PhysRevLett.79.1714,ponNag1999,PhysRevB.71.165309}. For Fermi-liquid-like leads, $g_\infty = 1$, independent of the value of the Luttinger parameter in the bulk. 
At sufficiently low voltage, the shot noise signal is therefore dominated by processes with the lowest number of backscattering particles. For the weakly correlated edge, the lowest number of particles that can be backscattered is $m = 1$. For the strongly correlated edge, there is a gap to one-electron scattering, and so the dominant contribution to the noise occurs from $m = 2$-electron scattering. 

\subsubsection{Backscattering of the weakly correlated edge.}
First, we will consider the backscattering action for the weakly correlated edge. 
The symmetry-allowed backscattering operators are 
\begin{equation}
    \begin{split}
        \psi_{1\uparrow\text{L}}^\dagger \psi_{1\uparrow\text{R}} + \psi_{1\downarrow\text{L}}^\dagger \psi_{1\downarrow\text{R}} +  \text{H.c.}\\
        \psi_{2\uparrow\text{L}}^\dagger \psi_{2\uparrow\text{R}} + \psi_{2\downarrow\text{L}}^\dagger \psi_{2\downarrow\text{R}} + \text{H.c.}\\
        \psi_{2\uparrow\text{L}}^\dagger \psi_{1\uparrow\text{R}} + \psi_{2\downarrow\text{L}}^\dagger \psi_{1\downarrow\text{R}} +  \text{H.c.}\\
        \psi_{1\uparrow\text{L}}^\dagger \psi_{2\uparrow\text{R}} + \psi_{1\downarrow\text{L}}^\dagger \psi_{2\downarrow\text{R}} + \text{H.c.}
    \end{split}
\label{eq:electronTun}\end{equation} 
In bosonized form, the backscattering action for the weakly correlated edge is (see Appendix~\ref{bsbosonized})
\begin{equation}
    \begin{split}
        S_{\text{b}} = \int &\d t \, \Big[ \lambda_1 \cos(\phi_{\text{c}})\cos(\phi_{\text{n}+})\cos(\phi_{\text{s}+})\cos(\phi_{\text{n}-})\\
    &+ \lambda_2 \sin(\phi_{\text{c}})\sin(\phi_{\text{n}+})\sin(\phi_{\text{s}+})\sin(\phi_{\text{n}-})\\
&+\lambda_3\cos(\phi_{\text{c}})\cos(\vartheta_{\text{n}+})\cos(\phi_{\text{s}+})\cos(\vartheta_{\text{n}-})\\
&+\lambda_4\sin(\phi_{\text{c}})\sin(\vartheta_{\text{n}+})\sin(\phi_{\text{s}+})\sin(\vartheta_{\text{n}-})\Big].
    \label{Sbnoninteracting}
    \end{split}
\end{equation}

All the fields in the above expression are evaluated at coordinates $\bm{x} = (t, x_0)$ where $x_0$ is the location of the QPC in the folded picture. $\lambda_i$ are the tunneling strengths of the different processes in Eq.~\ref{eq:electronTun}.

\subsubsection{Backscattering of the strongly correlated edge.}
For the strongly correlated edge, there is a gap to single-electron operators. However, there are still relevant two-electron tunneling operators at the edge,
\begin{equation}
    S_{\text{b}} =  \int \d t \left[ \lambda'_1  \cos(2\phi_{\text{c}}) +  \lambda'_2 \cos( 2 \phi_{\text{c}}) \cos(2 \phi_{\text{s}}) \right].
    \label{SbinteractingSC}
\end{equation}
The first term describes tunneling of the charge $2e$ spin $0$ modes in Eq.~\ref{eq:FermBL2}, while the second term describes the tunneling of the charge $2e$ spin $\pm 1$ modes in Eq.~\ref{eq:FermBL1}. Again, all fields are evaluated at $\bm{x} = (t, x_0)$, and $\lambda'_i$ are the strengths of the different tunneling processes. The difference in the coefficient of $\phi_\text{c}$ in the cosine terms in Eqs.~\ref{Sbnoninteracting} and~\ref{SbinteractingSC} is ultimately responsible for the difference in the Fano factors, as we will show in the following section.

\subsection{\label{noise} Shot Noise.}
We now calculate the Fano factor for the weakly correlated edge, and the strongly correlated edge. Since the charge is carried by the $\phi_\text{c}$ boson for both types of edges, the analysis of the weakly correlated edge and the strongly correlated edge is very similar in our current formulation.

\subsubsection{Shot noise of the weakly correlated edge.}\label{generalTreatment}
Our starting point is the action $S = S_0 + S_\text{b} + S_V$, where $S_0$ and $S_\text{b}$ are the bare and backscattering action of the weakly correlated edge discussed previously, and $S_V$ describes the coupling to the external leads of the system, which we will define below. In the folded picture, we assume the leads are at $x = \pm L/2$, where $L$ is the length of the edge, and that the QPC is somwhere between the leads, $-L/2 < x_0 < L/2$. 

When there is a voltage applied across the system, the edge is brought out of equilibrium. Thus, we must consider time evolution along the Keldysh contour~\cite{Keldysh:1964ud}, which we denote by $\Gamma$. This contour takes an excursion from $-\infty$ to $\infty$ above the real time axis, and then returns to $-\infty$ below it. In this formulation, the charge and spin part of the quadratic action is
\begin{equation}
    \L_{0,\text{cs}}  = \vec{\phi}_{\text{c}}^\intercal \hat{\vec{D}}_{\text{c}} \vec{\phi}_{\text{c}} + \vec{\phi}_{\text{s}}^\intercal \hat{\vec{D}}_{\text{s}} \vec{\phi}_{\text{s}} 
\end{equation}
where $\vec{\phi}_{\gamma} = [\phi_\gamma(t^+,x), \phi_\gamma(t^-, x)]^\intercal$, and the entries of the matrix $\hat{\vec{D}}_\gamma$ ($\gamma = \text{c, s}$) are 
\begin{equation}
    \hat{\vec{D}}_\gamma^{\alpha,\beta} = \left[G_{\gamma}^{\alpha,\beta}\right]\inv 
\end{equation}
where $G$ is the bare propagator time-ordered along the contour $\Gamma$,
\begin{equation}
    G_{\gamma}^{\alpha,\beta}(t, x,x')= \angled{\T_{\Gamma} \, \phi_{\gamma}(t^{\alpha}, x)\phi_{\gamma}(0^\beta, x')}_0,
    \label{greensfunc}
\end{equation}
where $\alpha,\beta = \pm$, and $t^{\pm} \equiv \lim_{\eps\to 0} t \pm i\eps$, and $\angled{\cdots}_0$ refers to an average in equilibrium (i.e., without any applied voltage). Throughout this subsection, the superscript $\pm$ refers to the Keldysh contours. The neutral sector Lagrangian takes on a similar form in the Keldysh formulation, but will not be important for our forthcoming analysis.

In the folded picture, the total current operator is $I = \partial_t \phi_\text{c}/2\pi$. In the Keldysh formulation, the part of the action which accounts for the electric field of the external leads is
\begin{equation}
    S_V = \frac{1}{2\pi}  \int_\Gamma \d t \int \d x \,  \phi_{\text{c}}(t,x)  \partial_x V(x).
\label{eq:startSNCalc}\end{equation}
This term shifts the average value of $\phi_{\text{c}}(x,t)$. We compensate for this by shifting the variables of the functional integration~\cite{ponNag1999},
\begin{equation}
    \phi_{\text{c}}(t^{\pm}, x) \equiv \eta_{\text{c}}(t^{\pm}, x) + \Delta\phi_{\text{c}}(t^{\pm},x).
\end{equation}
Here,
\begin{equation}
    \Delta \phi_{\text{c}}(t^\pm, x) = -\frac{1}{2\pi i} \int \d t' \d x' G^{\text{ret}}_\text{c}(t^\pm-t',x,x')\partial_{x'}V(x'),
    \label{eq:shiftSD}
\end{equation}
where $G_{c}^{\text{ret}}$ is the retarded propagator of the charge boson, $G_{c}^{\text{ret}} = G_{c}^{+,+} - G_{c}^{+,-} = G_{c}^{-,+}- G_{c}^{-,-}$.
After this shift, the action is
\begin{multline}
    S_0[\phi_\text{c}, \partial_t \phi_\text{c},...] + S_\text{b}[\phi_\text{c}, \partial_t \phi_\text{c},...] + S_V[\phi_{\text{c}}] \\\rightarrow  S_0[\eta_\text{c}, \partial_t \eta_\text{c},...] + S_\text{b}[\eta_\text{c}, \partial_t \eta_\text{c},...] + \text{const.},
\end{multline}
and the current at time $t$ and position $x$ is
\begin{equation}
    \angled{I(t,x)} = \sigma_0 V - \frac{1}{ 2\pi} \angled{\partial_t \eta_{\text{c}}(t,x)}.
\end{equation}
In this form, we can read off that the expectation value of the backscattering current is  $\langle I_{\text{b}} \rangle = \langle -\partial_t \eta_{\text{c}}/2\pi\rangle$. Using this definition, the fluctuations of the total current are 
\begin{equation}
\begin{split}
        \delta I^2(\omega, x,x') &= \int \d t \, e^{i\omega t} \frac{1}{4\pi^2}\angled{\{ \partial_t \eta_{\text{c}}(t,x), \partial_t \eta_{\text{c}}(0, x')\}} \\&\phantom{===}- 2\angled{I_\text{b}(t,x)}\angled{I_\text{b}(0,x')},
\end{split}
\label{eq:fluctuationDef}\end{equation}
where the expectation value is taken with respect to the out-of-equilibrium system.

As we show in Appendix~\ref{dressedPropag}, the expectation of value of the backscattering current, $I_{b}$, can also be written as
\begin{equation}
    \begin{split}
    \angled{I_{\text{b}}(t,x)}&=\frac{1}{2\pi i }\int \d t' \d x' \partial_t G^{\text{ret}}_c(t,t', x,x') \angled{\frac{\delta S_{\text{b}}}{\delta \eta_{\text{c}}(t', x'_{})}}\\ 
    &\overset{\omega \rightarrow 0}{=} -g_{\infty}e  \int \d t \, e^{i\omega t}\angled{ \frac{\delta S_{\text{b}}}{\delta \eta_{\text{c}}(t, x_0)}}
\label{eq:BSCurrentExp}\end{split}
\end{equation}
where we have used low-energy asymptotics of the retarded propagator in the second line (see Appendix~\ref{seclowenergyasymp}). Note that the final expression is independent of $x$, so we will drop any position dependence of the currents. 

Since we are considering the weak-backscattering limit, the expectation value on the right-hand-side can be calculated perturbatively in powers of the tunneling/backscattering amplitude $\lambda_i$~\cite{ponNag1999,dolcini2005transport}. The first non-trivial contribution comes in at order $\lambda_i^2$. Furthermore, at order $\lambda_i^2$ we have the following equality at low frequency (see Appendix~\ref{noiseweak})
\begin{multline}
        \int \d t \, e^{i\omega t}\angled{\frac{\delta S_{\text{b}}}{\delta \eta_{\text{c}}(t, x_0)}}\overset{\omega \rightarrow 0}{=} \\ -\frac{1}{8 \pi^2 g^2_\infty e^2 }\int \d t \, e^{i\omega t}\angled{\{ \partial_t \eta_{\text{c}}(t,x), \partial_t \eta_{\text{c}}(0, x')\}}
\label{eq:BSCurrentResult}
\end{multline}
for arbitrary values of $x$ and $x'$. Comparing Eq.~\ref{eq:fluctuationDef} and Eq.~\ref{eq:BSCurrentResult} we find that in the $\omega \rightarrow 0$ limit 
\begin{equation}
\begin{split}
   \delta I^2 (\omega) \overset{\omega \rightarrow 0}{=} 2g_{\infty} e \langle I_{b}(\omega)\rangle, 
\end{split}
\end{equation}
where we have dropped the surpurfluous position dependence from the left-hand-side. The Fano Factor is therefore $g_{\infty}$, which is equal to $1$ for Fermi liquid--like leads.

\subsubsection{\label{sectionshotnoiseint}Shot noise of the strongly correlated edge.}
We now consider shot noise for the strongly correlated edge. The analysis is largely the same as for the weakly correlated edge, as the current is still carried by the $\phi_\text{c}$ boson in the strongly correlated edge. The steps from Eqs.~\ref{eq:startSNCalc}-\ref{eq:BSCurrentExp} are unchanged. However, instead of Eq.~\ref{eq:BSCurrentResult} we have (see Appendix~\ref{shotnoiseintedgesection})
\begin{multline}
        \int \d t \, e^{i\omega t} \angled{\frac{\delta S_{\text{b}}}{\delta \eta_{\text{c}}(t, x_0)}}\overset{\omega \rightarrow 0}{=}  \\-\frac{1}{16 \pi^2 g^2_\infty e^2 }\int \d t \, e^{i\omega t}\angled{\{ \partial_t \eta_{\text{c}}(t, x),\partial_t \eta_{\text{c}}(0, x')\}}
\end{multline}
The current noise is then
\begin{equation}
\begin{split}
   \delta I^2(\omega) &\overset{\omega \rightarrow 0}{=}4g_{\infty} e \langle I_{b}(\omega)\rangle, 
\end{split}
\end{equation}
and we find that the Fano Factor is $2$ for Fermi liquid--like leads. 

\section{Discussion.}\label{sec:discussion}
In this work, we analyzed the role of interactions at the edge of the double quantum spin Hall insulators realized in twisted TMDs. We have shown that the edge has two symmetry preserving phases: a weakly correlated phase, a strongly correlated phase. The weakly correlated edge has two pairs of helical modes and is adiabatically connected to two decoupled quantum spin Hall edges. The strongly correlated edge has a single helical mode, and notably, is gapped to single electrons. In contrast, pairs of electrons remain gapless, leading the strongly correlated edge to have a Fano factor of $\mathcal{F}= 2$ in shot noise measurements. Since $\mathcal{F}= 1$ for the weakly correlated edge, shot noise measurements are a concrete method for distinguishing between the two possible edges.

While we have only considered the double quantum spin Hall insulator, it is straightforward to generalize these results to $N$-fold QSHIs. In this case, there is a non-trivial strongly correlated edge where there is a gap to single electrons, and the local degrees of freedom carry charge $Ne$. The Fano factor of the strongly correlated $N$-fold QSHI edge will therefore be $\mathcal{F}= N$. In particular, if the experimentally-observed triple quantum spin Hall insulator~\cite{kang2024evidence,kang2025time} realizes a strongly correlated edge, the Fano factor will be $\mathcal{F}= 3$. 

In general, the existence of the strongly correlated edge depends on the Luttinger parameter of the edge theory. The Luttinger parameter, in turn, depends on the velocity of the edge modes with a lower bare velocity producing a smaller Luttinger parameter and stronger correlations. There are two main factors that determine the bare velocity of the edge modes: the properties of the bulk band dispersion and the microscopic details of the edge termination. The bulk bands of twisted TMDs are quite flat, making them potential candidates for realizing the strongly correlated DQSHI edge. However, the physical boundaries of the twisted TMD must still be taken into account to confirm this expectation. To measure the doubled Fano factor of the strongly correlated edge, it is also important that the constriction used to create the QPC does not destabilize the strongly correlated edge itself. If the single electron gap vanishes at the QPC, then individual electrons can tunnel between the edges, as in the weakly correlated edge. More sophisticated modeling of the boundaries and constrictions of a realistic TMD-device is needed to determine if if the effects discussed here will occur or not.

\section*{Acknowledgments.}
We thank Srinivas Raghu, Kin Fai Mak, Ady Stern and Eslam Khalaf for helpful discussions. AT was supported by the National Science Foundation Graduate Research Fellowship Program under Grant No. DGE-2146755. TD and JMM were supported by a startup fund at Stanford University.

\nocite{*}
%\bibliography{refs}% Produces the bibliography via BibTeX.

\begin{thebibliography}{95}%
\makeatletter
\providecommand \@ifxundefined [1]{%
 \@ifx{#1\undefined}
}%
\providecommand \@ifnum [1]{%
 \ifnum #1\expandafter \@firstoftwo
 \else \expandafter \@secondoftwo
 \fi
}%
\providecommand \@ifx [1]{%
 \ifx #1\expandafter \@firstoftwo
 \else \expandafter \@secondoftwo
 \fi
}%
\providecommand \natexlab [1]{#1}%
\providecommand \enquote  [1]{``#1''}%
\providecommand \bibnamefont  [1]{#1}%
\providecommand \bibfnamefont [1]{#1}%
\providecommand \citenamefont [1]{#1}%
\providecommand \href@noop [0]{\@secondoftwo}%
\providecommand \href [0]{\begingroup \@sanitize@url \@href}%
\providecommand \@href[1]{\@@startlink{#1}\@@href}%
\providecommand \@@href[1]{\endgroup#1\@@endlink}%
\providecommand \@sanitize@url [0]{\catcode `\\12\catcode `\$12\catcode `\&12\catcode `\#12\catcode `\^12\catcode `\_12\catcode `\%12\relax}%
\providecommand \@@startlink[1]{}%
\providecommand \@@endlink[0]{}%
\providecommand \url  [0]{\begingroup\@sanitize@url \@url }%
\providecommand \@url [1]{\endgroup\@href {#1}{\urlprefix }}%
\providecommand \urlprefix  [0]{URL }%
\providecommand \Eprint [0]{\href }%
\providecommand \doibase [0]{https://doi.org/}%
\providecommand \selectlanguage [0]{\@gobble}%
\providecommand \bibinfo  [0]{\@secondoftwo}%
\providecommand \bibfield  [0]{\@secondoftwo}%
\providecommand \translation [1]{[#1]}%
\providecommand \BibitemOpen [0]{}%
\providecommand \bibitemStop [0]{}%
\providecommand \bibitemNoStop [0]{.\EOS\space}%
\providecommand \EOS [0]{\spacefactor3000\relax}%
\providecommand \BibitemShut  [1]{\csname bibitem#1\endcsname}%
\let\auto@bib@innerbib\@empty
%</preamble>
\bibitem [{\citenamefont {K{\"o}nig}\ \emph {et~al.}(2008)\citenamefont {K{\"o}nig}, \citenamefont {Buhmann}, \citenamefont {W.~Molenkamp}, \citenamefont {Hughes}, \citenamefont {Liu}, \citenamefont {Qi},\ and\ \citenamefont {Zhang}}]{konig2008quantum}%
  \BibitemOpen
  \bibfield  {author} {\bibinfo {author} {\bibfnamefont {M.}~\bibnamefont {K{\"o}nig}}, \bibinfo {author} {\bibfnamefont {H.}~\bibnamefont {Buhmann}}, \bibinfo {author} {\bibfnamefont {L.}~\bibnamefont {W.~Molenkamp}}, \bibinfo {author} {\bibfnamefont {T.}~\bibnamefont {Hughes}}, \bibinfo {author} {\bibfnamefont {C.-X.}\ \bibnamefont {Liu}}, \bibinfo {author} {\bibfnamefont {X.-L.}\ \bibnamefont {Qi}},\ and\ \bibinfo {author} {\bibfnamefont {S.-C.}\ \bibnamefont {Zhang}},\ }\bibfield  {title} {\bibinfo {title} {The quantum spin hall effect: theory and experiment},\ }\href@noop {} {\bibfield  {journal} {\bibinfo  {journal} {Journal of the Physical Society of Japan}\ }\textbf {\bibinfo {volume} {77}},\ \bibinfo {pages} {031007} (\bibinfo {year} {2008})}\BibitemShut {NoStop}%
\bibitem [{\citenamefont {Sinova}\ \emph {et~al.}(2004)\citenamefont {Sinova}, \citenamefont {Culcer}, \citenamefont {Niu}, \citenamefont {Sinitsyn}, \citenamefont {Jungwirth},\ and\ \citenamefont {MacDonald}}]{sinova2004universal}%
  \BibitemOpen
  \bibfield  {author} {\bibinfo {author} {\bibfnamefont {J.}~\bibnamefont {Sinova}}, \bibinfo {author} {\bibfnamefont {D.}~\bibnamefont {Culcer}}, \bibinfo {author} {\bibfnamefont {Q.}~\bibnamefont {Niu}}, \bibinfo {author} {\bibfnamefont {N.}~\bibnamefont {Sinitsyn}}, \bibinfo {author} {\bibfnamefont {T.}~\bibnamefont {Jungwirth}},\ and\ \bibinfo {author} {\bibfnamefont {A.~H.}\ \bibnamefont {MacDonald}},\ }\bibfield  {title} {\bibinfo {title} {Universal intrinsic spin hall effect},\ }\href@noop {} {\bibfield  {journal} {\bibinfo  {journal} {Physical review letters}\ }\textbf {\bibinfo {volume} {92}},\ \bibinfo {pages} {126603} (\bibinfo {year} {2004})}\BibitemShut {NoStop}%
\bibitem [{\citenamefont {Kane}\ and\ \citenamefont {Mele}(2005{\natexlab{a}})}]{kane2005quantum}%
  \BibitemOpen
  \bibfield  {author} {\bibinfo {author} {\bibfnamefont {C.~L.}\ \bibnamefont {Kane}}\ and\ \bibinfo {author} {\bibfnamefont {E.~J.}\ \bibnamefont {Mele}},\ }\bibfield  {title} {\bibinfo {title} {Quantum spin hall effect in graphene},\ }\href@noop {} {\bibfield  {journal} {\bibinfo  {journal} {Physical review letters}\ }\textbf {\bibinfo {volume} {95}},\ \bibinfo {pages} {226801} (\bibinfo {year} {2005}{\natexlab{a}})}\BibitemShut {NoStop}%
\bibitem [{\citenamefont {Bernevig}\ and\ \citenamefont {Zhang}(2006{\natexlab{a}})}]{bernevig2006quantum}%
  \BibitemOpen
  \bibfield  {author} {\bibinfo {author} {\bibfnamefont {B.~A.}\ \bibnamefont {Bernevig}}\ and\ \bibinfo {author} {\bibfnamefont {S.-C.}\ \bibnamefont {Zhang}},\ }\bibfield  {title} {\bibinfo {title} {Quantum spin hall effect},\ }\href@noop {} {\bibfield  {journal} {\bibinfo  {journal} {Physical review letters}\ }\textbf {\bibinfo {volume} {96}},\ \bibinfo {pages} {106802} (\bibinfo {year} {2006}{\natexlab{a}})}\BibitemShut {NoStop}%
\bibitem [{\citenamefont {Bernevig}\ \emph {et~al.}(2006)\citenamefont {Bernevig}, \citenamefont {Hughes},\ and\ \citenamefont {Zhang}}]{bernevig2006quantum2}%
  \BibitemOpen
  \bibfield  {author} {\bibinfo {author} {\bibfnamefont {B.~A.}\ \bibnamefont {Bernevig}}, \bibinfo {author} {\bibfnamefont {T.~L.}\ \bibnamefont {Hughes}},\ and\ \bibinfo {author} {\bibfnamefont {S.-C.}\ \bibnamefont {Zhang}},\ }\bibfield  {title} {\bibinfo {title} {Quantum spin hall effect and topological phase transition in hgte quantum wells},\ }\href@noop {} {\bibfield  {journal} {\bibinfo  {journal} {science}\ }\textbf {\bibinfo {volume} {314}},\ \bibinfo {pages} {1757} (\bibinfo {year} {2006})}\BibitemShut {NoStop}%
\bibitem [{\citenamefont {Konig}\ \emph {et~al.}(2007)\citenamefont {Konig}, \citenamefont {Wiedmann}, \citenamefont {Brune}, \citenamefont {Roth}, \citenamefont {Buhmann}, \citenamefont {Molenkamp}, \citenamefont {Qi},\ and\ \citenamefont {Zhang}}]{konig2007quantum}%
  \BibitemOpen
  \bibfield  {author} {\bibinfo {author} {\bibfnamefont {M.}~\bibnamefont {Konig}}, \bibinfo {author} {\bibfnamefont {S.}~\bibnamefont {Wiedmann}}, \bibinfo {author} {\bibfnamefont {C.}~\bibnamefont {Brune}}, \bibinfo {author} {\bibfnamefont {A.}~\bibnamefont {Roth}}, \bibinfo {author} {\bibfnamefont {H.}~\bibnamefont {Buhmann}}, \bibinfo {author} {\bibfnamefont {L.~W.}\ \bibnamefont {Molenkamp}}, \bibinfo {author} {\bibfnamefont {X.-L.}\ \bibnamefont {Qi}},\ and\ \bibinfo {author} {\bibfnamefont {S.-C.}\ \bibnamefont {Zhang}},\ }\bibfield  {title} {\bibinfo {title} {Quantum spin hall insulator state in hgte quantum wells},\ }\href@noop {} {\bibfield  {journal} {\bibinfo  {journal} {Science}\ }\textbf {\bibinfo {volume} {318}},\ \bibinfo {pages} {766} (\bibinfo {year} {2007})}\BibitemShut {NoStop}%
\bibitem [{\citenamefont {Wu}\ \emph {et~al.}(2006)\citenamefont {Wu}, \citenamefont {Bernevig},\ and\ \citenamefont {Zhang}}]{wu2006helical}%
  \BibitemOpen
  \bibfield  {author} {\bibinfo {author} {\bibfnamefont {C.}~\bibnamefont {Wu}}, \bibinfo {author} {\bibfnamefont {B.~A.}\ \bibnamefont {Bernevig}},\ and\ \bibinfo {author} {\bibfnamefont {S.-C.}\ \bibnamefont {Zhang}},\ }\bibfield  {title} {\bibinfo {title} {Helical liquid and the edge of quantum spin hall systems},\ }\href@noop {} {\bibfield  {journal} {\bibinfo  {journal} {Physical review letters}\ }\textbf {\bibinfo {volume} {96}},\ \bibinfo {pages} {106401} (\bibinfo {year} {2006})}\BibitemShut {NoStop}%
\bibitem [{\citenamefont {Giamarchi}(2003)}]{giamarchi2003quantum}%
  \BibitemOpen
  \bibfield  {author} {\bibinfo {author} {\bibfnamefont {T.}~\bibnamefont {Giamarchi}},\ }\href@noop {} {\emph {\bibinfo {title} {Quantum physics in one dimension}}},\ Vol.\ \bibinfo {volume} {121}\ (\bibinfo  {publisher} {Clarendon press},\ \bibinfo {year} {2003})\BibitemShut {NoStop}%
\bibitem [{\citenamefont {Haldane}(1981)}]{haldane1981luttinger}%
  \BibitemOpen
  \bibfield  {author} {\bibinfo {author} {\bibfnamefont {F.~D.~M.}\ \bibnamefont {Haldane}},\ }\bibfield  {title} {\bibinfo {title} {'luttinger liquid theory'of one-dimensional quantum fluids. i. properties of the luttinger model and their extension to the general 1d interacting spinless fermi gas},\ }\href@noop {} {\bibfield  {journal} {\bibinfo  {journal} {Journal of Physics C: Solid State Physics}\ }\textbf {\bibinfo {volume} {14}},\ \bibinfo {pages} {2585} (\bibinfo {year} {1981})}\BibitemShut {NoStop}%
\bibitem [{\citenamefont {Kane}\ and\ \citenamefont {Mele}(2005{\natexlab{b}})}]{kane2005z}%
  \BibitemOpen
  \bibfield  {author} {\bibinfo {author} {\bibfnamefont {C.~L.}\ \bibnamefont {Kane}}\ and\ \bibinfo {author} {\bibfnamefont {E.~J.}\ \bibnamefont {Mele}},\ }\bibfield  {title} {\bibinfo {title} {Z 2 topological order and the quantum spin hall effect},\ }\href@noop {} {\bibfield  {journal} {\bibinfo  {journal} {Physical review letters}\ }\textbf {\bibinfo {volume} {95}},\ \bibinfo {pages} {146802} (\bibinfo {year} {2005}{\natexlab{b}})}\BibitemShut {NoStop}%
\bibitem [{\citenamefont {Kitaev}(2009)}]{kitaev2009periodic}%
  \BibitemOpen
  \bibfield  {author} {\bibinfo {author} {\bibfnamefont {A.}~\bibnamefont {Kitaev}},\ }\bibfield  {title} {\bibinfo {title} {Periodic table for topological insulators and superconductors},\ }in\ \href@noop {} {\emph {\bibinfo {booktitle} {AIP conference proceedings}}},\ Vol.\ \bibinfo {volume} {1134}\ (\bibinfo {organization} {American Institute of Physics},\ \bibinfo {year} {2009})\ pp.\ \bibinfo {pages} {22--30}\BibitemShut {NoStop}%
\bibitem [{\citenamefont {Chiu}\ \emph {et~al.}(2016)\citenamefont {Chiu}, \citenamefont {Teo}, \citenamefont {Schnyder},\ and\ \citenamefont {Ryu}}]{chiu2016classification}%
  \BibitemOpen
  \bibfield  {author} {\bibinfo {author} {\bibfnamefont {C.-K.}\ \bibnamefont {Chiu}}, \bibinfo {author} {\bibfnamefont {J.~C.}\ \bibnamefont {Teo}}, \bibinfo {author} {\bibfnamefont {A.~P.}\ \bibnamefont {Schnyder}},\ and\ \bibinfo {author} {\bibfnamefont {S.}~\bibnamefont {Ryu}},\ }\bibfield  {title} {\bibinfo {title} {Classification of topological quantum matter with symmetries},\ }\href@noop {} {\bibfield  {journal} {\bibinfo  {journal} {Reviews of Modern Physics}\ }\textbf {\bibinfo {volume} {88}},\ \bibinfo {pages} {035005} (\bibinfo {year} {2016})}\BibitemShut {NoStop}%
\bibitem [{\citenamefont {Maciejko}\ \emph {et~al.}(2011)\citenamefont {Maciejko}, \citenamefont {Hughes},\ and\ \citenamefont {Zhang}}]{maciejko2011quantum}%
  \BibitemOpen
  \bibfield  {author} {\bibinfo {author} {\bibfnamefont {J.}~\bibnamefont {Maciejko}}, \bibinfo {author} {\bibfnamefont {T.~L.}\ \bibnamefont {Hughes}},\ and\ \bibinfo {author} {\bibfnamefont {S.-C.}\ \bibnamefont {Zhang}},\ }\bibfield  {title} {\bibinfo {title} {The quantum spin hall effect},\ }\href@noop {} {\bibfield  {journal} {\bibinfo  {journal} {Annu. Rev. Condens. Matter Phys.}\ }\textbf {\bibinfo {volume} {2}},\ \bibinfo {pages} {31} (\bibinfo {year} {2011})}\BibitemShut {NoStop}%
\bibitem [{\citenamefont {Qi}\ \emph {et~al.}(2006)\citenamefont {Qi}, \citenamefont {Wu},\ and\ \citenamefont {Zhang}}]{PhysRevB.74.085308}%
  \BibitemOpen
  \bibfield  {author} {\bibinfo {author} {\bibfnamefont {X.-L.}\ \bibnamefont {Qi}}, \bibinfo {author} {\bibfnamefont {Y.-S.}\ \bibnamefont {Wu}},\ and\ \bibinfo {author} {\bibfnamefont {S.-C.}\ \bibnamefont {Zhang}},\ }\bibfield  {title} {\bibinfo {title} {Topological quantization of the spin hall effect in two-dimensional paramagnetic semiconductors},\ }\href {https://doi.org/10.1103/PhysRevB.74.085308} {\bibfield  {journal} {\bibinfo  {journal} {Phys. Rev. B}\ }\textbf {\bibinfo {volume} {74}},\ \bibinfo {pages} {085308} (\bibinfo {year} {2006})}\BibitemShut {NoStop}%
\bibitem [{\citenamefont {Prodan}(2009)}]{PhysRevB.80.125327}%
  \BibitemOpen
  \bibfield  {author} {\bibinfo {author} {\bibfnamefont {E.}~\bibnamefont {Prodan}},\ }\bibfield  {title} {\bibinfo {title} {Robustness of the spin-chern number},\ }\href {https://doi.org/10.1103/PhysRevB.80.125327} {\bibfield  {journal} {\bibinfo  {journal} {Phys. Rev. B}\ }\textbf {\bibinfo {volume} {80}},\ \bibinfo {pages} {125327} (\bibinfo {year} {2009})}\BibitemShut {NoStop}%
\bibitem [{\citenamefont {Campos}\ \emph {et~al.}(2024)\citenamefont {Campos}, \citenamefont {Penteado}, \citenamefont {Zanon}, \citenamefont {Faria~Junior}, \citenamefont {Candido},\ and\ \citenamefont {Egues}}]{PhysRevB.110.195142}%
  \BibitemOpen
  \bibfield  {author} {\bibinfo {author} {\bibfnamefont {W.~H.}\ \bibnamefont {Campos}}, \bibinfo {author} {\bibfnamefont {P.~H.}\ \bibnamefont {Penteado}}, \bibinfo {author} {\bibfnamefont {J.}~\bibnamefont {Zanon}}, \bibinfo {author} {\bibfnamefont {P.~E.}\ \bibnamefont {Faria~Junior}}, \bibinfo {author} {\bibfnamefont {D.~R.}\ \bibnamefont {Candido}},\ and\ \bibinfo {author} {\bibfnamefont {J.~C.}\ \bibnamefont {Egues}},\ }\bibfield  {title} {\bibinfo {title} {Dual topological insulator with mirror symmetry protected helical edge states},\ }\href {https://doi.org/10.1103/PhysRevB.110.195142} {\bibfield  {journal} {\bibinfo  {journal} {Phys. Rev. B}\ }\textbf {\bibinfo {volume} {110}},\ \bibinfo {pages} {195142} (\bibinfo {year} {2024})}\BibitemShut {NoStop}%
\bibitem [{\citenamefont {Kang}\ \emph {et~al.}(2024{\natexlab{a}})\citenamefont {Kang}, \citenamefont {Qiu}, \citenamefont {Watanabe}, \citenamefont {Taniguchi}, \citenamefont {Shan},\ and\ \citenamefont {Mak}}]{kang2024observationdoublequantumspin}%
  \BibitemOpen
  \bibfield  {author} {\bibinfo {author} {\bibfnamefont {K.}~\bibnamefont {Kang}}, \bibinfo {author} {\bibfnamefont {Y.}~\bibnamefont {Qiu}}, \bibinfo {author} {\bibfnamefont {K.}~\bibnamefont {Watanabe}}, \bibinfo {author} {\bibfnamefont {T.}~\bibnamefont {Taniguchi}}, \bibinfo {author} {\bibfnamefont {J.}~\bibnamefont {Shan}},\ and\ \bibinfo {author} {\bibfnamefont {K.~F.}\ \bibnamefont {Mak}},\ }\href {https://arxiv.org/abs/2402.04196} {\bibinfo {title} {Observation of the double quantum spin hall phase in moir\'e wse2}} (\bibinfo {year} {2024}{\natexlab{a}}),\ \Eprint {https://arxiv.org/abs/2402.04196} {arXiv:2402.04196 [cond-mat.mes-hall]} \BibitemShut {NoStop}%
\bibitem [{\citenamefont {Kang}\ \emph {et~al.}(2024{\natexlab{b}})\citenamefont {Kang}, \citenamefont {Qiu}, \citenamefont {Watanabe}, \citenamefont {Taniguchi}, \citenamefont {Shan},\ and\ \citenamefont {Mak}}]{kang2024double}%
  \BibitemOpen
  \bibfield  {author} {\bibinfo {author} {\bibfnamefont {K.}~\bibnamefont {Kang}}, \bibinfo {author} {\bibfnamefont {Y.}~\bibnamefont {Qiu}}, \bibinfo {author} {\bibfnamefont {K.}~\bibnamefont {Watanabe}}, \bibinfo {author} {\bibfnamefont {T.}~\bibnamefont {Taniguchi}}, \bibinfo {author} {\bibfnamefont {J.}~\bibnamefont {Shan}},\ and\ \bibinfo {author} {\bibfnamefont {K.~F.}\ \bibnamefont {Mak}},\ }\bibfield  {title} {\bibinfo {title} {Double quantum spin hall phase in moir{\'e} wse2},\ }\href@noop {} {\bibfield  {journal} {\bibinfo  {journal} {Nano Letters}\ }\textbf {\bibinfo {volume} {24}},\ \bibinfo {pages} {14901} (\bibinfo {year} {2024}{\natexlab{b}})}\BibitemShut {NoStop}%
\bibitem [{\citenamefont {Kang}\ \emph {et~al.}(2024{\natexlab{c}})\citenamefont {Kang}, \citenamefont {Shen}, \citenamefont {Qiu}, \citenamefont {Zeng}, \citenamefont {Xia}, \citenamefont {Watanabe}, \citenamefont {Taniguchi}, \citenamefont {Shan},\ and\ \citenamefont {Mak}}]{kang2024evidence}%
  \BibitemOpen
  \bibfield  {author} {\bibinfo {author} {\bibfnamefont {K.}~\bibnamefont {Kang}}, \bibinfo {author} {\bibfnamefont {B.}~\bibnamefont {Shen}}, \bibinfo {author} {\bibfnamefont {Y.}~\bibnamefont {Qiu}}, \bibinfo {author} {\bibfnamefont {Y.}~\bibnamefont {Zeng}}, \bibinfo {author} {\bibfnamefont {Z.}~\bibnamefont {Xia}}, \bibinfo {author} {\bibfnamefont {K.}~\bibnamefont {Watanabe}}, \bibinfo {author} {\bibfnamefont {T.}~\bibnamefont {Taniguchi}}, \bibinfo {author} {\bibfnamefont {J.}~\bibnamefont {Shan}},\ and\ \bibinfo {author} {\bibfnamefont {K.~F.}\ \bibnamefont {Mak}},\ }\bibfield  {title} {\bibinfo {title} {Evidence of the fractional quantum spin hall effect in moir{\'e} mote2},\ }\href@noop {} {\bibfield  {journal} {\bibinfo  {journal} {Nature}\ }\textbf {\bibinfo {volume} {628}},\ \bibinfo {pages} {522} (\bibinfo {year} {2024}{\natexlab{c}})}\BibitemShut {NoStop}%
\bibitem [{\citenamefont {Xu}\ \emph {et~al.}(2024)\citenamefont {Xu}, \citenamefont {Chang}, \citenamefont {Xiao}, \citenamefont {Zhang}, \citenamefont {Liu}, \citenamefont {Sun}, \citenamefont {Mao}, \citenamefont {Peshcherenko}, \citenamefont {Li}, \citenamefont {Watanabe} \emph {et~al.}}]{xu2024interplay}%
  \BibitemOpen
  \bibfield  {author} {\bibinfo {author} {\bibfnamefont {F.}~\bibnamefont {Xu}}, \bibinfo {author} {\bibfnamefont {X.}~\bibnamefont {Chang}}, \bibinfo {author} {\bibfnamefont {J.}~\bibnamefont {Xiao}}, \bibinfo {author} {\bibfnamefont {Y.}~\bibnamefont {Zhang}}, \bibinfo {author} {\bibfnamefont {F.}~\bibnamefont {Liu}}, \bibinfo {author} {\bibfnamefont {Z.}~\bibnamefont {Sun}}, \bibinfo {author} {\bibfnamefont {N.}~\bibnamefont {Mao}}, \bibinfo {author} {\bibfnamefont {N.}~\bibnamefont {Peshcherenko}}, \bibinfo {author} {\bibfnamefont {J.}~\bibnamefont {Li}}, \bibinfo {author} {\bibfnamefont {K.}~\bibnamefont {Watanabe}}, \emph {et~al.},\ }\bibfield  {title} {\bibinfo {title} {Interplay between topology and correlations in the second moir$\backslash$'e band of twisted bilayer mote2},\ }\href@noop {} {\bibfield  {journal} {\bibinfo  {journal} {arXiv preprint arXiv:2406.09687}\ } (\bibinfo {year} {2024})}\BibitemShut {NoStop}%
\bibitem [{\citenamefont {Kang}\ \emph {et~al.}(2025)\citenamefont {Kang}, \citenamefont {Qiu}, \citenamefont {Shen}, \citenamefont {Lee}, \citenamefont {Xia}, \citenamefont {Zeng}, \citenamefont {Watanabe}, \citenamefont {Taniguchi}, \citenamefont {Shan},\ and\ \citenamefont {Mak}}]{kang2025time}%
  \BibitemOpen
  \bibfield  {author} {\bibinfo {author} {\bibfnamefont {K.}~\bibnamefont {Kang}}, \bibinfo {author} {\bibfnamefont {Y.}~\bibnamefont {Qiu}}, \bibinfo {author} {\bibfnamefont {B.}~\bibnamefont {Shen}}, \bibinfo {author} {\bibfnamefont {K.}~\bibnamefont {Lee}}, \bibinfo {author} {\bibfnamefont {Z.}~\bibnamefont {Xia}}, \bibinfo {author} {\bibfnamefont {Y.}~\bibnamefont {Zeng}}, \bibinfo {author} {\bibfnamefont {K.}~\bibnamefont {Watanabe}}, \bibinfo {author} {\bibfnamefont {T.}~\bibnamefont {Taniguchi}}, \bibinfo {author} {\bibfnamefont {J.}~\bibnamefont {Shan}},\ and\ \bibinfo {author} {\bibfnamefont {K.~F.}\ \bibnamefont {Mak}},\ }\bibfield  {title} {\bibinfo {title} {Time-reversal symmetry breaking fractional quantum spin hall insulator in moir$\backslash$'e mote2},\ }\href@noop {} {\bibfield  {journal} {\bibinfo  {journal} {arXiv preprint arXiv:2501.02525}\ } (\bibinfo {year} {2025})}\BibitemShut {NoStop}%
\bibitem [{\citenamefont {Schulz}(1996)}]{schulz1996phases}%
  \BibitemOpen
  \bibfield  {author} {\bibinfo {author} {\bibfnamefont {H.}~\bibnamefont {Schulz}},\ }\bibfield  {title} {\bibinfo {title} {Phases of two coupled luttinger liquids},\ }\href@noop {} {\bibfield  {journal} {\bibinfo  {journal} {Physical Review B}\ }\textbf {\bibinfo {volume} {53}},\ \bibinfo {pages} {R2959} (\bibinfo {year} {1996})}\BibitemShut {NoStop}%
\bibitem [{\citenamefont {Wu}\ \emph {et~al.}(2003)\citenamefont {Wu}, \citenamefont {Liu},\ and\ \citenamefont {Fradkin}}]{wu2003competing}%
  \BibitemOpen
  \bibfield  {author} {\bibinfo {author} {\bibfnamefont {C.}~\bibnamefont {Wu}}, \bibinfo {author} {\bibfnamefont {W.~V.}\ \bibnamefont {Liu}},\ and\ \bibinfo {author} {\bibfnamefont {E.}~\bibnamefont {Fradkin}},\ }\bibfield  {title} {\bibinfo {title} {Competing orders in coupled luttinger liquids},\ }\href@noop {} {\bibfield  {journal} {\bibinfo  {journal} {Physical Review B}\ }\textbf {\bibinfo {volume} {68}},\ \bibinfo {pages} {115104} (\bibinfo {year} {2003})}\BibitemShut {NoStop}%
\bibitem [{\citenamefont {Tanaka}\ and\ \citenamefont {Nagaosa}(2009)}]{tanaka2009two}%
  \BibitemOpen
  \bibfield  {author} {\bibinfo {author} {\bibfnamefont {Y.}~\bibnamefont {Tanaka}}\ and\ \bibinfo {author} {\bibfnamefont {N.}~\bibnamefont {Nagaosa}},\ }\bibfield  {title} {\bibinfo {title} {Two interacting helical edge modes in quantum spin hall systems},\ }\href@noop {} {\bibfield  {journal} {\bibinfo  {journal} {Physical review letters}\ }\textbf {\bibinfo {volume} {103}},\ \bibinfo {pages} {166403} (\bibinfo {year} {2009})}\BibitemShut {NoStop}%
\bibitem [{\citenamefont {Liu}\ \emph {et~al.}(2011)\citenamefont {Liu}, \citenamefont {Budich}, \citenamefont {Recher},\ and\ \citenamefont {Trauzettel}}]{liu2011charge}%
  \BibitemOpen
  \bibfield  {author} {\bibinfo {author} {\bibfnamefont {C.-X.}\ \bibnamefont {Liu}}, \bibinfo {author} {\bibfnamefont {J.~C.}\ \bibnamefont {Budich}}, \bibinfo {author} {\bibfnamefont {P.}~\bibnamefont {Recher}},\ and\ \bibinfo {author} {\bibfnamefont {B.}~\bibnamefont {Trauzettel}},\ }\bibfield  {title} {\bibinfo {title} {Charge-spin duality in nonequilibrium transport of helical liquids},\ }\href@noop {} {\bibfield  {journal} {\bibinfo  {journal} {Physical Review B—Condensed Matter and Materials Physics}\ }\textbf {\bibinfo {volume} {83}},\ \bibinfo {pages} {035407} (\bibinfo {year} {2011})}\BibitemShut {NoStop}%
\bibitem [{\citenamefont {Sela}\ \emph {et~al.}(2011)\citenamefont {Sela}, \citenamefont {Altland},\ and\ \citenamefont {Rosch}}]{sela2011majorana}%
  \BibitemOpen
  \bibfield  {author} {\bibinfo {author} {\bibfnamefont {E.}~\bibnamefont {Sela}}, \bibinfo {author} {\bibfnamefont {A.}~\bibnamefont {Altland}},\ and\ \bibinfo {author} {\bibfnamefont {A.}~\bibnamefont {Rosch}},\ }\bibfield  {title} {\bibinfo {title} {Majorana fermions in strongly interacting helical liquids},\ }\href@noop {} {\bibfield  {journal} {\bibinfo  {journal} {Physical Review B—Condensed Matter and Materials Physics}\ }\textbf {\bibinfo {volume} {84}},\ \bibinfo {pages} {085114} (\bibinfo {year} {2011})}\BibitemShut {NoStop}%
\bibitem [{\citenamefont {Schmidt}\ \emph {et~al.}(2012)\citenamefont {Schmidt}, \citenamefont {Rachel}, \citenamefont {von Oppen},\ and\ \citenamefont {Glazman}}]{schmidt2012inelastic}%
  \BibitemOpen
  \bibfield  {author} {\bibinfo {author} {\bibfnamefont {T.~L.}\ \bibnamefont {Schmidt}}, \bibinfo {author} {\bibfnamefont {S.}~\bibnamefont {Rachel}}, \bibinfo {author} {\bibfnamefont {F.}~\bibnamefont {von Oppen}},\ and\ \bibinfo {author} {\bibfnamefont {L.~I.}\ \bibnamefont {Glazman}},\ }\bibfield  {title} {\bibinfo {title} {Inelastic electron backscattering in a generic helical edge channel},\ }\href@noop {} {\bibfield  {journal} {\bibinfo  {journal} {Physical review letters}\ }\textbf {\bibinfo {volume} {108}},\ \bibinfo {pages} {156402} (\bibinfo {year} {2012})}\BibitemShut {NoStop}%
\bibitem [{\citenamefont {Keselman}\ and\ \citenamefont {Berg}(2015)}]{keselman2015gapless}%
  \BibitemOpen
  \bibfield  {author} {\bibinfo {author} {\bibfnamefont {A.}~\bibnamefont {Keselman}}\ and\ \bibinfo {author} {\bibfnamefont {E.}~\bibnamefont {Berg}},\ }\bibfield  {title} {\bibinfo {title} {Gapless symmetry-protected topological phase of fermions in one dimension},\ }\href@noop {} {\bibfield  {journal} {\bibinfo  {journal} {Physical Review B}\ }\textbf {\bibinfo {volume} {91}},\ \bibinfo {pages} {235309} (\bibinfo {year} {2015})}\BibitemShut {NoStop}%
\bibitem [{\citenamefont {Oreg}\ \emph {et~al.}(2014)\citenamefont {Oreg}, \citenamefont {Sela},\ and\ \citenamefont {Stern}}]{oreg2014fractional}%
  \BibitemOpen
  \bibfield  {author} {\bibinfo {author} {\bibfnamefont {Y.}~\bibnamefont {Oreg}}, \bibinfo {author} {\bibfnamefont {E.}~\bibnamefont {Sela}},\ and\ \bibinfo {author} {\bibfnamefont {A.}~\bibnamefont {Stern}},\ }\bibfield  {title} {\bibinfo {title} {Fractional helical liquids in quantum wires},\ }\href@noop {} {\bibfield  {journal} {\bibinfo  {journal} {Physical Review B}\ }\textbf {\bibinfo {volume} {89}},\ \bibinfo {pages} {115402} (\bibinfo {year} {2014})}\BibitemShut {NoStop}%
\bibitem [{\citenamefont {Werman}\ and\ \citenamefont {Berg}(2015)}]{werman2015exciton}%
  \BibitemOpen
  \bibfield  {author} {\bibinfo {author} {\bibfnamefont {Y.}~\bibnamefont {Werman}}\ and\ \bibinfo {author} {\bibfnamefont {E.}~\bibnamefont {Berg}},\ }\bibfield  {title} {\bibinfo {title} {Exciton quasicondensation in one-dimensional systems},\ }\href@noop {} {\bibfield  {journal} {\bibinfo  {journal} {Physical Review B}\ }\textbf {\bibinfo {volume} {91}},\ \bibinfo {pages} {245410} (\bibinfo {year} {2015})}\BibitemShut {NoStop}%
\bibitem [{\citenamefont {Santos}\ and\ \citenamefont {Gutman}(2015)}]{santos2015interaction}%
  \BibitemOpen
  \bibfield  {author} {\bibinfo {author} {\bibfnamefont {R.~A.}\ \bibnamefont {Santos}}\ and\ \bibinfo {author} {\bibfnamefont {D.}~\bibnamefont {Gutman}},\ }\bibfield  {title} {\bibinfo {title} {Interaction-protected topological insulators with time reversal symmetry},\ }\href@noop {} {\bibfield  {journal} {\bibinfo  {journal} {Physical Review B}\ }\textbf {\bibinfo {volume} {92}},\ \bibinfo {pages} {075135} (\bibinfo {year} {2015})}\BibitemShut {NoStop}%
\bibitem [{\citenamefont {Santos}\ \emph {et~al.}(2016)\citenamefont {Santos}, \citenamefont {Gutman},\ and\ \citenamefont {Carr}}]{santos2016phase}%
  \BibitemOpen
  \bibfield  {author} {\bibinfo {author} {\bibfnamefont {R.~A.}\ \bibnamefont {Santos}}, \bibinfo {author} {\bibfnamefont {D.}~\bibnamefont {Gutman}},\ and\ \bibinfo {author} {\bibfnamefont {S.~T.}\ \bibnamefont {Carr}},\ }\bibfield  {title} {\bibinfo {title} {Phase diagram of two interacting helical states},\ }\href@noop {} {\bibfield  {journal} {\bibinfo  {journal} {Physical Review B}\ }\textbf {\bibinfo {volume} {93}},\ \bibinfo {pages} {235436} (\bibinfo {year} {2016})}\BibitemShut {NoStop}%
\bibitem [{\citenamefont {May-Mann}\ \emph {et~al.}(2024)\citenamefont {May-Mann}, \citenamefont {Stern},\ and\ \citenamefont {Devakul}}]{maymann2024theoryhalfintegerfractionalquantum}%
  \BibitemOpen
  \bibfield  {author} {\bibinfo {author} {\bibfnamefont {J.}~\bibnamefont {May-Mann}}, \bibinfo {author} {\bibfnamefont {A.}~\bibnamefont {Stern}},\ and\ \bibinfo {author} {\bibfnamefont {T.}~\bibnamefont {Devakul}},\ }\href {https://arxiv.org/abs/2403.03964} {\bibinfo {title} {Theory of half-integer fractional quantum spin hall insulator edges}} (\bibinfo {year} {2024}),\ \Eprint {https://arxiv.org/abs/2403.03964} {arXiv:2403.03964 [cond-mat.mes-hall]} \BibitemShut {NoStop}%
\bibitem [{\citenamefont {Amaricci}\ \emph {et~al.}(2017)\citenamefont {Amaricci}, \citenamefont {Privitera}, \citenamefont {Petocchi}, \citenamefont {Capone}, \citenamefont {Sangiovanni},\ and\ \citenamefont {Trauzettel}}]{PhysRevB.95.205120}%
  \BibitemOpen
  \bibfield  {author} {\bibinfo {author} {\bibfnamefont {A.}~\bibnamefont {Amaricci}}, \bibinfo {author} {\bibfnamefont {L.}~\bibnamefont {Privitera}}, \bibinfo {author} {\bibfnamefont {F.}~\bibnamefont {Petocchi}}, \bibinfo {author} {\bibfnamefont {M.}~\bibnamefont {Capone}}, \bibinfo {author} {\bibfnamefont {G.}~\bibnamefont {Sangiovanni}},\ and\ \bibinfo {author} {\bibfnamefont {B.}~\bibnamefont {Trauzettel}},\ }\bibfield  {title} {\bibinfo {title} {Edge state reconstruction from strong correlations in quantum spin hall insulators},\ }\href {https://doi.org/10.1103/PhysRevB.95.205120} {\bibfield  {journal} {\bibinfo  {journal} {Phys. Rev. B}\ }\textbf {\bibinfo {volume} {95}},\ \bibinfo {pages} {205120} (\bibinfo {year} {2017})}\BibitemShut {NoStop}%
\bibitem [{\citenamefont {Yurkevich}\ and\ \citenamefont {Kagalovsky}(2021)}]{yurkevich2021superconducting}%
  \BibitemOpen
  \bibfield  {author} {\bibinfo {author} {\bibfnamefont {I.}~\bibnamefont {Yurkevich}}\ and\ \bibinfo {author} {\bibfnamefont {V.}~\bibnamefont {Kagalovsky}},\ }\bibfield  {title} {\bibinfo {title} {Superconducting edge states in a topological insulator},\ }\href@noop {} {\bibfield  {journal} {\bibinfo  {journal} {Scientific Reports}\ }\textbf {\bibinfo {volume} {11}},\ \bibinfo {pages} {18400} (\bibinfo {year} {2021})}\BibitemShut {NoStop}%
\bibitem [{\citenamefont {Hung}\ \emph {et~al.}(2025)\citenamefont {Hung}, \citenamefont {Hsu},\ and\ \citenamefont {Bansil}}]{9n85-r2xw}%
  \BibitemOpen
  \bibfield  {author} {\bibinfo {author} {\bibfnamefont {Y.-C.}\ \bibnamefont {Hung}}, \bibinfo {author} {\bibfnamefont {C.-H.}\ \bibnamefont {Hsu}},\ and\ \bibinfo {author} {\bibfnamefont {A.}~\bibnamefont {Bansil}},\ }\bibfield  {title} {\bibinfo {title} {Majorana kramers pairs in synthetic high-spin chern insulators},\ }\href {https://doi.org/10.1103/9n85-r2xw} {\bibfield  {journal} {\bibinfo  {journal} {Phys. Rev. B}\ }\textbf {\bibinfo {volume} {111}},\ \bibinfo {pages} {245145} (\bibinfo {year} {2025})}\BibitemShut {NoStop}%
\bibitem [{\citenamefont {Landauer}(1993)}]{landauer1993solid}%
  \BibitemOpen
  \bibfield  {author} {\bibinfo {author} {\bibfnamefont {R.}~\bibnamefont {Landauer}},\ }\bibfield  {title} {\bibinfo {title} {Solid-state shot noise},\ }\href@noop {} {\bibfield  {journal} {\bibinfo  {journal} {Physical Review B}\ }\textbf {\bibinfo {volume} {47}},\ \bibinfo {pages} {16427} (\bibinfo {year} {1993})}\BibitemShut {NoStop}%
\bibitem [{\citenamefont {Sp\aa{}nsl\"att}\ \emph {et~al.}(2019)\citenamefont {Sp\aa{}nsl\"att}, \citenamefont {Park}, \citenamefont {Gefen},\ and\ \citenamefont {Mirlin}}]{PhysRevLett.123.137701}%
  \BibitemOpen
  \bibfield  {author} {\bibinfo {author} {\bibfnamefont {C.}~\bibnamefont {Sp\aa{}nsl\"att}}, \bibinfo {author} {\bibfnamefont {J.}~\bibnamefont {Park}}, \bibinfo {author} {\bibfnamefont {Y.}~\bibnamefont {Gefen}},\ and\ \bibinfo {author} {\bibfnamefont {A.~D.}\ \bibnamefont {Mirlin}},\ }\bibfield  {title} {\bibinfo {title} {Topological classification of shot noise on fractional quantum hall edges},\ }\href {https://doi.org/10.1103/PhysRevLett.123.137701} {\bibfield  {journal} {\bibinfo  {journal} {Phys. Rev. Lett.}\ }\textbf {\bibinfo {volume} {123}},\ \bibinfo {pages} {137701} (\bibinfo {year} {2019})}\BibitemShut {NoStop}%
\bibitem [{\citenamefont {Aseev}\ and\ \citenamefont {Nagaev}(2016)}]{PhysRevB.94.045425}%
  \BibitemOpen
  \bibfield  {author} {\bibinfo {author} {\bibfnamefont {P.~P.}\ \bibnamefont {Aseev}}\ and\ \bibinfo {author} {\bibfnamefont {K.~E.}\ \bibnamefont {Nagaev}},\ }\bibfield  {title} {\bibinfo {title} {Shot noise in the edge states of two-dimensional topological insulators},\ }\href {https://doi.org/10.1103/PhysRevB.94.045425} {\bibfield  {journal} {\bibinfo  {journal} {Phys. Rev. B}\ }\textbf {\bibinfo {volume} {94}},\ \bibinfo {pages} {045425} (\bibinfo {year} {2016})}\BibitemShut {NoStop}%
\bibitem [{\citenamefont {Niyazov}\ \emph {et~al.}(2024{\natexlab{a}})\citenamefont {Niyazov}, \citenamefont {Krainov}, \citenamefont {Aristov},\ and\ \citenamefont {Kachorovskii}}]{niyazov2024shot}%
  \BibitemOpen
  \bibfield  {author} {\bibinfo {author} {\bibfnamefont {R.}~\bibnamefont {Niyazov}}, \bibinfo {author} {\bibfnamefont {I.}~\bibnamefont {Krainov}}, \bibinfo {author} {\bibfnamefont {D.~N.}\ \bibnamefont {Aristov}},\ and\ \bibinfo {author} {\bibfnamefont {V.~Y.}\ \bibnamefont {Kachorovskii}},\ }\bibfield  {title} {\bibinfo {title} {Shot noise in helical edge states in presence of a static magnetic defect},\ }\href@noop {} {\bibfield  {journal} {\bibinfo  {journal} {JETP Letters}\ }\textbf {\bibinfo {volume} {119}},\ \bibinfo {pages} {372} (\bibinfo {year} {2024}{\natexlab{a}})}\BibitemShut {NoStop}%
\bibitem [{\citenamefont {Niyazov}\ \emph {et~al.}(2024{\natexlab{b}})\citenamefont {Niyazov}, \citenamefont {Krainov}, \citenamefont {Aristov},\ and\ \citenamefont {Kachorovskii}}]{niyazov2024shotnoiseaharonovbohminterferometers}%
  \BibitemOpen
  \bibfield  {author} {\bibinfo {author} {\bibfnamefont {R.~A.}\ \bibnamefont {Niyazov}}, \bibinfo {author} {\bibfnamefont {I.~V.}\ \bibnamefont {Krainov}}, \bibinfo {author} {\bibfnamefont {D.~N.}\ \bibnamefont {Aristov}},\ and\ \bibinfo {author} {\bibfnamefont {V.~Y.}\ \bibnamefont {Kachorovskii}},\ }\href {https://arxiv.org/abs/2407.08329} {\bibinfo {title} {Shot noise in aharonov-bohm interferometers: Comparison of helical and conventional setups}} (\bibinfo {year} {2024}{\natexlab{b}}),\ \Eprint {https://arxiv.org/abs/2407.08329} {arXiv:2407.08329 [cond-mat.mes-hall]} \BibitemShut {NoStop}%
\bibitem [{\citenamefont {Fendley}\ \emph {et~al.}(1995)\citenamefont {Fendley}, \citenamefont {Ludwig},\ and\ \citenamefont {Saleur}}]{fendley1995exact}%
  \BibitemOpen
  \bibfield  {author} {\bibinfo {author} {\bibfnamefont {P.}~\bibnamefont {Fendley}}, \bibinfo {author} {\bibfnamefont {A.}~\bibnamefont {Ludwig}},\ and\ \bibinfo {author} {\bibfnamefont {H.}~\bibnamefont {Saleur}},\ }\bibfield  {title} {\bibinfo {title} {Exact nonequilibrium dc shot noise in luttinger liquids and fractional quantum hall devices},\ }\href@noop {} {\bibfield  {journal} {\bibinfo  {journal} {Physical review letters}\ }\textbf {\bibinfo {volume} {75}},\ \bibinfo {pages} {2196} (\bibinfo {year} {1995})}\BibitemShut {NoStop}%
\bibitem [{\citenamefont {Ponomarenko}\ and\ \citenamefont {Nagaosa}(1999)}]{ponNag1999}%
  \BibitemOpen
  \bibfield  {author} {\bibinfo {author} {\bibfnamefont {V.~V.}\ \bibnamefont {Ponomarenko}}\ and\ \bibinfo {author} {\bibfnamefont {N.}~\bibnamefont {Nagaosa}},\ }\bibfield  {title} {\bibinfo {title} {Features of renormalization induced by interaction in one-dimensional transport},\ }\href {https://doi.org/10.1103/PhysRevB.60.16865} {\bibfield  {journal} {\bibinfo  {journal} {Phys. Rev. B}\ }\textbf {\bibinfo {volume} {60}},\ \bibinfo {pages} {16865} (\bibinfo {year} {1999})}\BibitemShut {NoStop}%
\bibitem [{\citenamefont {Trauzettel}\ \emph {et~al.}(2004)\citenamefont {Trauzettel}, \citenamefont {Safi}, \citenamefont {Dolcini},\ and\ \citenamefont {Grabert}}]{trauzettel2004appearance}%
  \BibitemOpen
  \bibfield  {author} {\bibinfo {author} {\bibfnamefont {B.}~\bibnamefont {Trauzettel}}, \bibinfo {author} {\bibfnamefont {I.}~\bibnamefont {Safi}}, \bibinfo {author} {\bibfnamefont {F.}~\bibnamefont {Dolcini}},\ and\ \bibinfo {author} {\bibfnamefont {H.}~\bibnamefont {Grabert}},\ }\bibfield  {title} {\bibinfo {title} {Appearance of fractional charge in the noise of nonchiral luttinger liquids},\ }\href@noop {} {\bibfield  {journal} {\bibinfo  {journal} {Physical review letters}\ }\textbf {\bibinfo {volume} {92}},\ \bibinfo {pages} {226405} (\bibinfo {year} {2004})}\BibitemShut {NoStop}%
\bibitem [{\citenamefont {Dolcini}\ \emph {et~al.}(2005{\natexlab{a}})\citenamefont {Dolcini}, \citenamefont {Trauzettel}, \citenamefont {Safi},\ and\ \citenamefont {Grabert}}]{dolcini2005transport}%
  \BibitemOpen
  \bibfield  {author} {\bibinfo {author} {\bibfnamefont {F.}~\bibnamefont {Dolcini}}, \bibinfo {author} {\bibfnamefont {B.}~\bibnamefont {Trauzettel}}, \bibinfo {author} {\bibfnamefont {I.}~\bibnamefont {Safi}},\ and\ \bibinfo {author} {\bibfnamefont {H.}~\bibnamefont {Grabert}},\ }\bibfield  {title} {\bibinfo {title} {Transport properties of single-channel quantum wires with an impurity: Influence of finite length and temperature on average current and noise},\ }\href@noop {} {\bibfield  {journal} {\bibinfo  {journal} {Physical Review B—Condensed Matter and Materials Physics}\ }\textbf {\bibinfo {volume} {71}},\ \bibinfo {pages} {165309} (\bibinfo {year} {2005}{\natexlab{a}})}\BibitemShut {NoStop}%
\bibitem [{\citenamefont {Bi}\ \emph {et~al.}(2017)\citenamefont {Bi}, \citenamefont {Zhang}, \citenamefont {You}, \citenamefont {Young}, \citenamefont {Balents}, \citenamefont {Liu},\ and\ \citenamefont {Xu}}]{bi2017bilayer}%
  \BibitemOpen
  \bibfield  {author} {\bibinfo {author} {\bibfnamefont {Z.}~\bibnamefont {Bi}}, \bibinfo {author} {\bibfnamefont {R.}~\bibnamefont {Zhang}}, \bibinfo {author} {\bibfnamefont {Y.-Z.}\ \bibnamefont {You}}, \bibinfo {author} {\bibfnamefont {A.}~\bibnamefont {Young}}, \bibinfo {author} {\bibfnamefont {L.}~\bibnamefont {Balents}}, \bibinfo {author} {\bibfnamefont {C.-X.}\ \bibnamefont {Liu}},\ and\ \bibinfo {author} {\bibfnamefont {C.}~\bibnamefont {Xu}},\ }\bibfield  {title} {\bibinfo {title} {Bilayer graphene as a platform for bosonic symmetry-protected topological states},\ }\href@noop {} {\bibfield  {journal} {\bibinfo  {journal} {Physical review letters}\ }\textbf {\bibinfo {volume} {118}},\ \bibinfo {pages} {126801} (\bibinfo {year} {2017})}\BibitemShut {NoStop}%
\bibitem [{\citenamefont {Zhang}\ and\ \citenamefont {Liu}(2017)}]{zhang2017fingerprints}%
  \BibitemOpen
  \bibfield  {author} {\bibinfo {author} {\bibfnamefont {R.-X.}\ \bibnamefont {Zhang}}\ and\ \bibinfo {author} {\bibfnamefont {C.-X.}\ \bibnamefont {Liu}},\ }\bibfield  {title} {\bibinfo {title} {Fingerprints of a bosonic symmetry-protected topological state in a quantum point contact},\ }\href@noop {} {\bibfield  {journal} {\bibinfo  {journal} {Physical review letters}\ }\textbf {\bibinfo {volume} {118}},\ \bibinfo {pages} {216803} (\bibinfo {year} {2017})}\BibitemShut {NoStop}%
\bibitem [{\citenamefont {Wang}\ \emph {et~al.}(2021)\citenamefont {Wang}, \citenamefont {Xu},\ and\ \citenamefont {Duan}}]{wang2021ising}%
  \BibitemOpen
  \bibfield  {author} {\bibinfo {author} {\bibfnamefont {C.}~\bibnamefont {Wang}}, \bibinfo {author} {\bibfnamefont {Y.}~\bibnamefont {Xu}},\ and\ \bibinfo {author} {\bibfnamefont {W.}~\bibnamefont {Duan}},\ }\bibfield  {title} {\bibinfo {title} {Ising superconductivity and its hidden variants},\ }\href@noop {} {\bibfield  {journal} {\bibinfo  {journal} {Accounts of Materials Research}\ }\textbf {\bibinfo {volume} {2}},\ \bibinfo {pages} {526} (\bibinfo {year} {2021})}\BibitemShut {NoStop}%
\bibitem [{\citenamefont {Korm\'anyos}\ \emph {et~al.}(2014)\citenamefont {Korm\'anyos}, \citenamefont {Z\'olyomi}, \citenamefont {Drummond},\ and\ \citenamefont {Burkard}}]{PhysRevX.4.011034}%
  \BibitemOpen
  \bibfield  {author} {\bibinfo {author} {\bibfnamefont {A.}~\bibnamefont {Korm\'anyos}}, \bibinfo {author} {\bibfnamefont {V.}~\bibnamefont {Z\'olyomi}}, \bibinfo {author} {\bibfnamefont {N.~D.}\ \bibnamefont {Drummond}},\ and\ \bibinfo {author} {\bibfnamefont {G.}~\bibnamefont {Burkard}},\ }\bibfield  {title} {\bibinfo {title} {Spin-orbit coupling, quantum dots, and qubits in monolayer transition metal dichalcogenides},\ }\href {https://doi.org/10.1103/PhysRevX.4.011034} {\bibfield  {journal} {\bibinfo  {journal} {Phys. Rev. X}\ }\textbf {\bibinfo {volume} {4}},\ \bibinfo {pages} {011034} (\bibinfo {year} {2014})}\BibitemShut {NoStop}%
\bibitem [{\citenamefont {Xiao}\ \emph {et~al.}(2012)\citenamefont {Xiao}, \citenamefont {Liu}, \citenamefont {Feng}, \citenamefont {Xu},\ and\ \citenamefont {Yao}}]{xiao2012coupled}%
  \BibitemOpen
  \bibfield  {author} {\bibinfo {author} {\bibfnamefont {D.}~\bibnamefont {Xiao}}, \bibinfo {author} {\bibfnamefont {G.-B.}\ \bibnamefont {Liu}}, \bibinfo {author} {\bibfnamefont {W.}~\bibnamefont {Feng}}, \bibinfo {author} {\bibfnamefont {X.}~\bibnamefont {Xu}},\ and\ \bibinfo {author} {\bibfnamefont {W.}~\bibnamefont {Yao}},\ }\bibfield  {title} {\bibinfo {title} {Coupled spin and valley physics in monolayers of mos 2 and other group-vi dichalcogenides},\ }\href@noop {} {\bibfield  {journal} {\bibinfo  {journal} {Physical review letters}\ }\textbf {\bibinfo {volume} {108}},\ \bibinfo {pages} {196802} (\bibinfo {year} {2012})}\BibitemShut {NoStop}%
\bibitem [{\citenamefont {Kane}\ and\ \citenamefont {Fisher}(1994{\natexlab{a}})}]{kane1994nonequilibrium}%
  \BibitemOpen
  \bibfield  {author} {\bibinfo {author} {\bibfnamefont {C.}~\bibnamefont {Kane}}\ and\ \bibinfo {author} {\bibfnamefont {M.~P.}\ \bibnamefont {Fisher}},\ }\bibfield  {title} {\bibinfo {title} {Nonequilibrium noise and fractional charge in the quantum hall effect},\ }\href@noop {} {\bibfield  {journal} {\bibinfo  {journal} {Physical review letters}\ }\textbf {\bibinfo {volume} {72}},\ \bibinfo {pages} {724} (\bibinfo {year} {1994}{\natexlab{a}})}\BibitemShut {NoStop}%
\bibitem [{\citenamefont {Kane}\ and\ \citenamefont {Fisher}(1997)}]{kane1997shot}%
  \BibitemOpen
  \bibfield  {author} {\bibinfo {author} {\bibfnamefont {C.~L.}\ \bibnamefont {Kane}}\ and\ \bibinfo {author} {\bibfnamefont {M.~P.}\ \bibnamefont {Fisher}},\ }\bibfield  {title} {\bibinfo {title} {A shot in the arm for fractional charge},\ }\href@noop {} {\bibfield  {journal} {\bibinfo  {journal} {Nature}\ }\textbf {\bibinfo {volume} {389}},\ \bibinfo {pages} {119} (\bibinfo {year} {1997})}\BibitemShut {NoStop}%
\bibitem [{\citenamefont {Sandler}\ \emph {et~al.}(1999)\citenamefont {Sandler}, \citenamefont {Chamon},\ and\ \citenamefont {Fradkin}}]{sandler1999noise}%
  \BibitemOpen
  \bibfield  {author} {\bibinfo {author} {\bibfnamefont {N.~P.}\ \bibnamefont {Sandler}}, \bibinfo {author} {\bibfnamefont {C.~d.~C.}\ \bibnamefont {Chamon}},\ and\ \bibinfo {author} {\bibfnamefont {E.}~\bibnamefont {Fradkin}},\ }\bibfield  {title} {\bibinfo {title} {Noise measurements and fractional charge in fractional quantum hall liquids},\ }\href@noop {} {\bibfield  {journal} {\bibinfo  {journal} {Physical Review B}\ }\textbf {\bibinfo {volume} {59}},\ \bibinfo {pages} {12521} (\bibinfo {year} {1999})}\BibitemShut {NoStop}%
\bibitem [{\citenamefont {Kane}\ and\ \citenamefont {Fisher}(2003)}]{kane2003shot}%
  \BibitemOpen
  \bibfield  {author} {\bibinfo {author} {\bibfnamefont {C.}~\bibnamefont {Kane}}\ and\ \bibinfo {author} {\bibfnamefont {M.~P.}\ \bibnamefont {Fisher}},\ }\bibfield  {title} {\bibinfo {title} {Shot noise and the transmission of dilute laughlin quasiparticles},\ }\href@noop {} {\bibfield  {journal} {\bibinfo  {journal} {Physical Review B}\ }\textbf {\bibinfo {volume} {67}},\ \bibinfo {pages} {045307} (\bibinfo {year} {2003})}\BibitemShut {NoStop}%
\bibitem [{\citenamefont {Bid}\ \emph {et~al.}(2009)\citenamefont {Bid}, \citenamefont {Ofek}, \citenamefont {Heiblum}, \citenamefont {Umansky},\ and\ \citenamefont {Mahalu}}]{bid2009shot}%
  \BibitemOpen
  \bibfield  {author} {\bibinfo {author} {\bibfnamefont {A.}~\bibnamefont {Bid}}, \bibinfo {author} {\bibfnamefont {N.}~\bibnamefont {Ofek}}, \bibinfo {author} {\bibfnamefont {M.}~\bibnamefont {Heiblum}}, \bibinfo {author} {\bibfnamefont {V.}~\bibnamefont {Umansky}},\ and\ \bibinfo {author} {\bibfnamefont {D.}~\bibnamefont {Mahalu}},\ }\bibfield  {title} {\bibinfo {title} {Shot noise and charge at the 2/3 composite fractional quantum hall state},\ }\href@noop {} {\bibfield  {journal} {\bibinfo  {journal} {Physical Review Letters}\ }\textbf {\bibinfo {volume} {103}},\ \bibinfo {pages} {236802} (\bibinfo {year} {2009})}\BibitemShut {NoStop}%
\bibitem [{\citenamefont {Hashisaka}\ \emph {et~al.}(2015)\citenamefont {Hashisaka}, \citenamefont {Ota}, \citenamefont {Muraki},\ and\ \citenamefont {Fujisawa}}]{hashisaka2015shot}%
  \BibitemOpen
  \bibfield  {author} {\bibinfo {author} {\bibfnamefont {M.}~\bibnamefont {Hashisaka}}, \bibinfo {author} {\bibfnamefont {T.}~\bibnamefont {Ota}}, \bibinfo {author} {\bibfnamefont {K.}~\bibnamefont {Muraki}},\ and\ \bibinfo {author} {\bibfnamefont {T.}~\bibnamefont {Fujisawa}},\ }\bibfield  {title} {\bibinfo {title} {Shot-noise evidence of fractional quasiparticle creation in a local fractional quantum hall state},\ }\href@noop {} {\bibfield  {journal} {\bibinfo  {journal} {Physical review letters}\ }\textbf {\bibinfo {volume} {114}},\ \bibinfo {pages} {056802} (\bibinfo {year} {2015})}\BibitemShut {NoStop}%
\bibitem [{\citenamefont {Manna}\ and\ \citenamefont {Das}(2023)}]{manna2023experimentallymotivatedorderlength}%
  \BibitemOpen
  \bibfield  {author} {\bibinfo {author} {\bibfnamefont {S.}~\bibnamefont {Manna}}\ and\ \bibinfo {author} {\bibfnamefont {A.}~\bibnamefont {Das}},\ }\href {https://arxiv.org/abs/2307.08264} {\bibinfo {title} {Experimentally motivated order of length scales affect shot noise}} (\bibinfo {year} {2023}),\ \Eprint {https://arxiv.org/abs/2307.08264} {arXiv:2307.08264 [cond-mat.mes-hall]} \BibitemShut {NoStop}%
\bibitem [{\citenamefont {Manna}\ \emph {et~al.}(2024{\natexlab{a}})\citenamefont {Manna}, \citenamefont {Das}, \citenamefont {Gefen},\ and\ \citenamefont {Goldstein}}]{manna2024diagnosticsanomalousconductanceplateaus}%
  \BibitemOpen
  \bibfield  {author} {\bibinfo {author} {\bibfnamefont {S.}~\bibnamefont {Manna}}, \bibinfo {author} {\bibfnamefont {A.}~\bibnamefont {Das}}, \bibinfo {author} {\bibfnamefont {Y.}~\bibnamefont {Gefen}},\ and\ \bibinfo {author} {\bibfnamefont {M.}~\bibnamefont {Goldstein}},\ }\href {https://arxiv.org/abs/2307.05173} {\bibinfo {title} {Diagnostics of anomalous conductance plateaus in abelian quantum hall regime}} (\bibinfo {year} {2024}{\natexlab{a}}),\ \Eprint {https://arxiv.org/abs/2307.05173} {arXiv:2307.05173 [cond-mat.mes-hall]} \BibitemShut {NoStop}%
\bibitem [{\citenamefont {Manna}\ \emph {et~al.}(2024{\natexlab{b}})\citenamefont {Manna}, \citenamefont {Das}, \citenamefont {Goldstein},\ and\ \citenamefont {Gefen}}]{PhysRevLett.132.136502}%
  \BibitemOpen
  \bibfield  {author} {\bibinfo {author} {\bibfnamefont {S.}~\bibnamefont {Manna}}, \bibinfo {author} {\bibfnamefont {A.}~\bibnamefont {Das}}, \bibinfo {author} {\bibfnamefont {M.}~\bibnamefont {Goldstein}},\ and\ \bibinfo {author} {\bibfnamefont {Y.}~\bibnamefont {Gefen}},\ }\bibfield  {title} {\bibinfo {title} {Full classification of transport on an equilibrated $5/2$ edge via shot noise},\ }\href {https://doi.org/10.1103/PhysRevLett.132.136502} {\bibfield  {journal} {\bibinfo  {journal} {Phys. Rev. Lett.}\ }\textbf {\bibinfo {volume} {132}},\ \bibinfo {pages} {136502} (\bibinfo {year} {2024}{\natexlab{b}})}\BibitemShut {NoStop}%
\bibitem [{\citenamefont {Manna}\ \emph {et~al.}(2024{\natexlab{c}})\citenamefont {Manna}, \citenamefont {Das}, \citenamefont {Gefen},\ and\ \citenamefont {Goldstein}}]{10.1063/10.0034344}%
  \BibitemOpen
  \bibfield  {author} {\bibinfo {author} {\bibfnamefont {S.}~\bibnamefont {Manna}}, \bibinfo {author} {\bibfnamefont {A.}~\bibnamefont {Das}}, \bibinfo {author} {\bibfnamefont {Y.}~\bibnamefont {Gefen}},\ and\ \bibinfo {author} {\bibfnamefont {M.}~\bibnamefont {Goldstein}},\ }\bibfield  {title} {\bibinfo {title} {Shot noise as a diagnostic in the $\nu = 2/3$ fractional quantum hall edge zoo},\ }\href {https://doi.org/10.1063/10.0034344} {\bibfield  {journal} {\bibinfo  {journal} {Low Temperature Physics}\ }\textbf {\bibinfo {volume} {50}},\ \bibinfo {pages} {1113} (\bibinfo {year} {2024}{\natexlab{c}})},\ \Eprint {https://arxiv.org/abs/https://pubs.aip.org/aip/ltp/article-pdf/50/12/1113/20289499/1113\_1\_10.0034344.pdf} {https://pubs.aip.org/aip/ltp/article-pdf/50/12/1113/20289499/1113\_1\_10.0034344.pdf} \BibitemShut {NoStop}%
\bibitem [{\citenamefont {Fradkin}(2013)}]{fradkin2013field}%
  \BibitemOpen
  \bibfield  {author} {\bibinfo {author} {\bibfnamefont {E.}~\bibnamefont {Fradkin}},\ }\href@noop {} {\emph {\bibinfo {title} {Field theories of condensed matter physics}}}\ (\bibinfo  {publisher} {Cambridge University Press},\ \bibinfo {year} {2013})\BibitemShut {NoStop}%
\bibitem [{\citenamefont {Gogolin}\ \emph {et~al.}(2004)\citenamefont {Gogolin}, \citenamefont {Nersesyan},\ and\ \citenamefont {Tsvelik}}]{gogolin2004bosonization}%
  \BibitemOpen
  \bibfield  {author} {\bibinfo {author} {\bibfnamefont {A.~O.}\ \bibnamefont {Gogolin}}, \bibinfo {author} {\bibfnamefont {A.~A.}\ \bibnamefont {Nersesyan}},\ and\ \bibinfo {author} {\bibfnamefont {A.~M.}\ \bibnamefont {Tsvelik}},\ }\href@noop {} {\emph {\bibinfo {title} {Bosonization and strongly correlated systems}}}\ (\bibinfo  {publisher} {Cambridge university press},\ \bibinfo {year} {2004})\BibitemShut {NoStop}%
\bibitem [{\citenamefont {Devakul}\ \emph {et~al.}(2021)\citenamefont {Devakul}, \citenamefont {Cr{\'e}pel}, \citenamefont {Zhang},\ and\ \citenamefont {Fu}}]{devakul2021magic}%
  \BibitemOpen
  \bibfield  {author} {\bibinfo {author} {\bibfnamefont {T.}~\bibnamefont {Devakul}}, \bibinfo {author} {\bibfnamefont {V.}~\bibnamefont {Cr{\'e}pel}}, \bibinfo {author} {\bibfnamefont {Y.}~\bibnamefont {Zhang}},\ and\ \bibinfo {author} {\bibfnamefont {L.}~\bibnamefont {Fu}},\ }\bibfield  {title} {\bibinfo {title} {Magic in twisted transition metal dichalcogenide bilayers},\ }\href@noop {} {\bibfield  {journal} {\bibinfo  {journal} {Nature communications}\ }\textbf {\bibinfo {volume} {12}},\ \bibinfo {pages} {6730} (\bibinfo {year} {2021})}\BibitemShut {NoStop}%
\bibitem [{\citenamefont {Zhang}\ \emph {et~al.}(2025)\citenamefont {Zhang}, \citenamefont {Yang}, \citenamefont {Wang}, \citenamefont {Liu}, \citenamefont {Cao},\ and\ \citenamefont {Xiao}}]{zhang2025twist}%
  \BibitemOpen
  \bibfield  {author} {\bibinfo {author} {\bibfnamefont {X.-W.}\ \bibnamefont {Zhang}}, \bibinfo {author} {\bibfnamefont {K.}~\bibnamefont {Yang}}, \bibinfo {author} {\bibfnamefont {C.}~\bibnamefont {Wang}}, \bibinfo {author} {\bibfnamefont {X.}~\bibnamefont {Liu}}, \bibinfo {author} {\bibfnamefont {T.}~\bibnamefont {Cao}},\ and\ \bibinfo {author} {\bibfnamefont {D.}~\bibnamefont {Xiao}},\ }\bibfield  {title} {\bibinfo {title} {Twist-angle transferable continuum model and second flat chern band in twisted mote2 and wse2},\ }\href@noop {} {\bibfield  {journal} {\bibinfo  {journal} {npj Quantum Materials}\ } (\bibinfo {year} {2025})}\BibitemShut {NoStop}%
\bibitem [{\citenamefont {Chubukov}\ and\ \citenamefont {Varma}(2025)}]{chubukov2025quantum}%
  \BibitemOpen
  \bibfield  {author} {\bibinfo {author} {\bibfnamefont {A.}~\bibnamefont {Chubukov}}\ and\ \bibinfo {author} {\bibfnamefont {C.}~\bibnamefont {Varma}},\ }\bibfield  {title} {\bibinfo {title} {Quantum criticality and superconductivity in twisted transition metal dichalcogenides},\ }\href@noop {} {\bibfield  {journal} {\bibinfo  {journal} {Physical Review B}\ }\textbf {\bibinfo {volume} {111}},\ \bibinfo {pages} {014507} (\bibinfo {year} {2025})}\BibitemShut {NoStop}%
\bibitem [{\citenamefont {Chou}\ and\ \citenamefont {Das~Sarma}(2024)}]{PhysRevB.110.155117}%
  \BibitemOpen
  \bibfield  {author} {\bibinfo {author} {\bibfnamefont {Y.-Z.}\ \bibnamefont {Chou}}\ and\ \bibinfo {author} {\bibfnamefont {S.}~\bibnamefont {Das~Sarma}},\ }\bibfield  {title} {\bibinfo {title} {Composite helical edges from abelian fractional topological insulators},\ }\href {https://doi.org/10.1103/PhysRevB.110.155117} {\bibfield  {journal} {\bibinfo  {journal} {Phys. Rev. B}\ }\textbf {\bibinfo {volume} {110}},\ \bibinfo {pages} {155117} (\bibinfo {year} {2024})}\BibitemShut {NoStop}%
\bibitem [{\citenamefont {Maslov}\ and\ \citenamefont {Stone}(1995)}]{PhysRevB.52.R5539}%
  \BibitemOpen
  \bibfield  {author} {\bibinfo {author} {\bibfnamefont {D.~L.}\ \bibnamefont {Maslov}}\ and\ \bibinfo {author} {\bibfnamefont {M.}~\bibnamefont {Stone}},\ }\bibfield  {title} {\bibinfo {title} {Landauer conductance of luttinger liquids with leads},\ }\href {https://doi.org/10.1103/PhysRevB.52.R5539} {\bibfield  {journal} {\bibinfo  {journal} {Phys. Rev. B}\ }\textbf {\bibinfo {volume} {52}},\ \bibinfo {pages} {R5539} (\bibinfo {year} {1995})}\BibitemShut {NoStop}%
\bibitem [{\citenamefont {Hou}\ \emph {et~al.}(2009)\citenamefont {Hou}, \citenamefont {Kim},\ and\ \citenamefont {Chamon}}]{PhysRevLett.102.076602}%
  \BibitemOpen
  \bibfield  {author} {\bibinfo {author} {\bibfnamefont {C.-Y.}\ \bibnamefont {Hou}}, \bibinfo {author} {\bibfnamefont {E.-A.}\ \bibnamefont {Kim}},\ and\ \bibinfo {author} {\bibfnamefont {C.}~\bibnamefont {Chamon}},\ }\bibfield  {title} {\bibinfo {title} {Corner junction as a probe of helical edge states},\ }\href {https://doi.org/10.1103/PhysRevLett.102.076602} {\bibfield  {journal} {\bibinfo  {journal} {Phys. Rev. Lett.}\ }\textbf {\bibinfo {volume} {102}},\ \bibinfo {pages} {076602} (\bibinfo {year} {2009})}\BibitemShut {NoStop}%
\bibitem [{\citenamefont {Teo}\ and\ \citenamefont {Kane}(2009)}]{PhysRevB.79.235321}%
  \BibitemOpen
  \bibfield  {author} {\bibinfo {author} {\bibfnamefont {J.~C.~Y.}\ \bibnamefont {Teo}}\ and\ \bibinfo {author} {\bibfnamefont {C.~L.}\ \bibnamefont {Kane}},\ }\bibfield  {title} {\bibinfo {title} {Critical behavior of a point contact in a quantum spin hall insulator},\ }\href {https://doi.org/10.1103/PhysRevB.79.235321} {\bibfield  {journal} {\bibinfo  {journal} {Phys. Rev. B}\ }\textbf {\bibinfo {volume} {79}},\ \bibinfo {pages} {235321} (\bibinfo {year} {2009})}\BibitemShut {NoStop}%
\bibitem [{\citenamefont {Str{\"o}m}\ and\ \citenamefont {Johannesson}(2009)}]{strom2009tunneling}%
  \BibitemOpen
  \bibfield  {author} {\bibinfo {author} {\bibfnamefont {A.}~\bibnamefont {Str{\"o}m}}\ and\ \bibinfo {author} {\bibfnamefont {H.}~\bibnamefont {Johannesson}},\ }\bibfield  {title} {\bibinfo {title} {Tunneling between edge states in a quantum spin hall system},\ }\href@noop {} {\bibfield  {journal} {\bibinfo  {journal} {Physical review letters}\ }\textbf {\bibinfo {volume} {102}},\ \bibinfo {pages} {096806} (\bibinfo {year} {2009})}\BibitemShut {NoStop}%
\bibitem [{\citenamefont {Dolcini}\ \emph {et~al.}(2005{\natexlab{b}})\citenamefont {Dolcini}, \citenamefont {Trauzettel}, \citenamefont {Safi},\ and\ \citenamefont {Grabert}}]{PhysRevB.71.165309}%
  \BibitemOpen
  \bibfield  {author} {\bibinfo {author} {\bibfnamefont {F.}~\bibnamefont {Dolcini}}, \bibinfo {author} {\bibfnamefont {B.}~\bibnamefont {Trauzettel}}, \bibinfo {author} {\bibfnamefont {I.}~\bibnamefont {Safi}},\ and\ \bibinfo {author} {\bibfnamefont {H.}~\bibnamefont {Grabert}},\ }\bibfield  {title} {\bibinfo {title} {Transport properties of single-channel quantum wires with an impurity: Influence of finite length and temperature on average current and noise},\ }\href {https://doi.org/10.1103/PhysRevB.71.165309} {\bibfield  {journal} {\bibinfo  {journal} {Phys. Rev. B}\ }\textbf {\bibinfo {volume} {71}},\ \bibinfo {pages} {165309} (\bibinfo {year} {2005}{\natexlab{b}})}\BibitemShut {NoStop}%
\bibitem [{\citenamefont {Ponomarenko}\ and\ \citenamefont {Nagaosa}(1997)}]{PhysRevLett.79.1714}%
  \BibitemOpen
  \bibfield  {author} {\bibinfo {author} {\bibfnamefont {V.~V.}\ \bibnamefont {Ponomarenko}}\ and\ \bibinfo {author} {\bibfnamefont {N.}~\bibnamefont {Nagaosa}},\ }\bibfield  {title} {\bibinfo {title} {Threshold features in transport through a 1d constriction},\ }\href {https://doi.org/10.1103/PhysRevLett.79.1714} {\bibfield  {journal} {\bibinfo  {journal} {Phys. Rev. Lett.}\ }\textbf {\bibinfo {volume} {79}},\ \bibinfo {pages} {1714} (\bibinfo {year} {1997})}\BibitemShut {NoStop}%
\bibitem [{\citenamefont {Keldysh}(1964)}]{Keldysh:1964ud}%
  \BibitemOpen
  \bibfield  {author} {\bibinfo {author} {\bibfnamefont {L.~V.}\ \bibnamefont {Keldysh}},\ }\bibfield  {title} {\bibinfo {title} {{Diagram technique for nonequilibrium processes}},\ }\href@noop {} {\bibfield  {journal} {\bibinfo  {journal} {Zh. Eksp. Teor. Fiz.}\ }\textbf {\bibinfo {volume} {47}},\ \bibinfo {pages} {1515} (\bibinfo {year} {1964})}\BibitemShut {NoStop}%
\bibitem [{\citenamefont {Kane}\ and\ \citenamefont {Fisher}(1994{\natexlab{b}})}]{PhysRevLett.72.724}%
  \BibitemOpen
  \bibfield  {author} {\bibinfo {author} {\bibfnamefont {C.~L.}\ \bibnamefont {Kane}}\ and\ \bibinfo {author} {\bibfnamefont {M.~P.~A.}\ \bibnamefont {Fisher}},\ }\bibfield  {title} {\bibinfo {title} {Nonequilibrium noise and fractional charge in the quantum hall effect},\ }\href {https://doi.org/10.1103/PhysRevLett.72.724} {\bibfield  {journal} {\bibinfo  {journal} {Phys. Rev. Lett.}\ }\textbf {\bibinfo {volume} {72}},\ \bibinfo {pages} {724} (\bibinfo {year} {1994}{\natexlab{b}})}\BibitemShut {NoStop}%
\bibitem [{\citenamefont {Bernevig}\ and\ \citenamefont {Zhang}(2006{\natexlab{b}})}]{Bernevig_2006}%
  \BibitemOpen
  \bibfield  {author} {\bibinfo {author} {\bibfnamefont {B.~A.}\ \bibnamefont {Bernevig}}\ and\ \bibinfo {author} {\bibfnamefont {S.-C.}\ \bibnamefont {Zhang}},\ }\bibfield  {title} {\bibinfo {title} {Quantum spin hall effect},\ }\bibfield  {journal} {\bibinfo  {journal} {Physical Review Letters}\ }\textbf {\bibinfo {volume} {96}},\ \href {https://doi.org/10.1103/physrevlett.96.106802} {10.1103/physrevlett.96.106802} (\bibinfo {year} {2006}{\natexlab{b}})\BibitemShut {NoStop}%
\bibitem [{\citenamefont {Thouless}\ \emph {et~al.}(1982)\citenamefont {Thouless}, \citenamefont {Kohmoto}, \citenamefont {Nightingale},\ and\ \citenamefont {den Nijs}}]{thouless1982quantized}%
  \BibitemOpen
  \bibfield  {author} {\bibinfo {author} {\bibfnamefont {D.~J.}\ \bibnamefont {Thouless}}, \bibinfo {author} {\bibfnamefont {M.}~\bibnamefont {Kohmoto}}, \bibinfo {author} {\bibfnamefont {M.~P.}\ \bibnamefont {Nightingale}},\ and\ \bibinfo {author} {\bibfnamefont {M.}~\bibnamefont {den Nijs}},\ }\bibfield  {title} {\bibinfo {title} {Quantized hall conductance in a two-dimensional periodic potential},\ }\href@noop {} {\bibfield  {journal} {\bibinfo  {journal} {Physical review letters}\ }\textbf {\bibinfo {volume} {49}},\ \bibinfo {pages} {405} (\bibinfo {year} {1982})}\BibitemShut {NoStop}%
\bibitem [{\citenamefont {Kane}(2003)}]{kane2003telegraph}%
  \BibitemOpen
  \bibfield  {author} {\bibinfo {author} {\bibfnamefont {C.}~\bibnamefont {Kane}},\ }\bibfield  {title} {\bibinfo {title} {Telegraph noise and fractional statistics in the quantum hall effect},\ }\href@noop {} {\bibfield  {journal} {\bibinfo  {journal} {Physical review letters}\ }\textbf {\bibinfo {volume} {90}},\ \bibinfo {pages} {226802} (\bibinfo {year} {2003})}\BibitemShut {NoStop}%
\bibitem [{\citenamefont {Wu}\ \emph {et~al.}(2019)\citenamefont {Wu}, \citenamefont {Lovorn}, \citenamefont {Tutuc}, \citenamefont {Martin},\ and\ \citenamefont {MacDonald}}]{wu2019topological}%
  \BibitemOpen
  \bibfield  {author} {\bibinfo {author} {\bibfnamefont {F.}~\bibnamefont {Wu}}, \bibinfo {author} {\bibfnamefont {T.}~\bibnamefont {Lovorn}}, \bibinfo {author} {\bibfnamefont {E.}~\bibnamefont {Tutuc}}, \bibinfo {author} {\bibfnamefont {I.}~\bibnamefont {Martin}},\ and\ \bibinfo {author} {\bibfnamefont {A.}~\bibnamefont {MacDonald}},\ }\bibfield  {title} {\bibinfo {title} {Topological insulators in twisted transition metal dichalcogenide homobilayers},\ }\href@noop {} {\bibfield  {journal} {\bibinfo  {journal} {Physical review letters}\ }\textbf {\bibinfo {volume} {122}},\ \bibinfo {pages} {086402} (\bibinfo {year} {2019})}\BibitemShut {NoStop}%
\bibitem [{\citenamefont {Wu}\ \emph {et~al.}(2018)\citenamefont {Wu}, \citenamefont {Lovorn}, \citenamefont {Tutuc},\ and\ \citenamefont {MacDonald}}]{wu2018hubbard}%
  \BibitemOpen
  \bibfield  {author} {\bibinfo {author} {\bibfnamefont {F.}~\bibnamefont {Wu}}, \bibinfo {author} {\bibfnamefont {T.}~\bibnamefont {Lovorn}}, \bibinfo {author} {\bibfnamefont {E.}~\bibnamefont {Tutuc}},\ and\ \bibinfo {author} {\bibfnamefont {A.~H.}\ \bibnamefont {MacDonald}},\ }\bibfield  {title} {\bibinfo {title} {Hubbard model physics in transition metal dichalcogenide moir{\'e} bands},\ }\href@noop {} {\bibfield  {journal} {\bibinfo  {journal} {Physical review letters}\ }\textbf {\bibinfo {volume} {121}},\ \bibinfo {pages} {026402} (\bibinfo {year} {2018})}\BibitemShut {NoStop}%
\bibitem [{\citenamefont {Shabani}\ \emph {et~al.}(2021)\citenamefont {Shabani}, \citenamefont {Halbertal}, \citenamefont {Wu}, \citenamefont {Chen}, \citenamefont {Liu}, \citenamefont {Hone}, \citenamefont {Yao}, \citenamefont {Basov}, \citenamefont {Zhu},\ and\ \citenamefont {Pasupathy}}]{shabani2021deep}%
  \BibitemOpen
  \bibfield  {author} {\bibinfo {author} {\bibfnamefont {S.}~\bibnamefont {Shabani}}, \bibinfo {author} {\bibfnamefont {D.}~\bibnamefont {Halbertal}}, \bibinfo {author} {\bibfnamefont {W.}~\bibnamefont {Wu}}, \bibinfo {author} {\bibfnamefont {M.}~\bibnamefont {Chen}}, \bibinfo {author} {\bibfnamefont {S.}~\bibnamefont {Liu}}, \bibinfo {author} {\bibfnamefont {J.}~\bibnamefont {Hone}}, \bibinfo {author} {\bibfnamefont {W.}~\bibnamefont {Yao}}, \bibinfo {author} {\bibfnamefont {D.~N.}\ \bibnamefont {Basov}}, \bibinfo {author} {\bibfnamefont {X.}~\bibnamefont {Zhu}},\ and\ \bibinfo {author} {\bibfnamefont {A.~N.}\ \bibnamefont {Pasupathy}},\ }\bibfield  {title} {\bibinfo {title} {Deep moir{\'e} potentials in twisted transition metal dichalcogenide bilayers},\ }\href@noop {} {\bibfield  {journal} {\bibinfo  {journal} {Nature Physics}\ }\textbf {\bibinfo {volume} {17}},\ \bibinfo {pages} {720} (\bibinfo {year} {2021})}\BibitemShut {NoStop}%
\bibitem [{\citenamefont {Wu}\ \emph {et~al.}(2024)\citenamefont {Wu}, \citenamefont {Shaffer}, \citenamefont {Wu},\ and\ \citenamefont {Santos}}]{wu2024time}%
  \BibitemOpen
  \bibfield  {author} {\bibinfo {author} {\bibfnamefont {Y.-M.}\ \bibnamefont {Wu}}, \bibinfo {author} {\bibfnamefont {D.}~\bibnamefont {Shaffer}}, \bibinfo {author} {\bibfnamefont {Z.}~\bibnamefont {Wu}},\ and\ \bibinfo {author} {\bibfnamefont {L.~H.}\ \bibnamefont {Santos}},\ }\bibfield  {title} {\bibinfo {title} {Time-reversal invariant topological moir{\'e} flat band: A platform for the fractional quantum spin hall effect},\ }\href@noop {} {\bibfield  {journal} {\bibinfo  {journal} {Physical Review B}\ }\textbf {\bibinfo {volume} {109}},\ \bibinfo {pages} {115111} (\bibinfo {year} {2024})}\BibitemShut {NoStop}%
\bibitem [{\citenamefont {Pan}\ \emph {et~al.}(2020)\citenamefont {Pan}, \citenamefont {Wu},\ and\ \citenamefont {Das~Sarma}}]{pan2020band}%
  \BibitemOpen
  \bibfield  {author} {\bibinfo {author} {\bibfnamefont {H.}~\bibnamefont {Pan}}, \bibinfo {author} {\bibfnamefont {F.}~\bibnamefont {Wu}},\ and\ \bibinfo {author} {\bibfnamefont {S.}~\bibnamefont {Das~Sarma}},\ }\bibfield  {title} {\bibinfo {title} {Band topology, hubbard model, heisenberg model, and dzyaloshinskii-moriya interaction in twisted bilayer wse 2},\ }\href@noop {} {\bibfield  {journal} {\bibinfo  {journal} {Physical Review Research}\ }\textbf {\bibinfo {volume} {2}},\ \bibinfo {pages} {033087} (\bibinfo {year} {2020})}\BibitemShut {NoStop}%
\bibitem [{\citenamefont {Zang}\ \emph {et~al.}(2021)\citenamefont {Zang}, \citenamefont {Wang}, \citenamefont {Cano},\ and\ \citenamefont {Millis}}]{zang2021hartree}%
  \BibitemOpen
  \bibfield  {author} {\bibinfo {author} {\bibfnamefont {J.}~\bibnamefont {Zang}}, \bibinfo {author} {\bibfnamefont {J.}~\bibnamefont {Wang}}, \bibinfo {author} {\bibfnamefont {J.}~\bibnamefont {Cano}},\ and\ \bibinfo {author} {\bibfnamefont {A.~J.}\ \bibnamefont {Millis}},\ }\bibfield  {title} {\bibinfo {title} {Hartree-fock study of the moir{\'e} hubbard model for twisted bilayer transition metal dichalcogenides},\ }\href@noop {} {\bibfield  {journal} {\bibinfo  {journal} {Physical Review B}\ }\textbf {\bibinfo {volume} {104}},\ \bibinfo {pages} {075150} (\bibinfo {year} {2021})}\BibitemShut {NoStop}%
\bibitem [{\citenamefont {Senthil}(2015)}]{senthil2015symmetry}%
  \BibitemOpen
  \bibfield  {author} {\bibinfo {author} {\bibfnamefont {T.}~\bibnamefont {Senthil}},\ }\bibfield  {title} {\bibinfo {title} {Symmetry-protected topological phases of quantum matter},\ }\href@noop {} {\bibfield  {journal} {\bibinfo  {journal} {Annu. Rev. Condens. Matter Phys.}\ }\textbf {\bibinfo {volume} {6}},\ \bibinfo {pages} {299} (\bibinfo {year} {2015})}\BibitemShut {NoStop}%
\bibitem [{\citenamefont {Cr{\'e}pel}\ \emph {et~al.}(2024)\citenamefont {Cr{\'e}pel}, \citenamefont {Regnault},\ and\ \citenamefont {Queiroz}}]{crepel2024chiral}%
  \BibitemOpen
  \bibfield  {author} {\bibinfo {author} {\bibfnamefont {V.}~\bibnamefont {Cr{\'e}pel}}, \bibinfo {author} {\bibfnamefont {N.}~\bibnamefont {Regnault}},\ and\ \bibinfo {author} {\bibfnamefont {R.}~\bibnamefont {Queiroz}},\ }\bibfield  {title} {\bibinfo {title} {Chiral limit and origin of topological flat bands in twisted transition metal dichalcogenide homobilayers},\ }\href@noop {} {\bibfield  {journal} {\bibinfo  {journal} {Communications Physics}\ }\textbf {\bibinfo {volume} {7}},\ \bibinfo {pages} {146} (\bibinfo {year} {2024})}\BibitemShut {NoStop}%
\bibitem [{\citenamefont {Yu}\ \emph {et~al.}(2020)\citenamefont {Yu}, \citenamefont {Chen},\ and\ \citenamefont {Yao}}]{yu2020giant}%
  \BibitemOpen
  \bibfield  {author} {\bibinfo {author} {\bibfnamefont {H.}~\bibnamefont {Yu}}, \bibinfo {author} {\bibfnamefont {M.}~\bibnamefont {Chen}},\ and\ \bibinfo {author} {\bibfnamefont {W.}~\bibnamefont {Yao}},\ }\bibfield  {title} {\bibinfo {title} {Giant magnetic field from moir{\'e} induced berry phase in homobilayer semiconductors},\ }\href@noop {} {\bibfield  {journal} {\bibinfo  {journal} {National Science Review}\ }\textbf {\bibinfo {volume} {7}},\ \bibinfo {pages} {12} (\bibinfo {year} {2020})}\BibitemShut {NoStop}%
\bibitem [{\citenamefont {Morales-Dur{\'a}n}\ \emph {et~al.}(2024{\natexlab{a}})\citenamefont {Morales-Dur{\'a}n}, \citenamefont {Shi},\ and\ \citenamefont {MacDonald}}]{morales2024fractionalized}%
  \BibitemOpen
  \bibfield  {author} {\bibinfo {author} {\bibfnamefont {N.}~\bibnamefont {Morales-Dur{\'a}n}}, \bibinfo {author} {\bibfnamefont {J.}~\bibnamefont {Shi}},\ and\ \bibinfo {author} {\bibfnamefont {A.}~\bibnamefont {MacDonald}},\ }\bibfield  {title} {\bibinfo {title} {Fractionalized electrons in moir{\'e} materials},\ }\href@noop {} {\bibfield  {journal} {\bibinfo  {journal} {Nature Reviews Physics}\ }\textbf {\bibinfo {volume} {6}},\ \bibinfo {pages} {349} (\bibinfo {year} {2024}{\natexlab{a}})}\BibitemShut {NoStop}%
\bibitem [{\citenamefont {Ahn}\ \emph {et~al.}(2024)\citenamefont {Ahn}, \citenamefont {Lee}, \citenamefont {Yananose}, \citenamefont {Kim},\ and\ \citenamefont {Cho}}]{ahn2024non}%
  \BibitemOpen
  \bibfield  {author} {\bibinfo {author} {\bibfnamefont {C.-E.}\ \bibnamefont {Ahn}}, \bibinfo {author} {\bibfnamefont {W.}~\bibnamefont {Lee}}, \bibinfo {author} {\bibfnamefont {K.}~\bibnamefont {Yananose}}, \bibinfo {author} {\bibfnamefont {Y.}~\bibnamefont {Kim}},\ and\ \bibinfo {author} {\bibfnamefont {G.~Y.}\ \bibnamefont {Cho}},\ }\bibfield  {title} {\bibinfo {title} {Non-abelian fractional quantum anomalous hall states and first landau level physics of the second moir{\'e} band of twisted bilayer mote 2},\ }\href@noop {} {\bibfield  {journal} {\bibinfo  {journal} {Physical Review B}\ }\textbf {\bibinfo {volume} {110}},\ \bibinfo {pages} {L161109} (\bibinfo {year} {2024})}\BibitemShut {NoStop}%
\bibitem [{\citenamefont {Xu}\ \emph {et~al.}(2025)\citenamefont {Xu}, \citenamefont {Mao}, \citenamefont {Zeng},\ and\ \citenamefont {Zhang}}]{xu2025multiple}%
  \BibitemOpen
  \bibfield  {author} {\bibinfo {author} {\bibfnamefont {C.}~\bibnamefont {Xu}}, \bibinfo {author} {\bibfnamefont {N.}~\bibnamefont {Mao}}, \bibinfo {author} {\bibfnamefont {T.}~\bibnamefont {Zeng}},\ and\ \bibinfo {author} {\bibfnamefont {Y.}~\bibnamefont {Zhang}},\ }\bibfield  {title} {\bibinfo {title} {Multiple chern bands in twisted mote 2 and possible non-abelian states},\ }\href@noop {} {\bibfield  {journal} {\bibinfo  {journal} {Physical Review Letters}\ }\textbf {\bibinfo {volume} {134}},\ \bibinfo {pages} {066601} (\bibinfo {year} {2025})}\BibitemShut {NoStop}%
\bibitem [{\citenamefont {Wang}\ \emph {et~al.}(2025)\citenamefont {Wang}, \citenamefont {Zhang}, \citenamefont {Liu}, \citenamefont {Wang}, \citenamefont {Cao},\ and\ \citenamefont {Xiao}}]{wang2025higher}%
  \BibitemOpen
  \bibfield  {author} {\bibinfo {author} {\bibfnamefont {C.}~\bibnamefont {Wang}}, \bibinfo {author} {\bibfnamefont {X.-W.}\ \bibnamefont {Zhang}}, \bibinfo {author} {\bibfnamefont {X.}~\bibnamefont {Liu}}, \bibinfo {author} {\bibfnamefont {J.}~\bibnamefont {Wang}}, \bibinfo {author} {\bibfnamefont {T.}~\bibnamefont {Cao}},\ and\ \bibinfo {author} {\bibfnamefont {D.}~\bibnamefont {Xiao}},\ }\bibfield  {title} {\bibinfo {title} {Higher landau-level analogs and signatures of non-abelian states in twisted bilayer mote 2},\ }\href@noop {} {\bibfield  {journal} {\bibinfo  {journal} {Physical Review Letters}\ }\textbf {\bibinfo {volume} {134}},\ \bibinfo {pages} {076503} (\bibinfo {year} {2025})}\BibitemShut {NoStop}%
\bibitem [{\citenamefont {Zhang}\ \emph {et~al.}(2024)\citenamefont {Zhang}, \citenamefont {Pi}, \citenamefont {Liu}, \citenamefont {Miao}, \citenamefont {Qi}, \citenamefont {Regnault}, \citenamefont {Weng}, \citenamefont {Dai}, \citenamefont {Bernevig}, \citenamefont {Wu} \emph {et~al.}}]{zhang2024universal}%
  \BibitemOpen
  \bibfield  {author} {\bibinfo {author} {\bibfnamefont {Y.}~\bibnamefont {Zhang}}, \bibinfo {author} {\bibfnamefont {H.}~\bibnamefont {Pi}}, \bibinfo {author} {\bibfnamefont {J.}~\bibnamefont {Liu}}, \bibinfo {author} {\bibfnamefont {W.}~\bibnamefont {Miao}}, \bibinfo {author} {\bibfnamefont {Z.}~\bibnamefont {Qi}}, \bibinfo {author} {\bibfnamefont {N.}~\bibnamefont {Regnault}}, \bibinfo {author} {\bibfnamefont {H.}~\bibnamefont {Weng}}, \bibinfo {author} {\bibfnamefont {X.}~\bibnamefont {Dai}}, \bibinfo {author} {\bibfnamefont {B.~A.}\ \bibnamefont {Bernevig}}, \bibinfo {author} {\bibfnamefont {Q.}~\bibnamefont {Wu}}, \emph {et~al.},\ }\bibfield  {title} {\bibinfo {title} {Universal moir$\backslash$'e-model-building method without fitting: Application to twisted mote $ \_2 $ and wse $ \_2$},\ }\href@noop {} {\bibfield  {journal} {\bibinfo  {journal} {arXiv preprint arXiv:2411.08108}\ } (\bibinfo {year} {2024})}\BibitemShut {NoStop}%
\bibitem [{\citenamefont {Morales-Dur{\'a}n}\ \emph {et~al.}(2024{\natexlab{b}})\citenamefont {Morales-Dur{\'a}n}, \citenamefont {Wei}, \citenamefont {Shi},\ and\ \citenamefont {MacDonald}}]{morales2024magic}%
  \BibitemOpen
  \bibfield  {author} {\bibinfo {author} {\bibfnamefont {N.}~\bibnamefont {Morales-Dur{\'a}n}}, \bibinfo {author} {\bibfnamefont {N.}~\bibnamefont {Wei}}, \bibinfo {author} {\bibfnamefont {J.}~\bibnamefont {Shi}},\ and\ \bibinfo {author} {\bibfnamefont {A.~H.}\ \bibnamefont {MacDonald}},\ }\bibfield  {title} {\bibinfo {title} {Magic angles and fractional chern insulators in twisted homobilayer transition metal dichalcogenides},\ }\href@noop {} {\bibfield  {journal} {\bibinfo  {journal} {Physical Review Letters}\ }\textbf {\bibinfo {volume} {132}},\ \bibinfo {pages} {096602} (\bibinfo {year} {2024}{\natexlab{b}})}\BibitemShut {NoStop}%
\bibitem [{\citenamefont {Jia}\ \emph {et~al.}(2024)\citenamefont {Jia}, \citenamefont {Yu}, \citenamefont {Liu}, \citenamefont {Herzog-Arbeitman}, \citenamefont {Qi}, \citenamefont {Pi}, \citenamefont {Regnault}, \citenamefont {Weng}, \citenamefont {Bernevig},\ and\ \citenamefont {Wu}}]{jia2024moire}%
  \BibitemOpen
  \bibfield  {author} {\bibinfo {author} {\bibfnamefont {Y.}~\bibnamefont {Jia}}, \bibinfo {author} {\bibfnamefont {J.}~\bibnamefont {Yu}}, \bibinfo {author} {\bibfnamefont {J.}~\bibnamefont {Liu}}, \bibinfo {author} {\bibfnamefont {J.}~\bibnamefont {Herzog-Arbeitman}}, \bibinfo {author} {\bibfnamefont {Z.}~\bibnamefont {Qi}}, \bibinfo {author} {\bibfnamefont {H.}~\bibnamefont {Pi}}, \bibinfo {author} {\bibfnamefont {N.}~\bibnamefont {Regnault}}, \bibinfo {author} {\bibfnamefont {H.}~\bibnamefont {Weng}}, \bibinfo {author} {\bibfnamefont {B.~A.}\ \bibnamefont {Bernevig}},\ and\ \bibinfo {author} {\bibfnamefont {Q.}~\bibnamefont {Wu}},\ }\bibfield  {title} {\bibinfo {title} {Moir{\'e} fractional chern insulators. i. first-principles calculations and continuum models of twisted bilayer mote 2},\ }\href@noop {} {\bibfield  {journal} {\bibinfo  {journal} {Physical Review B}\ }\textbf {\bibinfo {volume} {109}},\ \bibinfo {pages} {205121} (\bibinfo {year} {2024})}\BibitemShut {NoStop}%
\bibitem [{\citenamefont {Qian}\ \emph {et~al.}(2014)\citenamefont {Qian}, \citenamefont {Liu}, \citenamefont {Fu},\ and\ \citenamefont {Li}}]{qian2014quantum}%
  \BibitemOpen
  \bibfield  {author} {\bibinfo {author} {\bibfnamefont {X.}~\bibnamefont {Qian}}, \bibinfo {author} {\bibfnamefont {J.}~\bibnamefont {Liu}}, \bibinfo {author} {\bibfnamefont {L.}~\bibnamefont {Fu}},\ and\ \bibinfo {author} {\bibfnamefont {J.}~\bibnamefont {Li}},\ }\bibfield  {title} {\bibinfo {title} {Quantum spin hall effect in two-dimensional transition metal dichalcogenides},\ }\href@noop {} {\bibfield  {journal} {\bibinfo  {journal} {Science}\ }\textbf {\bibinfo {volume} {346}},\ \bibinfo {pages} {1344} (\bibinfo {year} {2014})}\BibitemShut {NoStop}%
\bibitem [{\citenamefont {Chen}\ \emph {et~al.}(2010)\citenamefont {Chen}, \citenamefont {Gu},\ and\ \citenamefont {Wen}}]{chen2010local}%
  \BibitemOpen
  \bibfield  {author} {\bibinfo {author} {\bibfnamefont {X.}~\bibnamefont {Chen}}, \bibinfo {author} {\bibfnamefont {Z.-C.}\ \bibnamefont {Gu}},\ and\ \bibinfo {author} {\bibfnamefont {X.-G.}\ \bibnamefont {Wen}},\ }\bibfield  {title} {\bibinfo {title} {Local unitary transformation, long-range quantum entanglement, wave function renormalization, and topological order},\ }\href@noop {} {\bibfield  {journal} {\bibinfo  {journal} {Physical Review B—Condensed Matter and Materials Physics}\ }\textbf {\bibinfo {volume} {82}},\ \bibinfo {pages} {155138} (\bibinfo {year} {2010})}\BibitemShut {NoStop}%
\end{thebibliography}

\newpage
\onecolumngrid
\appendix 
\section{Mapping the edge theory to two Luttinger liquids.}
In this section, we specify how the change of variables (\ref{newbasis}) maps the Lagrangian (\ref{initLagrangian}) to two decoupled Luttinger liquids. The change of basis may be written in matrix form:

\begin{equation}
    \vec{\phi}' \equiv \vec{W}\inv \vec{\phi}
\end{equation}
where
\begin{equation}
    \vec{W} = \frac{1}{2}\begin{pmatrix}
        1 & 1 & 1 & 1\\
        -1 & -1 & 1 & 1\\
        1 & -1 & -1 & 1 \\
        -1 & 1  & -1 & 1
    \end{pmatrix}, \text{ and } \vec{W}\inv = \vec{W}^\intercal = \frac{1}{2}\begin{pmatrix}
        1 & -1 & 1 & -1\\
        1 & -1 & -1 & 1 \\
        1 & 1 & -1 & -1 \\
        1 & 1 & 1 & 1
    \end{pmatrix}
\end{equation}
such that 
\begin{equation}
    \vec{\phi'} =  \begin{pmatrix}
            \phi_+ \\ \phi_- \\ \theta_- \\ \theta_+
        \end{pmatrix}.
\end{equation}
Our free Lagrangian (\ref{initLagrangian}) becomes 
\begin{equation}
    \begin{split}
        \L &= -\frac{1}{4\pi} \partial_x (\vec{W}\vec{\phi}')^\intercal \vec{K} \partial_t (\vec{W}\vec{\phi}') + \frac{1}{4\pi} \partial_x (\vec{W}\vec{\phi}')^\intercal \vec{V} \partial_x (\vec{W}\vec{\phi}')\\
        & = -\frac{1}{4\pi} \partial_x \vec{\phi}'^\intercal \underbrace{(\vec{W}^\intercal \vec{K} \vec{W})}_{\mathclap{\vec{K}'}} \partial_t \vec{\phi}' + \frac{1}{4\pi} \partial_x \vec{\phi}'^\intercal \underbrace{(\vec{W}^\intercal \vec{V} \vec{W})}_{\mathclap{\vec{V}'}}\partial_x \vec{\phi}'
    \end{split}
\end{equation}
where the transformed matrix $\vec{K}'$ is now
\begin{equation}
    \vec{K} = \begin{pmatrix}
        0 & 0 & 0 & 1 \\
        0 & 0 & 1 & 0 \\
        0 & 1 & 0 & 0 \\
        1 & 0 & 0 & 0 \\
    \end{pmatrix}
\end{equation}
And the velocity matrix is now diagonal:

\begin{equation}
    \vec{V}' = \vec{W}^\intercal \vec{V}\vec{W} = \begin{pmatrix}
        v - v' + u - u' & & & \\
        & v - v' - u + u' & & \\
        & & v + v' - u - u' & \\
        & & & v + v' + u + u'
    \end{pmatrix}.
\end{equation}
We may expand the Lagrangian in full:
\begin{equation}
    \begin{split}
        \L = & \frac{1}{2\pi}\left[(\partial_x \phi_+)(\partial_t\vartheta_+) + (\partial_x \phi_-)(\partial_t\vartheta_-)\right]\\
        &+\frac{1}{4\pi} \left[(v + v' - u - u')(\partial_x \vartheta_-)^2  + (v - v' - u + u')(\partial_x \phi_-)^2\right]\\
        & +\frac{1}{4\pi} \left[(v + v' + u + u')(\partial_x \vartheta_+)^2  + (v - v' + u - u')(\partial_x\phi_+)^2\right]\\
    \end{split}
\end{equation}
Defining the following quantities
\begin{align*}
    v_\pm &= \sqrt{[(v - v)' \pm (u - u') ][(v + v)' \pm (u + u') ]}\\
    g_\pm &= \sqrt{\frac{(v - v)' \pm (u - u') }{(v + v)' \pm (u + u') }},
\end{align*}
our theory becomes 
\begin{equation}
\begin{split}
        \L =& \frac{1}{2\pi}\left[(\partial_x \phi_+)(\partial_t\vartheta_+) + (\partial_x \phi_-)(\partial_t\vartheta_-)\right]\\
            &+ \frac{v_+}{4\pi}\left[g_+ (\partial_x \vartheta_+)^2 + \frac{1}{g_+} (\partial_x \phi_+)^2\right]\\
        &+ \frac{v_-}{4\pi}\left[g_- (\partial_x \vartheta_-)^2 + \frac{1}{g_-} (\partial_x \phi_-)^2\right]
\end{split}
\end{equation}

In general, the edge velocities are not equal to one another, and each channel will have different scattering parameters. One can consider a more general velocity matrix, e.g. of the form 
\begin{equation}
    \vec{V} = \begin{pmatrix}
        v_1 & -v_1' & u & -u' \\
        -v_1' & v_1 & -u' & u\\
        u & -u' & v_2 & -v_2'\\
        -u' & u & -v_2' & v_2
    \end{pmatrix}
    \label{generalVmat}
\end{equation}
which leads to extra interactions of the form
\begin{equation}
    w(\partial_x \phi_+) (\partial_x \phi_-) + w'(\partial_x \vartheta_+)(\partial_x \vartheta_-)
\end{equation} 
where $w = v_1 - v_2 -v_1' + v_2'$ and $w' = v_1 - v_2 + v_1' -v_2'$.
However, we show in Appendix \ref{seclowenergyasymp} that including these interactions do not affect the main results, since the propagators between the $+$ and $-$ sectors vanish in the limit as $\omega\to 0$.

\newpage
\section{\label{bsbosonized}Backscattering action of the non-interacting edge.}
In this section, we will derive the backscattering action for the non-interacting theory (\ref{Sbnoninteracting}). Once the Luttinger liquid has been folded as described in section \ref{foldedpicture}, the possible backscattering operators in the non-interacting theory are 
\begin{equation}
    \begin{split}
        \psi_{1\uparrow\text{L}}^\dagger \psi_{1\uparrow\text{R}} + \psi_{1\downarrow\text{L}}^\dagger \psi_{1\downarrow\text{R}} +  \text{H.c.}\\
        \psi_{2\uparrow\text{L}}^\dagger \psi_{2\uparrow\text{R}} + \psi_{2\downarrow\text{L}}^\dagger \psi_{2\downarrow\text{R}} + \text{H.c.}\\
        \psi_{2\uparrow\text{L}}^\dagger \psi_{1\uparrow\text{R}} + \psi_{2\downarrow\text{L}}^\dagger \psi_{1\downarrow\text{R}} +  \text{H.c.}\\
        \psi_{1\uparrow\text{L}}^\dagger \psi_{2\uparrow\text{R}} + \psi_{1\downarrow\text{L}}^\dagger \psi_{2\downarrow\text{R}} + \text{H.c.}
    \end{split}
    \label{fermionbsappdx}
\end{equation} 
We may first look at the intra-channel backscattering terms. These are
\begin{align}
    \psi_{1\uparrow\text{L}}^\dagger \psi_{1\uparrow\text{R}} + \text{H.c.} &\sim \cos\left(\phi_{\text{c}} + \phi_{\text{n}+}+\phi_{\text{s}+} + \phi_{\text{n}-}\right)\\
    \psi_{1\downarrow\text{L}}^\dagger \psi_{1\downarrow\text{R}} + \text{H.c.} &\sim \cos\left(\phi_{\text{c}} + \phi_{\text{n}+}-\phi_{\text{s}+} - \phi_{\text{n}-}\right)
\end{align}
Hence, using the cosine sum-to-product formula, we find that the term which controls backscattering solely between channel 1 is 
\begin{equation}
    \psi_{1\uparrow\text{L}}^\dagger \psi_{1\uparrow\text{R}} + \psi_{1\downarrow\text{L}}^\dagger \psi_{1\downarrow\text{R}} + \text{H.c.} \sim \cos\left(\phi_{\text{c}} + \phi_{\text{n}+}\right)\cos\left(\phi_{\text{s}+} + \phi_{\text{n}-}\right).
\end{equation}
We can work out a similar expression with copy $a=2$ instead. In this instance,
\begin{equation}
    \psi_{2\uparrow\text{L}}^\dagger \psi_{2\uparrow\text{R}} + \psi_{2\downarrow\text{L}}^\dagger \psi_{2\downarrow\text{R}} + \text{H.c.} \sim \cos\left(\phi_{\text{c}} - \phi_{\text{n}+}\right)\cos\left(\phi_{\text{s}+} - \phi_{\text{n}-}\right).
\end{equation}
In a similar fashion, we may write the inter-channel backscattering terms, viz. the backscattering events which transfer an electron from channel 1 to channel 2 or vice versa. We find 
\begin{equation}
    \begin{split}
        \psi_{2\uparrow\text{L}}^\dagger \psi_{1\uparrow\text{R}} + \psi_{2\downarrow\text{L}}^\dagger \psi_{1\downarrow\text{R}} +  \text{H.c.} &\sim \cos(\phi_{\text{c}} + \vartheta_{\text{n}+})\cos(\phi_{\text{s}+} + \vartheta_{\text{n}-})\\
        \psi_{1\uparrow\text{L}}^\dagger \psi_{2\uparrow\text{R}} + \psi_{1\downarrow\text{L}}^\dagger \psi_{2\downarrow\text{R}} +  \text{H.c.} &\sim \cos(\phi_{\text{c}} - \vartheta_{\text{n}-})\cos(\phi_{\text{s}+} - \vartheta_{\text{n}-})
    \end{split}
\end{equation}

Therefore, 
\begin{multline}
    \L_{\text{b}} \sim \cos\left(\phi_{\text{c}} + \phi_{\text{n}+}\right)\cos\left(\phi_{\text{s}+} + \phi_{\text{n}-}\right)
    +\cos\left(\phi_{\text{c}} - \phi_{\text{n}+}\right)\cos\left(\phi_{\text{s}+} - \phi_{\text{n}-}\right)\\
    +\cos\left(\phi_{\text{c}} + \vartheta_{\text{n}+}\right)\cos\left(\phi_{\text{s}+} + \vartheta_{\text{n}-}\right)
    +\cos\left(\phi_{\text{c}} - \vartheta_{\text{n}+}\right)\cos\left(\phi_{\text{s}+} - \vartheta_{\text{n}-}\right).
\end{multline}
We can now use the formula 
\begin{equation}
    \cos(A + B)\cos(C+D) + \cos(A - B)\cos(C-D) = 2 \sin(A) \sin(B) \sin(C) \sin(D) + 2 \cos(A) \cos(B) \cos(C) \cos(D)
\end{equation}
To write the full backscattering Lagrangian as 
\begin{equation}
    \begin{split}
        \L_{\text{b}} \sim \cos(\phi_{\text{c}})\cos(\phi_{\text{n}+})\cos(\phi_{\text{s}+})\cos(\phi_{\text{n}-})
        + \sin(\phi_{\text{c}})\sin(\phi_{\text{n}+})\sin(\phi_{\text{s}+})\sin(\phi_{\text{n}-})\\
        + \cos(\phi_{\text{c}})\cos(\vartheta_{\text{n}+})\cos(\phi_{\text{s}+})\cos(\vartheta_{\text{n}-})
        + \sin(\phi_{\text{c}})\sin(\vartheta_{\text{n}+})\sin(\phi_{\text{s}+})\sin(\vartheta_{\text{n}-}).
    \end{split}
\end{equation}

\newpage 
\section{\label{seclowenergyasymp}Low-energy asymptotics of the retarded propagator}
In this section, we slightly generalize the calculation found in \cite{PhysRevB.52.R5539} to the case of two coupled inhomogeneous Luttinger liquids with unequal velocities and scattering parameters. We will obtain the low-energy asymptotic form of the various correlators, which we show are determined by the values of the parameters in the leads. We demonstrate that the original result of \cite{PhysRevB.52.R5539} still holds, and that the propagators between the $+$ and $-$ sectors vanish as $\omega \to 0$.  This means that we may ignore this interaction term for the purposes of this analysis, which is concerned exclusively with the zero-frequency limit of all quantities. 

\subsection{Formulation.}
Given the velocity matrix in equation \ref{generalVmat}, in the limit that the velocities and backscattering parameters $v_1 = v_2$ and $v_1' = v_2'$, the two luttinger liquids fully decouple from one another and can be treated separately. Failing that, we must instead consider the matrix differential equation 
\begin{equation}
\begin{split}
    \hat{\vec{D}}\vec{G} &= \begin{pmatrix}
        D_+ & D' \\ D' & D_-
    \end{pmatrix}\begin{pmatrix}
        G_+ & G' \\ G'' & G_-
    \end{pmatrix}\\
    &= \delta(x-x')\mathds{1}_{2\times 2} \\ 
\end{split}
    \label{matrixdiffeq}
\end{equation}
The entries of the matrix $\hat{\vec{D}}$ are the following differential operators:
\begin{align}
    D_{+} &= \partial_x \left(\frac{v_+}{K_+}\partial_x\right) +  \frac{1}{v_+K_+}\omega^2\\
    D_{-} &= \partial_x \left(\frac{v_-}{K_-}\partial_x\right) +  \frac{1}{v_-K_-}\omega^2\\
    D' &= \partial_x\left(h \partial_x\right) + h' \omega^2.
\end{align}
\begin{comment}
\begin{align}
    D_{+} &= -\partial_x \left(\frac{v_+}{K_+}\partial_x\right) + \frac{\omega^2}{v_+ K_+}\\
    D_{-} &= -\partial_x \left(\frac{v_-}{K_-}\partial_x\right) + \frac{\omega^2}{v_- K_-}\\
    D' &= -\partial_x\left(\lambda \partial_x\right) + \lambda'\omega^2.
\end{align}
\end{comment}
Assume that outsisde the wire of length $L$, these parameters assume the boundary values of $v_\pm \to  v$, $K_\pm \to  K_\infty$, and $h,h'\to 0$. Written out, we obtain a set of four differential equations:
\begin{align}
    D_{+}G_{+} + D' G'' &= \delta(x-x') \label{DG1}\\
    D_{+} G' + D' G_{-} &= 0 \label{DG2}\\
    D'G_{+} + D_{-} G'' &=0 \label{DG3}\\
    D'G' + D_{-}G_{-} &= \delta(x-x'). \label{DG4}
\end{align}
In the following sections, we will solve the differential equations for $x'\in [0,L]$, since all backscattering events occur within the wire.
\subsection{Region I: $x<0$}

In the region $x < 0$, the parameters $h$ and $h'$ are both 0. Hence, $D'=0$ and the differential operators $D_+$ and $D_-$ coincide and we obtain the same homogeneous differential equation for all four propagators. We impose outgoing wave boundary conditions and obtain plane-wave solutions for $x<0$,
\begin{align}
    G_{+}(x,x') &= A^+ e^{-i\omega x/v} \\
    G'(x,x') &= A'e^{-i\omega x/v} \\
    G''(x,x') &= A''e^{-i\omega x/v} \\
    G_{-}(x,x') &= A^-e^{-i\omega x/v} 
\end{align}
 Similarly, for values of $x>L$, 
\begin{align}
    G_{+}(x,x') &= B^+ e^{i\omega x/v} \\
    G'(x,x') &= B'e^{i\omega x/v} \\
    G''(x,x') &= B''e^{i\omega x/v} \\
    G_{-}(x,x') &= B^{-} e^{i\omega x/v}
\end{align}

\begin{comment}
    Hence, $D'=0$ and (\ref{matrixdiffeq}) becomes 
\begin{align}
    D_{+}G_{+} &= 0 \label{DG1leq0}\\
    D_{+} G' &= 0 \label{DG2leq0}\\
    D_{-} G'' &=0 \label{DG3leq0}\\
    D_{-}G_{-} &= 0. \label{DG4leq0}
\end{align}
In this region, $K_+ = K_- = K_{\infty }$, and $v_+ = v_- = v$. Hence, the differential operators coincide and we obtain the same homogeneous differential equation for all three propagators. Thus,
\begin{align}
    G_{+}(x,x') &= A^+ e^{-i\omega x/v} \\
    G'(x,x') &= A'e^{-i\omega x/v} \\
    G''(x,x') &= A''e^{-i\omega x/v} \\
    G_{-}(x,x') &= A^-e^{-i\omega x/v} 
\end{align}
\begin{comment}
    \begin{align}
    G_{+}(x,x') &= A^+ e^{\omegax/v} \\
    G'(x,x') &= A'e^{\omegax/v} \\
    G''(x,x') &= A''e^{\omegax/v} \\
    G_{-}(x,x') &= A^- e^{\omegax/v} 
\end{align}
where we impose outgoing wave boundary conditions.
\end{comment}

\subsection{Region II: $0<x<x'$}
In this region, equation (\ref{matrixdiffeq}) may be written as
\begin{align}
    D_{+}G_{+} + D' G'' &= 0 \label{DG1xx'}\\
    D_{+} G' + D' G_{-} &= 0 \label{DG2xx'}\\
    D'G_{+} + D_{-} G'' &=0 \label{DG3xx'}\\
    D'G' + D_{-}G_{-} &= 0. \label{DG4xx'}
\end{align}
By acting with the appropriate differential operators on the equations (for instance, $D'$ on (\ref{DG1xx'}) and with $D_{-}$ on (\ref{DG3xx'})), one finds that $G_\pm$, $G'$ and $G''$ are all solutions to the same homogeneous, fourth-order, linear differential equation:
\begin{equation}
\begin{split}
    \left\{D_-D_+ - D'D'\right\}G'(x,x') &= 0\\
    \left\{D_-D_+ - D'D'\right\}G''(x,x') &= 0
\end{split}
\end{equation}
Explicitly, we have that
\begin{equation}
        D_- D_+ = \frac{v_- v_+}{K_- K_+}\partial_x^4 - \omega^2 \left(\frac{v_+^2 + v_-^2}{v_- K_- v_+ K_+}\right)+ \frac{\omega^4}{v_- K_- v_+ K_+}\\
\end{equation}
and 
\begin{equation}
        D'D' = h^2 \partial_x^4 + 2h h' \omega^2 \partial_x^2 + h'^2 \omega^4
\end{equation}
Hence, 
\begin{equation}
    D_- D_+ - D'D' = \left(\frac{v_-v_+}{K_-K_+} - h^2\right)\partial_x^4 - \left(\frac{v_+^2 + v_-^2}{v_- K_- v_+ K_+} - 2h h'\right)\omega^2 \partial_x^2 + \left(\frac{\omega^4}{v_- K_- v_+ K_+}- h'^2\right) \omega^4
\end{equation}
We may solve this the usual way by finding the roots of the following characteristic equation 
\begin{equation}
    \begin{split}
        p(r) &= \left(\frac{v_-v_+}{K_-K_+} - \lambda^2\right) r^4 + \left(\frac{v_+^2 + v_-^2}{v_- K_- v_+ K_+} - 2\lambda \lambda'\right)\omega^2 r^2 + \left(\frac{\omega^4}{v_- K_- v_+ K_+}- \lambda'^2\right) \omega^4\\
        &= a r^4 + b r^2 + c.
    \end{split}
\end{equation}
We find that the roots to this equation are $\{\lambda_1, -\lambda_1,\lambda_2,-\lambda_2\}$. Moreover, defining $\rho_1$ and $\rho_2$ such that $\lambda_1 = \omega \rho_1$ and $\lambda_2 = \omega\rho_2$, we obtain the general solution for the propagators inside the wire:
\begin{equation}
\begin{split}
        G_+(x,x') &= A_1^+ e^{\omega\rho_1 x} + A_2^+ e^{-\omega\rho_1 x} + A_3^+ e^{\omega\rho_2 x} + A_4^+ e^{-\omega\rho_2 x}\\
        G'(x,x') &= A_1' e^{\omega\rho_1 x} + A_2' e^{-\omega\rho_1 x} + A_3' e^{\omega\rho_2 x} + A_4' e^{-\omega\rho_2 x}\\
        G''(x,x') &= A_1'' e^{\omega\rho_1 x} + A_2'' e^{-\omega\rho_1 x} + A_3'' e^{\omega\rho_2 x} + A_4'' e^{-\omega\rho_2 x}\\
        G_-(x,x') &= A_1^- e^{\omega\rho_1 x} + A_2^- e^{-\omega\rho_1 x} + A_3^- e^{\omega\rho_2 x} + A_4^- e^{-\omega\rho_2 x}.
\end{split}
\end{equation}
\begin{comment}
    As a quick check, one can show that if $\lambda = \lambda' = 0$, then 
\begin{equation}
    \begin{split}
        r_1 &= \omega\left(\frac{-|v_+^2 - v_-^2| + v_+^2 + v_-^2}{2v_+^2v_-^2}\right)^{1/2}\\
        r_2 &= \omega\left(\frac{|v_+^2 - v_-^2| + v_+^2 + v_-^2}{2v_+^2v_-^2}\right)^{1/2}.
    \end{split}
\end{equation}
The numerator of $r_1$ is $2v_+^2$ if $v_+ \geq v_-$ and $2v_-^2$ if $v_ + < v_-$. Similarly, the numerator of $r_2$ is $2v_-^2$ if $v_+ \geq v_-$ and $2v_+^2$ if $v_+ < v_-$. Hence, we find that there are always two roots which become to $\omega/v_+$ and two which become $\omega/v_-$ as $\lambda$ and $\lambda'$ go to zero.
\end{comment}

\subsection{Region III: $x' < x < L$}
Our analysis in the previous section applies to this region as well. Thus, 
\begin{align}
    G_+(x,x') &= B_1^+ e^{\omega\rho_1 x} + B_2+ e^{-\omega\rho_1 x} + B_3^+ e^{\omega\rho_2 x} + B_4^+ e^{-\omega\rho_2 x}\\
    G''(x,x') &= B_1'' e^{\omega\rho_1 x} + B_2'' e^{-\omega\rho_1 x} + B_3'' e^{\omega\rho_2 x} + B_4'' e^{-\omega\rho_2 x}\\
    G'(x,x') &= B_1' e^{\omega\rho_1 x} + B_2' e^{-\omega\rho_1 x} + B_3' e^{\omega\rho_2 x} + B_4' e^{-\omega\rho_2 x}\\
    G_-(x,x') &= B_1^- e^{\omega\rho_1 x} + B_2^- e^{-\omega\rho_1 x} + B_3^- e^{\omega\rho_2 x} + B_4^- e^{-\omega\rho_2 x}.
\end{align}

\subsection{Matching boundary conditions.}
We will now determine the coefficients of these solutions by imposing boundary conditions.

\begin{comment}
    First, we will take stock of how many unknown quantities are present, and how many linearly independent equations we have relating them. In total, there are 40 unknown quantities: four from $A^\pm, A', A''$, plus sixteen from $A_i^\pm, A_i', A_i''$, multiplied by 2 for the corresponding $B$ coefficients. Our set of four differential equations already constrains the problem, providing two relationships for two coefficients per equation, multiplied by 2 for the $B$ coefficients, yielding $4\times 2 \times 2 = 16$. Lastly, we can determine the remaining variables by imposing boundary conditions. Since there are three regions, we obtain 3 equations by demanding continuity of the propagators $G_\pm, G', G''$, for a total of $3\times 4=12$ more equations. The remaining 12 equations are provided by integrating the four equations (\ref{matrixdiffeq}) around the regions $0,L$, and $x'$, yielding $16 + 12 + 12 = 40$. 
\end{comment}
The equation (\ref{matrixdiffeq}) implies that at $x=0, x'$, and $L$, the propagators are all continuous. Additionally, we may freely set $\omega$ to 0 in all these conditions, leading us to the following equations. Continuity at $x=0$ yields
\begin{align}
    A^+ &=  A_1^+ + A_2^+ + A_3^+ + A_4^+ \label{contpropag1} \\
    A'' &= A_1'' + A_2'' + A_3'' + A_4'' \label{contpropag2}\\
    A' &= A_1' + A_2' + A_3' + A_4' \\
    A^- &= A_1^- + A_2^- + A_3^- + A_4^- \label{contpropag4}
\end{align}
Continuity at $x=L$ yields
\begin{align}
    B^+ &= B_1^+ + B_2^+ + B_3^+ + B_4^+ \\
    B' &= B_1' + B_2' + B_3' + B_4'\\
    B'' &= B_1'' + B_2'' + B_3'' + B_4''\\
    B^- &= B_1^- + B_2^- + B_3^- + B_4^-
\end{align}
Lastly, continuity of the propagators at $x = x'$ yields
\begin{align}
     A_1^+ + A_2^+ + A_3^+ + A_4^+ &= B_1^+ + B_2^+ + B_3^+ + B_4^+\\
     A_1' + A_2' + A_3' + A_4' &=B_1' + B_2' + B_3' + B_4' \\
     A_1'' + A_2'' + A_3'' + A_4'' &=B_1'' + B_2'' + B_3'' + B_4'' \\
     A_1^- + A_2^- + A_3^- + A_4^- &= B_1^- + B_2^- + B_3^- + B_4^-.
\end{align}
Combining these, we find that the coefficients of the propagators must be equal outside the leads.
\begin{align}
    A^+ &= B^+\\
    A' &= B'\\
    A'' &= B''\\
    A^- &= B^-.
\end{align}

\subsubsection{Conditions on first derivatives at  $x= 0, x', L$}
Next, we turn our attention to boundary conditions involving the first derivatives of the propagators, which we obtain by integrating in an $\eps$-ball around the points $x= 0,x',L$, and taking the limit as $\eps\to 0$. When the differential operators are expanded out, we obtain that \ref{matrixdiffeq} yields the following four equations:
\begin{align}
    \left[\partial_x \left(\frac{v_+}{K_+}\partial_x \right) + \frac{\omega^2}{v_+ K_+} \right] G_+ + \left[h \partial_x^2  + \lambda' \omega^2 \right] G'' &= \delta(x-x')\\
    \left[\partial_x \left(\frac{v_+}{K_+}\partial_x \right) + \frac{\omega^2}{v_+ K_+} \right] G' + \left][ h \partial_x^2  + h' \omega^2 \right] G_- &= 0\\
    \left[\partial_x \left(\frac{v_-}{K_-}\partial_x \right) + \frac{\omega^2}{v_- K_-} \right] G'' + \left[ h \partial_x^2  + h' \omega^2 \right] G_+ &= 0\\
    \left[\partial_x \left(\frac{v_-}{K_-}\partial_x \right) + \frac{\omega^2}{v_- K_-} \right] G_- + \left[ h \partial_x^2  + h' \omega^2 \right] G' &= \delta(x-x').
\end{align}
In order for the equations to be well-behaved at $x=0,L$, we require continuity of the first derivatives at these two points. Hence, the following combinations of first derivatives of the propagator 
\begin{align}
    \frac{v_+}{K_+}\partial_x G_+ &+ h \partial_x G'' \label{contder1} \\
    \frac{v_+}{K_+}\partial_x G' &+ h \partial_x G_- \\
    \frac{v_-}{K_-}\partial_x G'' &+ h \partial_x G_+ \\
    \frac{v_-}{K_-}\partial_x G_- &+ h \partial_x G'
\end{align}
are all continuous at $x=0,L$. Moreover, for the middle two equations, we have that the combinations
\begin{align}
    \left.\frac{v_+}{K_+}\partial_x G' + h \partial_x G_- \right|^{x=x'+0}_{x=x'-0} &=0\\
    \left.\frac{v_-}{K_-}\partial_x G' + h \partial_x G_+\right|^{x=x'+0}_{x=x'-0} &=0.
\end{align}
are also continuous at $x=x'$. However, the first and the last equations imply that the following combinations of first derivatives undergo a jump of unit height at $x=x'$:
\begin{align}
    \left.\frac{v_+}{K_+}\partial_x G_+ + h \partial_x G'' \right|^{x=x'+0}_{x=x'-0} &=1\\
    \left.\frac{v_-}{K_-}\partial_x G_- + h \partial_x G' \right|^{x=x'+0}_{x=x'-0} &=1.
\end{align}
\subsubsection{Equations at $x = 0, L$}
From continuity of \ref{contder1} at $x = 0$, we have 
\begin{equation}
    \left(\frac{v_+}{K_+}\partial_x G_+ + h \partial_x G''\right) = \frac{v}{K_{\infty}}\partial_x G_+ + 0.
\end{equation}
We can eliminate the factors of $\omega$ which pop out from the derivative on both sides, and summarily set $\omega = 0$ to obtain
\begin{equation}
    \frac{v_+}{K_+} \left[\rho_1 A_1^+ - \rho_1A_2^+ + \rho_2 A_3^+ - \rho_2 A_4^+\right] + h\left[\rho_1 A_1'' - \rho_1A_2'' + \rho_2 A_3'' - \rho_2 A_4''\right] = \frac{1}{K_{\infty}}A^+
\end{equation}
The rest of the continuity equations follow similarly, yielding 
\begin{align}
    \frac{v_+}{K_+} \left[\rho_1 A_1^+ - \rho_1A_2^+ + \rho_2 A_3^+ - \rho_2 A_4^+\right] + h\left[\rho_1 A_1'' - \rho_1A_2'' + \rho_2 A_3'' - \rho_2 A_4''\right] &= \frac{1}{K_{\infty}}A^+ \label{cont1a} \\
    \frac{v_-}{K_-} \left[\rho_1 A_1'' - \rho_1A_2'' + \rho_2 A_3'' - \rho_2 A_4''\right] + h \left[\rho_1 A_1^+ - \rho_1A_2^+ + \rho_2 A_3^+ - \rho_2 A_4^+\right] &= \frac{1}{K_{\infty}}A'' \label{cont2a} \\
    \frac{v_+}{K_+} \left[\rho_1 A_1' - \rho_1A_2' + \rho_2 A_3' - \rho_2 A_4'\right] + h \left[\rho_1 A_1^- - \rho_1A_2^- + \rho_2 A_3^- - \rho_2 A_4^-\right] &= \frac{1}{K_{\infty}} A'\\
    \frac{v_-}{K_-} \left[\rho_1 A_1^- - \rho_1A_2^- + \rho_2 A_3^- - \rho_2 A_4^-\right] + h\left[\rho_1 A_1' - \rho_1A_2' + \rho_2 A_3' - \rho_2 A_4'\right] &= \frac{1}{K_{\infty}}A^-
\end{align}
And exactly analogously for the $B$ variables:
\begin{align}
    \frac{v_+}{K_+} \left[\rho_1 B_1^+ - \rho_1B_2^+ + \rho_2 B_3^+ - \rho_2 B_4^+\right] + h\left[\rho_1 B_1'' - \rho_1 B_2'' + \rho_2 B_3'' - \rho_2 B_4''\right] &= -\frac{1}{K_{\infty}}B^+ \label{cont1b} \\
    \frac{v_-}{K_-} \left[\rho_1 B_1'' - \rho_1B_2'' + \rho_2 B_3'' - \rho_2 B_4''\right] + h \left[\rho_1 B_1^+ - \rho_1B_2^+ + \rho_2 B_3^+ - \rho_2 B_4^+\right] &= -\frac{1}{K_{\infty}}B'' \label{cont2b}\\
    \frac{v_+}{K_+} \left[\rho_1 B_1' - \rho_1 B_2' + \rho_2 B_3' - \rho_2 B_4'\right] + h \left[\rho_1 B_1^- - \rho_1 B_2^- + \rho_2 B_3^- - \rho_2 B_4^-\right] &= -\frac{1}{K_{\infty}} B'\\
    \frac{v_-}{K_-} \left[\rho_1 B_1^- - \rho_1 B_2^- + \rho_2 B_3^- - \rho_2 B_4^-\right] + h\left[\rho_1 B_1' - \rho_1 B_2' + \rho_2 B_3' - \rho_2 B_4'\right] &= -\frac{1}{K_{\infty}}B^-
\end{align}
\subsubsection{Equations at $x=x'$}
Finally, two of our equations undergo a jump at $x = x'$, and two remain continuous:
\begin{align}
    \left.\frac{v_+}{K_+}\partial_x G_+ + h \partial_x G'' \right|^{x=x'+0}_{x=x'-0} &=1\\
    \left.\frac{v_+}{K_+}\partial_x G' + h \partial_x G_- \right|^{x=x'+0}_{x=x'-0} &=0\\
    \left.\frac{v_-}{K_-}\partial_x G'' + h \partial_x G_+\right|^{x=x'+0}_{x=x'-0} &=0\\
    \left.\frac{v_-}{K_-}\partial_x G_- + h \partial_x G' \right|^{x=x'+0}_{x=x'-0} &=1.
\end{align}
We will tackle the first equation, and obtain the rest by analogy. Because we are interested in the $\omega\to 0$ limit, we cannot set $\omega=0$ outright. However, 
\begin{align*}
    \partial_x A_1 e^{|\omega\rho_1 x} &= \rho_1 \omega A_1 e^{\omega\rho_1 x} \\
    &= \rho_1 \omega A_1 \left(1 + \rho_1 \omega + \cdots\right) \\
    &= \rho_1 \omega A_1 + \O(\omega^2).
\end{align*}
Therefore, to lowest order in $\omega$, we obtain 
\begin{multline}
     \frac{v_+}{K_+} \left[\rho_1 A_1^+ - \rho_1A_2^+ + \rho_2 A_3^+ - \rho_2 A_4^+\right] + h\left[\rho_1 A_1'' - \rho_1 A_2'' + \rho_2 A_3'' - \rho_2 A_4''\right] \\- \frac{v_+}{K_+} \left[\rho_1 B_1^+ - \rho_1B_2^+ + \rho_2 B_3^+ - \rho_2 B_4^+\right] - h\left[\rho_1 B_1'' - \rho_1 B_2'' + \rho_2 B_3'' - \rho_2 B_4''\right] =  \frac{1}{\omega}.
\end{multline}
Using the continuity conditions (\ref{cont1a}) and (\ref{cont1b}), this becomes
\begin{equation}
    \frac{1}{K_{\infty}}A^+ + \frac{1}{K_{\infty}} B^+ = \frac{1}{\omega}.
\end{equation}
We also have from before that $A^+ = B^+$, so the propagator $G_+(x,x')=K_{\infty}/2\omega$ outside of the wire, and the same goes for $G_-$. We can look at the equations for the $G'$ and $G''$. We get that 
\begin{multline}
    \frac{v_-}{K_-} \left[\rho_1 A_1'' - \rho_1A_2'' + \rho_2 A_3'' - \rho_2 A_4''\right] + h \left[\rho_1 A_1^+ - \rho_1A_2^+ + \rho_2 A_3^+ - \rho_2 A_4^+\right]\\
    - \frac{v_-}{K_-} \left[\rho_1 B_1'' - \rho_1B_2'' + \rho_2 B_3'' - \rho_2 B_4''\right] - h \left[\rho_1 B_1^+ - \rho_1B_2^+ + \rho_2 B_3^+ - \rho_2 B_4^+\right] =0.
\end{multline}
Using the continuity conditions (\ref{cont2a}) and (\ref{cont2b}), we get that 
\begin{equation}
    \frac{1}{K_{\infty}}A'' + \frac{1}{K_{\infty}}B'' = 0
\end{equation}
And since $A'' = B''$, this implies that the propagator $G''$ (and similarly, $G'$) vanish outside the wire. At this point, we have enough information to determine the propagator inside the wire as well. Observe that as $\omega\to 0$,
\begin{equation}
    G_+(x,x',\omega\to 0) \simeq A_1^+ + A_2^+ + A_3^+ + A_4^+
\end{equation}
which, by equation (\ref{contpropag1}), implies that 
\begin{equation}
    G_+(x,x', \omega\to 0) \simeq \frac{K_{\infty}}{2\omega}
\end{equation}
In all regions of the wire. Similarly, for $G''$, by (\ref{contpropag2}), 
\begin{equation}
\begin{split}
        G''(x,x', \omega\to 0) &\simeq A_1'' + A_2'' + A_3'' + A_4''\\
        &= 0.
\end{split}
\end{equation}
The same arguments apply for $G'$ and $G_-$. Thus, we have shown that in all regions of the wire,
\begin{equation}
    \vec{G}^\text{ret}(x,x',\omega\to0) \simeq \begin{pmatrix}
        \frac{K_{\infty}}{2\omega} & 0 \\
        0 &  \frac{K_{\infty}}{2\omega}
    \end{pmatrix}.
\end{equation}
In the main text, $K_{\infty}= 4\pi g_{\infty}$, and so we obtain the desired result:
\begin{equation}
    G(\omega, x, x') \overset{\omega \to 0}{=} \frac{2\pi g_\infty}{\omega}
\end{equation}

\newpage 
\section{Schwinger-Dyson equations for exact correlation functions}
\label{dressedPropag}
In section \ref{noise}, we wish to compute various exact correlation functions of the fields. It is of interest to us to express these exact correlation functions in terms of the retarded propagators, since that allows us to readily examin their low-energy asymptotics. We can do this by employing the Schwinger-Dyson equations,
\begin{equation}
    \angled{\frac{\delta}{\delta \phi_k}\F[\phi_k]} = -i \angled{F[\phi_k]\frac{\delta}{\delta \phi_k} S[\phi_k]}
    \label{eq:SD}
\end{equation}
where $\F[\phi_k]$ is a sufficiently regular functional of a collection of fields $\phi_k$. In our following discussion, we will consider an action of the form 
\begin{equation}
    S = S_0 + S_\text{b}
\end{equation}
where $S_0$ is the kinetic terms of the field, 
\begin{equation}
    S = \frac{1}{2} \phi_\alpha D^{\alpha\beta}\phi_\beta
\end{equation}
and $S_\text{b}$ is any generic interacting term. 

\subsection{Expression for the exact backscattering current}
First, consider the case where $F = \text{const}$. Then the left-hand side of eq. \ref{eq:SD} vanishes, and so we obtain 
\begin{equation}
\begin{split}
    0 &= \angled{\frac{\delta }{\delta \phi_\beta} [S_0 + S_\text{b}]} \\
    &\Rightarrow \angled{D^{\beta\gamma} \phi_\gamma} = - \angled{\frac{\delta S_{\text{b}}}{\delta \phi_\beta}}\\
    &\Rightarrow \angled{G^{\alpha\beta} D^{\beta\gamma} \phi_\gamma} = - \angled{G^{\alpha\beta}\frac{\delta S_{\text{b}}}{\delta \phi_\beta}}
\end{split}
\end{equation}
where 
\begin{equation}
    D^{\alpha\beta}G^{\beta\gamma} = \delta^{\alpha \gamma}\delta^{(2)}(\vec{x} - \vec{x'})
\end{equation}
Integrating both sides, we obtain 
\begin{equation}
    \angled{\phi_\alpha(x,t)} = \int \d t' \d x' \, G^{\alpha\beta}(x, x',t-t')\angled{\frac{\delta S_{\text{b}}}{\delta \phi_\beta(t', x')}}.
\end{equation}
In the main text, the fields are indexed by their branch of the Keldysh contour, $\alpha, \beta = \pm$. Since we are working with the shifted charge fields $\eta_{\text{c}}$, the correlators are evaluated in equilibrium $(V = 0)$, so we may select either $\alpha = +$ or $\alpha = -$. To conform to the convention of \cite{ponNag1999}, we will choose $\alpha = -$. Noting that the $\eta_\text{c}$ and $\phi_{\text{c}}$ propagators coincide, we have 
\begin{equation}
    \angled{\eta_{\text{c}}(x,t)} = \sum_{\beta = \pm} \int \d t'  \d x' \, G^{-,\beta}(x, x',t-t')\angled{\frac{\delta S_{\text{b}}}{\delta \eta(t'^\beta, x')}}.
\end{equation}
In our case, we are interested in the backscattering operator, so we take a time derivative of the above equation to obtain
\begin{equation}
    \angled{I_{\text{b}}(t)} = -\frac{1}{2\pi}\angled{\partial_t \phi(x,t^-)} = -\frac{1}{2\pi }\sum_{\beta = \pm} \int \d t' \d x' \, \partial_t G^{-, \beta}(x, x',t-t')\angled{\frac{\delta S_{\text{b}}}{\delta \phi(t'^-, x')}}.
\end{equation}
Moreover, we have that 
\begin{equation}
    S_{\text{b}}[\phi(t,x)] \to S_{\text{b}}[\phi(t^+, x)] - S_{\text{b}}[\phi(t^-, x)]. 
\end{equation}
and that the expectation values in equilibrium are independent of the time branch. Hence, we have that
\begin{equation}
    \angled{\frac{\delta S_{\text{b}}}{\delta \eta_\text{c}(t'^+, x')}} = \angled{\frac{\delta S_{\text{b}}}{\delta \eta_\text{c}(t'^-, x')}}.
\end{equation}
This means we may collect the terms in the above equation into the retarded propagator, since $G^{\text{ret}} = G^{-,+} - G^{-,-}$. We then have
\begin{equation}
    \angled{I_{\text{b}}(t)} = -\frac{1}{2\pi}\angled{\partial_t \eta_\alpha(x,t)} = -\frac{1}{2\pi }\int \d t' \d x' \, \partial_t G^{\text{ret}}(x, x',t-t')\angled{\frac{\delta S_{\text{b}}}{\delta \eta_{\text{c}}(t', x')}}.
\end{equation}
In the zero-frequency limit, we may insert the low-energy asymptotics of the correlation functions and obtain 
\begin{equation}
    \angled{I_b(x,t)} \overset{\omega \to 0}{=} -g_\infty e  \angled{\frac{\delta S_{\text{b}}}{\delta \eta_c (t,x_0)}}.
\end{equation}
\subsection{Expression for the shot noise}
The shot noise involves a current-current correlator of the form $\angled{\partial_t \eta (t, x) \partial_t \eta (t', x')}$. We may thus consider \ref{eq:SD} when $\F = \phi_{\alpha}$:
\begin{equation}
    \begin{split}
        \angled{\F D^{\gamma \delta} \phi_\delta} &= \angled{i\frac{\delta}{\delta \phi_\gamma} \F - \F \frac{\delta S_{\text{b}}}{\delta \phi_\gamma}} \\
        & \Rightarrow \angled{\phi_\alpha D^{\gamma \delta} \phi_\delta} = i\delta_{\alpha\gamma} \delta^{(2)}(\vec{x}' - \vec{x}'') - \angled{\phi_\alpha \frac{\delta S_{\text{b}}}{\delta \phi_\gamma}}.
    \end{split}
\end{equation}
Now, we may apply the Schwinger-Dyson equation again to the last term, with $\F = \delta S_{\text{b}}/\delta \phi_\beta$, and obtain an expression for the dressed propagator in terms of bare propagators functions and of the currents. Our fields are indexed by their branch on the Keldysh contour. Noting that $ \angled{\{\eta(t,x) , \eta(t, x')\}}= \angled{\{\eta(t^+,x) , \eta(t^-, x')\}}$, we obtain
\begin{multline}
    \angled{\{\eta(t,x) , \eta(t, x')\}} =  G^{\text{K}}(t-t', x, x') \\+i \sum_{\alpha = \pm} \int \d t'' \angled{\frac{\delta^2 S_{\text{b}}}{\delta \eta_{\text{c}}^2(t', x_0)}} '\left[\alpha G^{\text{K}} + G^{\text{ret}}\right](t'-t'', x, x_0) \left[ G^{\text{K}} + \alpha G^{\text{ret}}\right] (t-t'',x',x_0)\\
    + \int \d t'' \d t''' \sum_{\alpha, \beta = \pm} \left[\alpha G^{\text{K}} + G^{\text{ret}}\right](t-t'', x, x_0) \left[\beta G^{\text{K}} + G^{\text{ret}}\right](t'-t''', x', x_0)\angled{\frac{\delta S_{\text{b}}}{\delta \eta_{\text{c}}(t'',x_0)}\frac{\delta S_{\text{b}}}{\delta \eta_{\text{c}}(t''', x_0)}}
\end{multline}
where $G^{\text{K}}$ is the Keldysh correlation function. Substituting in the zero-frequency asymptotics of the correlators, one may readily obtain the standard result \cite{ponNag1999, PhysRevB.71.165309}, viz. that the fluctuations of the total current precisely equal the fluctuations in the backscattering current:
\begin{equation}
\begin{split}
        \delta I^2 &= \frac{1}{4\pi^2}\int \d t e^{i\omega t} \angled{\{\partial_t\eta(t,x) , \partial_t\eta(t, x')\}} - 2\angled{\partial_t \eta(t,x)}\angled{\partial_t \eta(t', x')} \\ 
        &\overset{\omega, T \to 0}{=} (g_{\infty}e)^2\int \d t \, e^{i\omega t} \angled{\left\{\frac{\delta S_{\text{b}}}{\delta \eta_{\text{c}}(t)}, \frac{\delta S_{\text{b}}}{\delta \eta_{\text{c}}(0)}\right\}} - 2\angled{\frac{\delta S_{\text{b}}}{\delta \eta_{\text{c}}(t)}}\angled{\frac{\delta S_{\text{b}}}{\delta \eta_{\text{c}}(0)}}\\
        &= \delta I^2_{\text{b}}.
        \label{eq:deltaI2deltaIb2}
\end{split}
\end{equation}

\newpage
\section{\label{noiseweak}Shot noise of the weakly correlated edge.}
In this section, we demonstrate that the shot noise is indeed proportional to the average backscattering current for the backscattering term in the weakly correlated theory. We have that the total average backscattering current is given by 
\begin{align}
    \angled{I_{\text{b}}} &\overset{\omega\to 0}{=} -g_\infty e\int \d t e^{i\omega t} \angled{\frac{\delta S_{\text{b}}}{\delta \eta_{\text{c}}(t)}}\\
    &\simeq -ig_{\infty} e \int \d t \angled{\L_{\text{b}}[\eta_{c}(t) + g_{\infty}eV t]\frac{\delta S_{\text{b}}}{\delta \eta_{c}(0)}}_0
\end{align}
Where for the non-interacting theory, recall that 
\begin{equation}
    \begin{split}
        \L_{\text{b}} &= \lambda_1\cos(\eta_{\text{c}} + g_{\infty}e V t)\cos(\phi_{\text{n}+})\cos(\phi_{\text{s}+})\cos(\phi_{\text{n}-})\\
        &+ \lambda_2 \sin(\eta_{\text{c}} + g_{\infty}e V t)\sin(\phi_{\text{n}+})\sin(\phi_{\text{s}+})\sin(\phi_{\text{n}-})\\
        &+ \lambda_3 \cos(\eta_{\text{c}} + g_{\infty}e V t)\cos(\vartheta_{\text{n}+})\cos(\phi_{\text{s}+})\cos(\vartheta_{\text{n}-})\\
        &+ \lambda_4\sin(\eta_{\text{c}} + g_{\infty}e V t)\sin(\vartheta_{\text{n}+})\sin(\phi_{\text{s}+})\sin(\vartheta_{\text{n}-})\Big].
    \end{split}
\end{equation}
And so 
\begin{equation}
    \begin{split}
        \frac{\delta S_{\text{b}}}{\delta \eta_{\text{c}}(0)} &= -\lambda_1\sin(\eta_{\text{c}})\cos(\phi_{\text{n}+})\cos(\phi_{\text{s}+})\cos(\phi_{\text{n}-})\\
        &+ \lambda_2 \cos(\eta_{\text{c}})\sin(\phi_{\text{n}+})\sin(\phi_{\text{s}+})\sin(\phi_{\text{n}-})\\
        &-\lambda_3 \sin(\eta_{\text{c}})\cos(\vartheta_{\text{n}+})\cos(\phi_{\text{s}+})\cos(\vartheta_{\text{n}-})\\
        &+ \lambda_4\cos(\eta_{\text{c}})\sin(\vartheta_{\text{n}+})\sin(\phi_{\text{s}+})\sin(\vartheta_{\text{n}-})\Big].
    \end{split}
\end{equation}
Due to the charge-neutrality condition, the cross-terms between $\L_\text{b}$ and $\delta S_{\text{b}}/\delta \eta_{c}$ vanish. We can then write the backscattering current as 
\begin{multline}
    \angled{I_{\text{b}}} = i  g_{\infty} e \int \d t \  \lambda_1^2 \angled{\cos(\eta_{\text{c}}(t) + g_{\infty}e V t)\sin(\eta_{\text{c}})}_0 \A[\phi_{\text{n}+}, \phi_{\text{s}+}, \phi_{\text{n}-}](t)\\
    - \lambda_2^2\angled{\sin(\eta_{\text{c}}(t) + g_{\infty}e V t)\cos(\eta_{\text{c}})}_0  \A[\phi_{\text{n}+}, \phi_{\text{s}+}, \phi_{\text{n}-}](t)\\
    + \lambda_3^2 \angled{\cos(\eta_{\text{c}}(t) + g_{\infty}e V t)\sin(\eta_{\text{c}})}_0  \A[\vartheta_{\text{n}+}, \phi_{\text{s}+}, \phi_{\text{n}-}](t)\\
    - \lambda_4^2\angled{\sin(\eta_{\text{c}}(t) + g_{\infty}e V t)\cos(\eta_{\text{c}})}_0 \A[\vartheta_{\text{n}+}, \phi_{\text{s}+}, \theta_{\text{n}-}](t)\\
\end{multline}

where we have introduced the following shorthand for the correlator of the fields $\phi_{\text{n}+}, \phi_{\text{s}+}, \phi_{\text{n}-}$:
\begin{equation}
    \begin{split}
        \A\left[\Phi_1,\cdots, \Phi_n\right](t) &\equiv \angled{\cos(\Phi_1(t))\cos(\Phi_1(0))}_0 \cdots \angled{\cos(\Phi_n(t))\cos(\Phi_n(0))}_0\\
        &=\angled{\sin(\Phi_1(t))\sin(\Phi_1(0))}_0 \cdots \angled{\sin(\Phi_n(t))\sin(\Phi_n(0))}_0
    \end{split}
\end{equation}
Now, we may turn our attention to the correlators. It can be shown that 
\begin{equation}
    \angled{\cos(\eta_{\text{c}}(t) + g_{\infty}e V t)\sin(\eta_{\text{c}})}_0 = -\angled{\sin(\eta_{\text{c}}(t) + g_{\infty}e V t)\cos(\eta_{\text{c}})}_0 = \frac{1}{4i}\left[e^{-ig_\infty e V t}e^{G_{\text{c}}(t)} - e^{i g_{\infty}e V t}e^{G_\text{c}(t)}\right]
\end{equation}
and thus, 
\begin{equation}
    \angled{I_{\text{b}}} \overset{\omega\to 0}{=} \frac{1}{4} g_{\infty} e \int \d t \,e^{ig_\infty e V t} \sum_{m = \pm 1} m e^{G_\text{c}(mt)} \A(mt)
    \label{eq:currentfinalform}
\end{equation}

where we have defined
\begin{equation}
    \A(t) \equiv \lambda_1^2\A[\phi_{\text{n}+}, \phi_{\text{s}+}, \phi_{\text{n}-}](t)  + \lambda_2^2\A[\phi_{\text{n}+}, \phi_{\text{s}+}, \vartheta_{\text{n}-}](t) + \lambda_3^2 \A[\vartheta_{\text{n}+}, \phi_{\text{s}+}, \vartheta_{\text{n}-}](t) + \lambda_4^2 \A[\vartheta_{\text{n}+}, \phi_{\text{s}+}, \phi_{\text{n}-}](t) 
\end{equation}

\subsection{Shot noise}
In order to compute the shot noise $\delta I^2$, we may use equation \ref{eq:deltaI2deltaIb2}. We have that at order $\lambda^2$,
\begin{equation}
    \delta I^2 \overset{\omega, T\to 0}{=} (g_\infty e)^2\int \d t \, e^{i\omega t} \angled{\left\{\frac{\delta S_{\text{b}}}{\delta \eta_{\text{c}}(t)}, \frac{\delta S_{\text{b}}}{\delta \eta_{\text{c}}(0)}\right\}}_0
\end{equation}
Hence,
\begin{equation}
    \begin{split}
            \delta I^2 &\overset{\omega, T\to 0}{=} (g_{\infty}e)^2 \int \d t \ \sum_{m = \pm1}\!\begin{multlined}[t]
            \lambda_1^2 \angled{\sin\left(\eta_{\text{c}}(mt) + g_{\infty}e V t\right)\sin\left(\eta_{\text{c}}(0)\right)}_0 \A [\phi_{\text{n}+}, \phi_{\text{s},+}, \phi_{\text{n},-} ](mt)\\
            + \lambda_2^2 \angled{\cos\left(\eta_{\text{c}}(mt) + g_{\infty}e V t\right)\cos\left(\eta_{\text{c}}(0)\right)}_0 \A [\phi_{\text{n}+}, \phi_{\text{s},+}, \phi_{\text{n},-} ](mt)\\
            + \lambda_3^2 \angled{\sin\left(\eta_{\text{c}}(mt) + g_{\infty}e V t\right)\sin\left(\eta_{\text{c}}(0)\right)}_0 \A [\vartheta_{\text{n}+}, \phi_{\text{s},+}, \vartheta_{\text{n}-} ](mt)\\
            + \lambda_4^2 \angled{\cos\left(\eta_{\text{c}}(mt) + g_{\infty}e V t\right)\cos\left(\eta_{\text{c}}(0)\right)}_0 \A [\vartheta_{\text{n}+}, \phi_{\text{s},+}, \vartheta_{\text{n}-} ](mt)\\
     \end{multlined}\\
     &= (g_{\infty}e)^2 \int \d t \sum_{m = \pm1} 
     \angled{\sin\left(\eta_{\text{c}}(mt) +g_{\infty}e V  t\right)\sin\left(\eta_{\text{c}}(0)\right)}_0 \times \A(mt)\\
     &= \frac{(g_{\infty}e)^2 }{4} \int \d t \sum_{m = \pm}
         \angled{e^{ig_{\infty}e Vmt} e^{i\eta_{\text{c}}(mt)} e^{-i\eta_{\text{c}}(0)} + e^{-ig_{\infty}e V mt} e^{-i\eta_{\text{c}}(mt)} e^{i\eta_{\text{c}}(0)}} \times \A(mt)\\
     & = \frac{(g_{\infty}e)^2 }{4} \int \d t \left(e^{i g_{\infty}e Vt} + e^{-ig_{\infty}e V t}\right) \sum_{m=\pm1} e^{G(mt)}\A(mt) \\
     &=  \frac{(g_{\infty}e)^2 }{4} \sum_{q = \pm1} \int \d t\  e^{i qg_{\infty}e V t}  \sum_{m=\pm1} e^{G(mt)}\A(mt)\\
    \end{split}
\end{equation}
 At this stage, we notice that the sum over $m$ amounts to a symmetrized correlation function. Therefore, we can employ the fluctuation-dissipation theorem, which holds in equilibrium, to turn it into an anti-symmetrized correlation function:
\begin{equation}
    \begin{split}
        \delta I^2 &= \frac{(g_{\infty}e)^2 }{4} \sum_{q = \pm1} \coth\left(\frac{qg_{\infty}e V}{2\kb T}\right)\int \d t\  e^{i qg_{\infty}e Vt}  \sum_{m=\pm1}m e^{G(mt)} \A(mt)\\
        &\overset{T\to 0}{=} \frac{(g_{\infty}e)^2 }{4} \sum_{q = \pm1} q \int \d t\  e^{i qg_{\infty}e V t}  \sum_{m=\pm}m e^{G(mt)}\A(mt)\\
    \end{split}
\end{equation}
Changing variables from $t\to -t$ in the second term of the $q$-sum yields
\begin{equation}
    \begin{split}
          \delta I^2 & = \frac{(g_{\infty}e)^2 }{2} \int \d t\ e^{ig_{\infty}e Vt}  \sum_{m=\pm1}m e^{G(mt)}\A(mt)
    \end{split}
    \label{eq:noisefinalform}
\end{equation}

By comparing our expressions for noise and current (eqs. \ref{eq:noisefinalform} and \ref{eq:currentfinalform}), we see that 
\begin{equation}
    \begin{split}
        \delta I^2 &= 2 g_{\infty}e \angled{I_{\text{b}}}.
    \end{split}
\end{equation}

\newpage
\section{\label{shotnoiseintedgesection}Shot noise of the interacting edge}
As before, 
\begin{equation}
        \angled{I_{\text{b}}} \overset{\omega\to 0}{\simeq} -ig_{\infty} e \int \d t \angled{\L_{\text{b}}[\eta_{c}(t) + g_{\infty}eV t]\frac{\delta S_{\text{b}}}{\delta \eta_{c}(0)}}_0
\end{equation}

For the interacting edge, we have that 
\begin{equation}
\begin{split}
    \L_\text{b} &= \lambda_2 \cos(2\eta_\text{c}(t) + 2g_{\infty}eVt) + \lambda_2' \cos(2\eta_\text{c}(t) + 2g_{\infty}eVt) \cos(2\phi_\text{s}(t))\\
    \frac{\delta S_{\text{b}}}{\delta \eta_{\text{c}}(0)} &= -2\left(\lambda_2\sin(2\eta_\text{c}(0)) + \lambda_2'\sin(2\eta_\text{c}(0) ) \cos(2\phi_\text{s}(0))\right)
\end{split}
\end{equation}
Hence, their product is 
\begin{multline}
        \angled{\L_{\text{b}}[\eta_{\text{c}}(t) + g_{\infty}eVt] \ \frac{\delta S_{\text{b}}}{\delta \eta_{\text{c}}(0)}}_0 \\= -2 \Big\langle \lambda_2^2\cos(2\eta_\text{c}(t) + 2g_{\infty}eVt) \sin(2\eta_{\text{c}}(0)) + \lambda_2 \lambda_2'\cos(2\eta_\text{c}(t) + 2g_{\infty}eVt) \sin(2\eta_{\text{c}}(0))\cos(2\phi_\text{s}) \\ + \lambda_2 \lambda_2'\cos(2\eta_\text{c}(t) + 2g_{\infty}eVt) \sin(2\eta_{\text{c}}(0))\cos(2\phi_\text{s})\\
        + \lambda_2'^2\cos(2\eta_\text{c}(t) + 2g_{\infty}eVt) \sin(2\eta_{\text{c}}(0))\cos(2\phi_\text{s})\cos(2\phi_\text{s})\Big\rangle_0
\end{multline}
The middle two terms vanish due to the charge-neutrality condition, since they have a dangling cosine term. Therefore, we may simplify our expression: 
\begin{multline}
        \angled{\L_{\text{b}}[\eta_{\text{c}}(t) + g_{\infty}eVt] \ \frac{\delta S_{\text{b}}}{\delta \eta_{\text{c}}(0)}}_0 = - 2 \Big\langle\lambda_2^2\cos(2\eta_\text{c}(t) + 2g_{\infty}eVt) \sin(2\eta_{\text{c}}(0)) \\
        + \lambda_2'^2\cos(2\eta_\text{c}(t) + 2g_{\infty}eVt) \sin(2\eta_{\text{c}}(0))\cos(2\phi_\text{s})\cos(2\phi_\text{s})\Big\rangle_0
\end{multline}
Proceeding as in the previous section, we find that 

\begin{equation}
        \angled{I_{\text{b}}} \overset{\omega\to 0}{=} \frac{g_{\infty} e }{2}\int \d t e^{2ig_\infty e V t} \sum_{m=\pm1}me^{2G(mt)}\left(\lambda_2^2 + \lambda_2'^2\A[2\phi_{\text{s}}](mt)\right)
        \label{eq:currentfinalform2}
\end{equation}

\subsection{Shot noise}
We will now compute the shot noise at zero temperature and zero frequency, utilizing eq. \ref{eq:deltaI2deltaIb2}:
\begin{equation}
    \begin{split}
        \delta I^2 
        &\overset{\omega\to 0}{=} 4 (g_{\infty}e)^2 \int \d t \sum_{m = \pm} \angled{\sin\left(2\phi_{\text{c}}(mt)\right)\sin\left(2\phi_{\text{c}}(0)\right)}_0 \left(\lambda_2^2 + \lambda_2'^2\A[2\phi_{\text{s}}](mt)\right)\\
        &= (g_{\infty}e)^2 \sum_{q=\pm} \int \d t\, e^{2iq g_{\infty }e V t}\sum_{m=\pm} e^{2G(mt)} \left(\lambda_2^2 + \lambda_2'^2\A[2\phi_{\text{s}}](mt)\right)\\
        &=  (g_{\infty}e)^2 \sum_{q = \pm} \coth\left(\frac{2qg_{\infty }e V }{2\kb T}\right)\int \d t \ e^{2iq g_{\infty }e V t}\sum_{m=\pm} m e^{2G(mt)}\left(\lambda_2^2 + \lambda_2'^2\A[2\phi_{\text{s}}](mt)\right)\\
        & \overset{T\to 0}{=} (g_{\infty}e)^2 \sum_{q = \pm} q\int \d t \ e^{2iq g_\infty e Vt}\sum_{m=\pm} m e^{2G(mt)}\left(\lambda_2^2 + \lambda_2'^2\A[2\phi_{\text{s}}](mt)\right)\\
        &= 2 g_{\infty}^2 e^2 \int \d t \, e^{2ig_\infty e t}\sum_{m=\pm} m e^{2G(mt)}\left(\lambda_2^2 + \lambda_2'^2\A[2\phi_{\text{s}}](mt)\right)\\ 
    \end{split}
\end{equation}
We can compare this to \ref{eq:currentfinalform2}, which establishes the desired result:
\begin{equation}
    \delta I^2 \overset{\omega \to 0}{=} 4 g_{\infty} e \angled{I_{\text{b}}}.
\end{equation}

\end{document}